\documentclass[11pt]{cernrep}
\usepackage{graphicx}
\usepackage{here}
\usepackage{latexsym}
\usepackage{axodraw}
\usepackage{amssymb}
\usepackage{hyperref}
\usepackage{tabularx}
\usepackage[vcentermath]{youngtab}

\makeatletter
\newcommand\numberwithin[2]{\@addtoreset{#1}{#2}}
\makeatother

\numberwithin{equation}{section}

\def\url#1{\mbox{\href{#1}{\sf #1}}}

\newcommand{\ie}{{\em i.e.}}

\newcommand{\gev}{\hbox{ GeV}}

\newcommand{\mev}{\hbox{ MeV}}
\newcommand{\mevcc}{\hbox{ MeV}\!/\!c^2}
\newcommand{\tev}{\hbox{ TeV}}

\newcommand{\cm}{\hbox{ cm}}
\newcommand{\mm}{\hbox{ mm}}
\newcommand{\nb}{\hbox{ nb}}

\newcommand{\eqn}[1]{(\ref{#1})}

\newcommand{\repr}[3]{\ensuremath{(\mathbf{#1},\mathbf{#2})_{#3}}}

\DeclareMathSymbol{\lll}          {\mathrel}{AMSa}{"6E}

\def\tr#1{\mathrm{tr}#1}

\newcommand{\M}{\ensuremath{\mathcal{M}}}
\newcommand{\D}{\ensuremath{\mathcal{D}}}

\newcommand{\nucl}[2]{\mbox{\ensuremath{^{#1}\mathrm{#2}}}}

\def\ltap{\mathop{\raisebox{-.4ex}{\rlap{$\sim$}} 
\raisebox{.4ex}{$<$}}}
\def\gtap{\mathop{\raisebox{-.4ex}{\rlap{$\sim$}} 
\raisebox{.4ex}{$>$}}}

\newcommand{\cfrac}[2]{\textstyle \frac{#1}{#2}}

\newcommand{\m}{\hbox{ m}}

\def\half{{\scriptstyle \frac{1}{2}}}     

\newcommand{\smgg}{\ensuremath{\mathrm{SU(3)_c} \otimes \mathrm{SU(2)}_{\mathrm{L}} \otimes \mathrm{U(1)}_Y}}
\newcommand{\ewgg}{\ensuremath{\mathrm{SU(2)}_{\mathrm{L}} \otimes \mathrm{U(1)}_Y}}
\newcommand{\suf}{\ensuremath{\mathrm{SU(5)}}}
\newcommand{\onetev}{1-TeV scale}

\def\vev#1{\langle #1\rangle_0}
\def\abs#1{\left| #1\right|}

\def\slashii#1{\setbox0=\hbox{$#1$}             
   \dimen0=\wd0                                 
   \setbox1=\hbox{\sl/} \dimen1=\wd1            
   \ifdim\dimen0>\dimen1                        
      \rlap{\hbox to \dimen0{\hfil\sl/\hfil}}   
      #1                                        
   \else                                        
      \rlap{\hbox to \dimen1{\hfil$#1$\hfil}}   
      \hbox{\sl/}                               
   \fi}                                         %
\def\slashiii#1{\setbox0=\hbox{$#1$}#1\hskip-\wd0\hbox to\wd0{\hss\sl/\/\hss}}
\def\slashiv#1{#1\llap{\sl/}}

\newtheorem{problem}{Problem}

\newcounter{bean}
\newenvironment{questions}[1]
{\bigskip\centerline{\underline{#1 Harvest of Questions}}\smallskip\begin{list}
    {Q--\arabic{bean}}{\usecounter{bean}\setlength{\rightmargin}{\leftmargin}}}
{\end{list}}

\renewcommand{\topmargin}{-2cm}
\begin{document}
 \title{BEYOND THE STANDARD MODEL IN MANY 
 DIRECTIONS\footnote{\phantom{X}Lectures presented at the 2003 Latin-American School of
 High-Energy Physics, San Miguel Regla (Hidalgo), Mexico. Slides and 
 animations are available at \url{http://boudin.fnal.gov/CQSanMiguel.tgz}.
 \hfill \fbox{\textsf{FERMILAB-Conf-04/049-T}}}}
\author{Chris Quigg}
\institute{Fermi National Accelerator Laboratory \\ P.O. Box 500, 
Batavia, Illinois 60510 USA}
\maketitle
\begin{abstract}
These four lectures constitute a gentle introduction to what may lie
beyond the standard model of quarks and leptons interacting through 
\smgg\ gauge bosons, prepared for an audience of 
graduate students in experimental particle physics. In
the first lecture, I introduce  a novel graphical representation of 
the particles and interactions, \textit{the double simplex,} to elicit questions 
that motivate our interest in physics beyond the standard model, 
without recourse to equations and formalism. Lecture~2 is devoted to 
a short review of the current status of the standard model, 
especially the electroweak theory, which serves as the point of 
departure for our explorations. The third lecture is concerned with 
unified theories of the strong, weak, and electromagnetic 
interactions. In the fourth lecture, I survey some attempts to extend 
and complete the electroweak theory, emphasizing some of the promise 
and challenges of supersymmetry. A short concluding section looks 
forward.
\end{abstract}

\section{QUESTIONS, QUESTIONS, QUESTIONS \label{sec:questions}}

When I told my colleague Andreas Kronfeld that I intended to begin 
this course of lectures by posing many questions, he agreed 
enthusiastically, saying, ``A summer school should provide a
lifetime of homework!'' I am sure that his comment is true for the 
lecturers, and I hope that it will be true for the students at this 
CERN--CLAF school as well.

These are revolutionary times for particle physics. Many enduring 
questions, including
$\Box$ Why are there atoms?
$\Box$ Why chemistry?
$\Box$	Why complex structures?
$\Box$	Why is our world the way it is?
$\Box$	Why is life possible?
are coming within the reach of our science. The answers will be landmarks 
in our understanding of nature. We should never forget that science is 
not the veneration of a corpus of approved knowledge. Science is organic, 
tentative; over time more and more questions enter the realm of 
scientific inquiry.

\subsection{A Decade of Discovery Past\label{subsec:discovery}}
We particle physicists are impatient and ambitious people, and so we tend to regard 
the decade just past as one of consolidation, as opposed to stunning 
breakthroughs. But a look at the headlines of the past ten years gives 
us a very impressive list of discoveries.

\begin{itemize}
    \item[$\rhd$] The electroweak theory has been elevated from a very 
    promising description to a \textit{law of nature.} This 
    achievement is truly the work of many hands; it has involved 
    experiments at the $Z^{0}$ pole, the study of $e^{+}e^{-}$, 
    $\bar{p}p$, and $\nu N$ interactions, and supremely precise 
    measurements such as the determination of $(g-2)_{\mu}$.
    
    \item[$\rhd$] Electroweak experiments have observed what we may 
    reasonably interpret as the influence of the Higgs boson in the 
    vacuum.
    
    \item[$\rhd$] Experiments using neutrinos generated by cosmic-ray 
    interactions in the atmosphere, by nuclear fusion in the Sun, and 
    by nuclear fission in reactors, have established neutrino flavor 
    oscillations: $\nu_{\mu} \to \nu_{\tau}$ and  $\nu_{e} \to 
    \nu_{\mu}/\nu_{\tau}$. 
    
    \item[$\rhd$] Aided by experiments on heavy quarks, studies of 
    $Z^{0}$,  investigations of high-energy $\bar{p}p$, $\nu N$, and $ep$ 
    collisions, and by developments in lattice field theory, we have 
    made remarkable strides in understanding quantum chromodynamics
    as the theory of the strong interactions.
    
    \item[$\rhd$] The top quark, a remarkable apparently elementary 
    fermion with the mass of an osmium atom, was discovered in 
    $\bar{p}p$ collisions.
    
    \item[$\rhd$] Direct $\mathcal{CP}$ violation has been observed in $K \to \pi\pi$ decay. 
    
    \item[$\rhd$] Experiments at asymmetric-energy $e^{+}e^{-} \to 
    B\bar{B}$ factories have established that $B^{0}$-meson decays do 
    not respect $\mathcal{CP}$ invariance.
    
    \item[$\rhd$] The study of type-Ia supernovae and detailed thermal 
    maps of the cosmic microwave background reveal that we live in a 
    flat universe dominated by dark matter and energy.
    
    \item[$\rhd$] A ``three-neutrino'' experiment has detected the 
    interactions of tau neutrinos.
    
    \item[$\rhd$] Many experiments, mainly those at the highest-energy 
    colliders, indicate that quarks and leptons are structureless on 
    the \onetev.
\end{itemize}

We have learned an impressive amount in ten years, and I find quite 
striking the diversity of experimental and observational approaches 
that have brought us new knowledge, as well as the richness of the 
interplay between theory and experiment. Let us turn now to the way 
the quark--lepton--gauge-symmetry revolution has taught us to view the 
world.

\subsection{How the world is made\label{subsec:how}}

\begin{figure}[b!]
\begin{center}
\includegraphics[width=6.3cm]{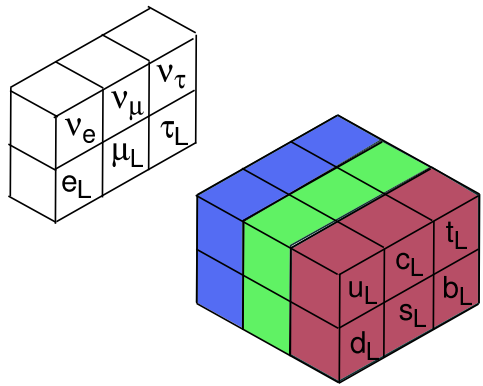}
\caption{The left-handed doublets of quarks and leptons that inspire 
the structure of the electroweak theory. \label{fig:DumbL}}
\end{center}
\end{figure}

Our picture of matter is based on the recognition of a set of pointlike 
constituents: the quarks,
\begin{equation}
\left(
		\begin{array}{c}
			u  \\
			d
		\end{array}
		 \right)_{\mathrm{L}} \;\;\;\;\;\;
		\left(
		\begin{array}{c}
			c  \\
			s
		\end{array}
		 \right)_{\mathrm{L}} \;\;\;\;\;\;
		\left(
		\begin{array}{c}
			t  \\
			b
		\end{array}
		 \right)_{\mathrm{L}}	\;,
		 \label{eq:quarks}
	\end{equation}	
and the leptons,
\begin{equation}
\left(
		\begin{array}{c}
			\nu_{e}  \\
			e^{-}
		\end{array}
		 \right)_{\mathrm{L}} \;\;\;\;\;\;
		\left(
		\begin{array}{c}
			\nu_{\mu}  \\
			\mu^{-}
		\end{array}
		 \right)_{\mathrm{L}} \;\;\;\;\;\;
		\left(
		\begin{array}{c}
			\nu_{\tau}  \\
			\tau^{-}
		\end{array}
		\right)_{\mathrm{L}}	\;,
		\label{eq:leptons}
	\end{equation}
as depicted in Figure~\ref{fig:DumbL}, 
and a few fundamental forces derived from gauge symmetries. The quarks 
are influenced by the strong interaction, and so carry \textit{color}, 
the strong-interaction charge, whereas the leptons do not feel the 
strong interaction, and are colorless. By pointlike, we understand 
that the quarks and leptons show no evidence of internal structure at 
the current limit of our resolution,  ($r \ltap 10^{-18}\m$).

\begin{figure}[t!]
\begin{center}
\includegraphics[width=15cm]{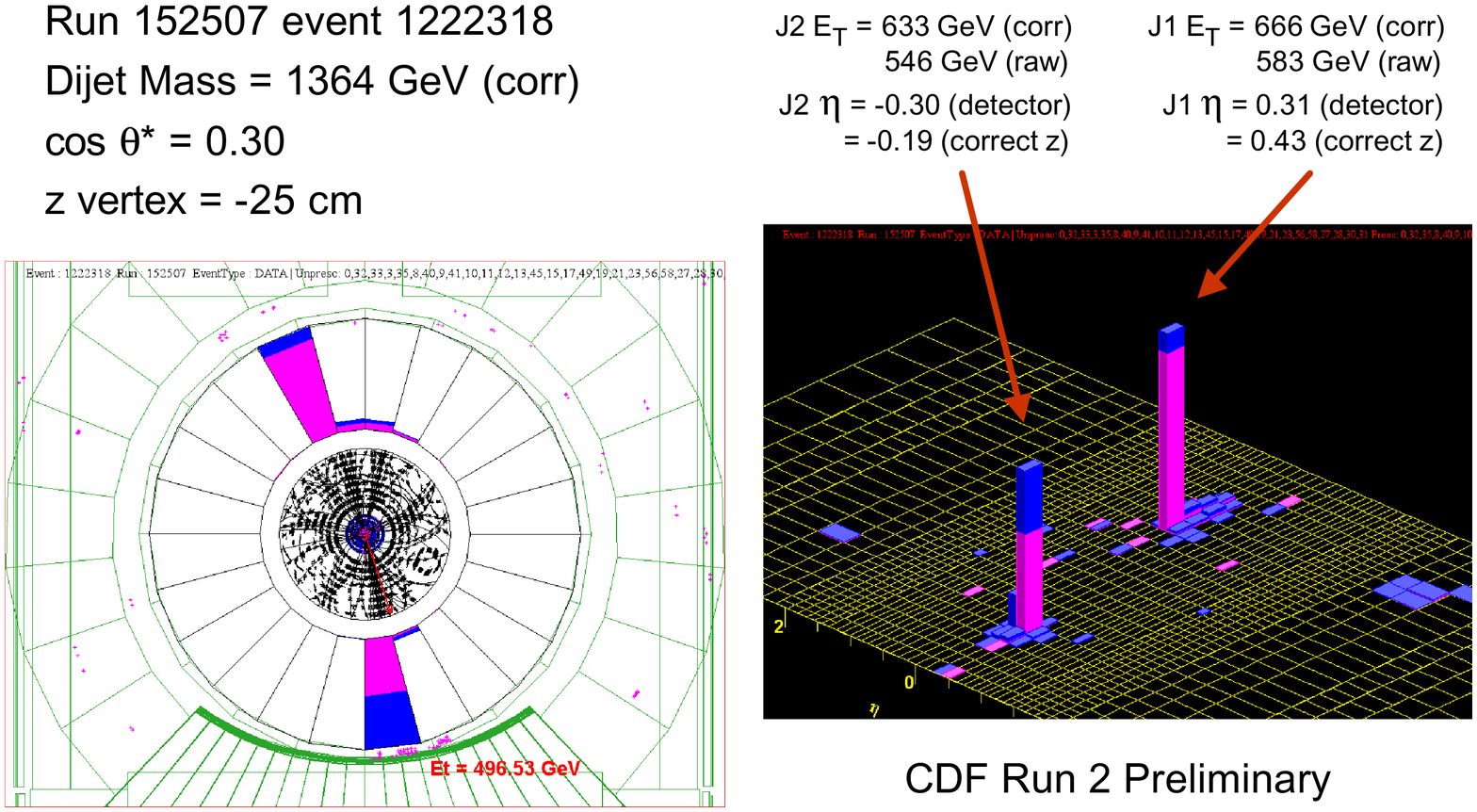}
\caption{A Tevatron Collider event with $1364\gev$ of transverse 
energy, recorded in the CDF detector. The left panel shows an end view 
of the detector, with tracking chambers at the center and calorimeter 
segments at medium and large radii. The right panel shows the 
Lego$^{\mathrm{TM}}$ 
plot of energy deposited in cells of the cylindrical detector, 
unrolled. See Ref.~\cite{Gallinaro:2003qr}.\label{fig:CDF1364}}
\end{center}
\end{figure}

The notion that the quarks and leptons are elementary---structureless 
and indivisible---is necessarily provisional. \textit{Elementarity} 
is one of the aspects of our picture of matter that we test ever more 
stringently as we improve the resolution with which we can examine 
the quarks and leptons. For the moment, the world's most powerful 
microscope is the Tevatron Collider at Fermilab, where collisions of 
980-GeV protons with 980-GeV antiprotons are studied in the CDF and 
D\O\ detectors. The most spectacular collision recorded so far, which 
is to say the closest look humans have ever had at anything, is the 
CDF two-jet event shown in Figure~\ref{fig:CDF1364}.
This event almost certainly corresponds to the collision of a quark 
from the proton with an antiquark from the antiproton. Remarkably, 
70\% of the energy carried into the collision by proton and 
antiproton emerges perpendicular to the incident beams. 
At a given transverse energy $E_{\perp}$, we may roughly estimate 
the resolution as $r \approx (\hbar c)/E_{\perp} \approx 2 \times 
10^{-19}\tev\m/E_{\perp}$. \footnote{See the note on ``Searches for Quark and 
Lepton Compositeness on p.~935 of Ref.~\cite{PDBook} for a more detailed 
discussion.}

\begin{figure}[bth]
\begin{center}
\includegraphics[width=10cm]{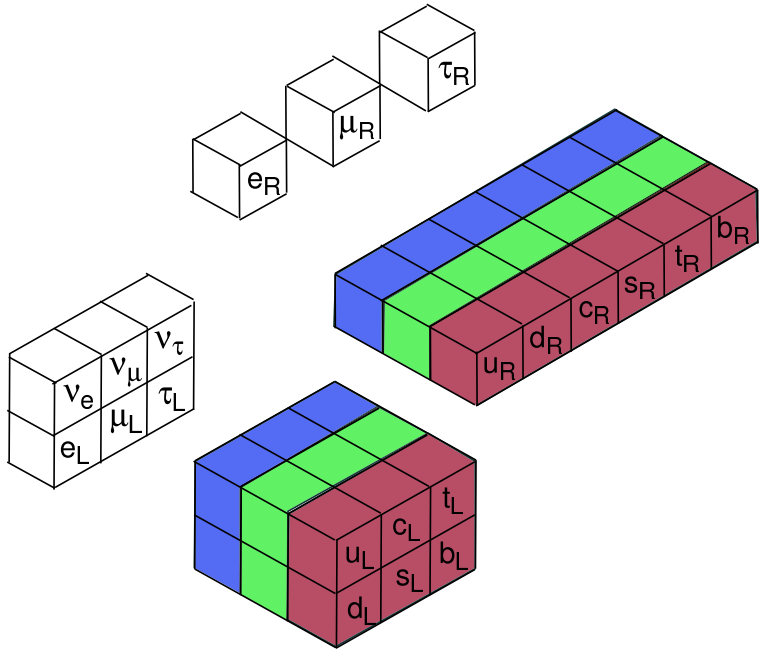}
\caption{The left-handed doublets and right-handed singlets of quarks 
and leptons.\label{fig:DumbSM}}
\end{center}
\end{figure}

Looking a little more closely at the constituents of matter, we find 
that our world is not as neat as the simple cartoon vision of 
Figure~\ref{fig:DumbL}. The left-handed and right-handed fermions 
behave very differently under the influence of the charged-current weak 
interactions. A more complete picture is given in Figure~\ref{fig:DumbSM}.
This figure represents the way we looked at the world before the 
discovery of neutrino oscillations that require neutrino mass and 
almost surely imply the existence of right-handed neutrinos. Neutrinos 
aside, the striking fact is the asymmetry between left-handed fermion 
doublets and right-handed fermion singlets, which is manifested in 
\textit{parity violation} in the charged-current weak interactions. 
What does this distinction mean?

All of us in San Miguel Regla have learned about parity violation at school, 
but it came as a stunning surprise to our scientific ancestors. In 
1956, Wu and collaborators~\cite{Wu:1957my} studied the $\beta$-decay 
$\nucl{60}{Co} \to \nucl{60}{Ni} \;e^{-} \bar{\nu}_{e}$ and observed a correlation 
between the direction $\hat{p}_{e}$ of the
outgoing electron and the spin vector $\vec{J}$ of the polarized
$^{60}\mathrm{Co}$  nucleus. Spatial reflection, or parity, leaves 
the (axial vector) spin unchanged, $\mathcal{P}: \vec{J} \to \vec{J}$, 
but reverses the electron direction, $\mathcal{P}: \hat{p}_{e} \to - 
\hat{p}_{e}$. Accordingly, the correlation $\vec{J} \cdot \hat{p}_{e}$ is 
manifestly \textit{parity violating}. 
Experiments in the late 1950s established that (charged-current) weak 
interactions are left-handed, and motivated the construction of a 
manifestly parity-violating theory of the weak interactions with 
only a left-handed neutrino $\nu_{\mathrm{L}}$. The left-handed doublets are an 
important element of the electroweak theory that 
I will review in Lecture~\ref{sec:electroweak}

Perhaps our familiarity with parity violation in the weak 
interactions has dulled our senses a bit. It seems to me that nature's 
broken mirror---the distinction between left-handed and right-handed 
fermions---qualifies as one of the great mysteries. Even if we will 
not get to the bottom of this mystery next week or next year, it 
should be prominent in our consciousness---and among the goals we 
present to others as the aspirations of our science.

There is more to our understanding of the world than
Figure~\ref{fig:DumbSM} reveals. The electroweak gauge symmetry is hidden,
$\ewgg \rightarrow \mathrm{U(1)_{em}}.$ If it were not, the world 
would be very different:
$\Box$~All the quarks and leptons would be massless and move at the 
speed of light.
$\Box$ Electromagnetism as we know it would not exist, but there would 
be a long-range hypercharge force.
$\Box$ The strong interaction, QCD, would confine quarks and generate 
baryon masses roughly as we know them.
$\Box$ The Bohr radius of ``atoms'' consisting of an electron or 
neutrino attracted by the hypercharge interaction to the nucleons would 
be infinite.
$\Box$ Beta decay, inhibited in our world by the great mass of the 
$W$ boson, would not be weak.
$\Box$ The unbroken $\mathrm{SU(2)}_{\mathrm{L}}$ interaction would confine 
objects that carry weak isospin. 

It is fair to say that electroweak symmetry breaking shapes our world! 
In fact, when we take into account every aspect of the influence of 
the strong interactions, the analysis of how the world would be is 
very subtle and fascinating. Please take time to think about
\begin{problem}
    What would the everyday world be like if the \ewgg\
	    electroweak symmetry were exact? Consider the effects of all 
	    of the  
	    \smgg\ gauge interactions.
\end{problem}
\subsection{Toward the double simplex \label{subsec:DS}}
We have seen that both quarks and leptons are spin-$\cfrac{1}{2}$,
pointlike fermions that occur in $\mathrm{SU(2)}_{\mathrm{L}}$ doublets.  The
obvious difference is that quarks carry $\mathrm{SU(3)_{c}}$ color
charge whereas leptons do not, so we could imagine that quarks and
leptons are simply distinct and unrelated species.  But we have reason
to believe otherwise.  The proton's electric charge very closely
balances the electron's, $(Q_{p}+Q_{e})/e < 10^{-21}$ \cite{PDBook},
suggesting that there must be a link between protons---hence,
quarks---and electrons---hence, leptons.  Moreover, quarks and leptons are
required, in matched pairs, for the electroweak theory to be 
anomaly-free, so that quantum corrections respect the symmetries on 
which the theory is based. Before we examine the connection between 
quarks and leptons, take a moment to consider the implications of 
ordinary matter that is not exactly neutral:

\begin{problem}
    How large would the imbalance between proton and electron charges 
    need to be for the resulting electrostatic repulsion of 
    un-ionized (\textit{nearly}  neutral) hydrogen atoms to account for 
    the expansion of the Universe? To make your estimate, compare the 
    electrostatic repulsion with the gravitational attraction of two 
    hydrogen atoms. See Ref.~\cite{LitBon}.
\end{problem}

\begin{figure}[t!]
\begin{center}
\includegraphics[width=4cm]{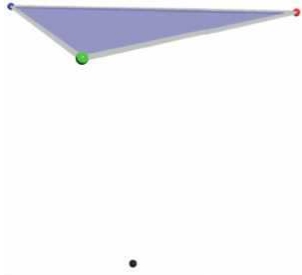}
\caption{A color triplet of quarks and a color singlet lepton, 
arrayed to explore lepton number as a fourth 
color. \label{fig:LLQ}}
\end{center}
\end{figure}

It is fruitful to display the color-triplet red, green, and blue quarks 
in the equilateral triangle weight diagram for the \textbf{3} 
representation of $\mathrm{SU(3)}_{\mathrm{c}}$, as shown in 
Figure~\ref{fig:LLQ}. There I have filled in the plane between them to 
indicate the transitions mediated by gluons.
The equality of proton and (anti)electron charges and the need to 
cancel anomalies in the electroweak theory suggest that we join the 
quarks and leptons in an extended family, or multiplet. Pati and 
Salam~\cite{Pati:1974yy} provided an apt metaphor when they proposed that we 
regard lepton number as a fourth color. To explore that possibility, 
I have placed the lepton in Figure~\ref{fig:LLQ} at the apex of a 
tetrahedron that corresponds to the fundamental \textbf{4} 
representation of $\mathrm{SU(4)}$.

\begin{figure}[b!]
\begin{center}
\includegraphics[width=4cm]{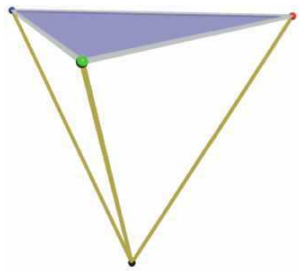}
\caption{Tetrahedron representing the \textbf{4} representation of 
$\mathrm{SU(4)}$, showing the hypothetical leptoquark transitions. 
\label{fig:LQtrans}}
\end{center}
\end{figure}

If $\mathrm{SU(4)}$ is not merely a useful \textit{classification 
symmetry} for the quarks and leptons, but a \textit{gauge symmetry,} 
then there must be new interactions that transform quarks into 
leptons, as indicated by the gold lines in Figure~\ref{fig:LQtrans}. 
If leptoquark transitions exist, they can mediate reactions that 
change baryon and lepton number, such as proton decay. The long 
proton lifetime~\cite{PDBook} tells us that, if leptoquark transitions 
do exist, they must be far weaker than the strong, weak, and 
electromagnetic interactions of the standard model. What accounts for 
the feebleness of leptoquark transitions?

\begin{figure}[t!]
\begin{center}
    \includegraphics[width=4.7cm]{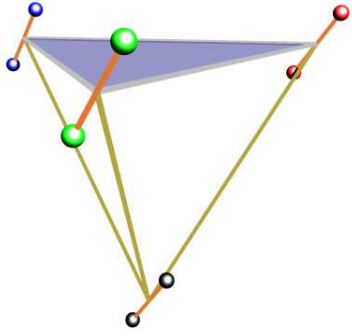}
\caption{The $\mathrm{SU(4)}$ tetrahedron, decorated with left-handed 
fermions.  \label{fig:LHdec}}
\end{center}
\end{figure}

Our world isn't built of a single quark flavor and a single lepton 
flavor. The left-handed quark and lepton doublets offer a key clue to 
the structure of the weak interactions. We can represent the $(u_{\mathrm{L}}, 
d_{\mathrm{L}})$ and $(\nu_{\mathrm{L}}, e_{\mathrm{L}})$ doublets by decorating the tetrahedron, as 
shown in Figure~\ref{fig:LHdec}. 
The orange stalks connecting $u_{\mathrm{L}} \leftrightarrow d_{\mathrm{L}}$ 
and $\nu_{\mathrm{L}} 
\leftrightarrow e_{\mathrm{L}}$ represent the $W$-bosons that 
mediate the charged-current weak interactions.

\begin{figure}[b!]
\begin{center}
\includegraphics[width=4.6cm]{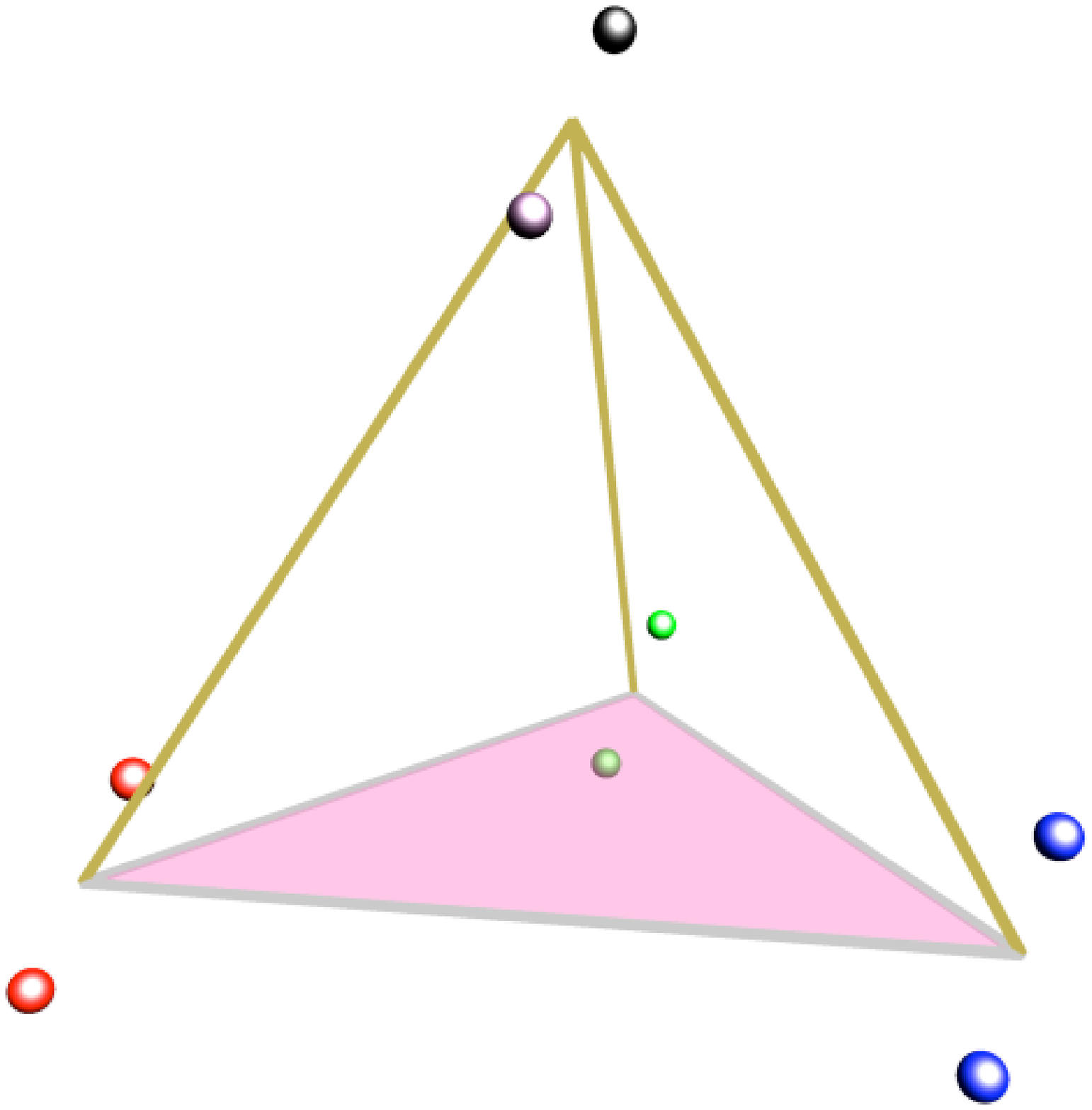}
\qquad\qquad
\includegraphics[width=4.6cm]{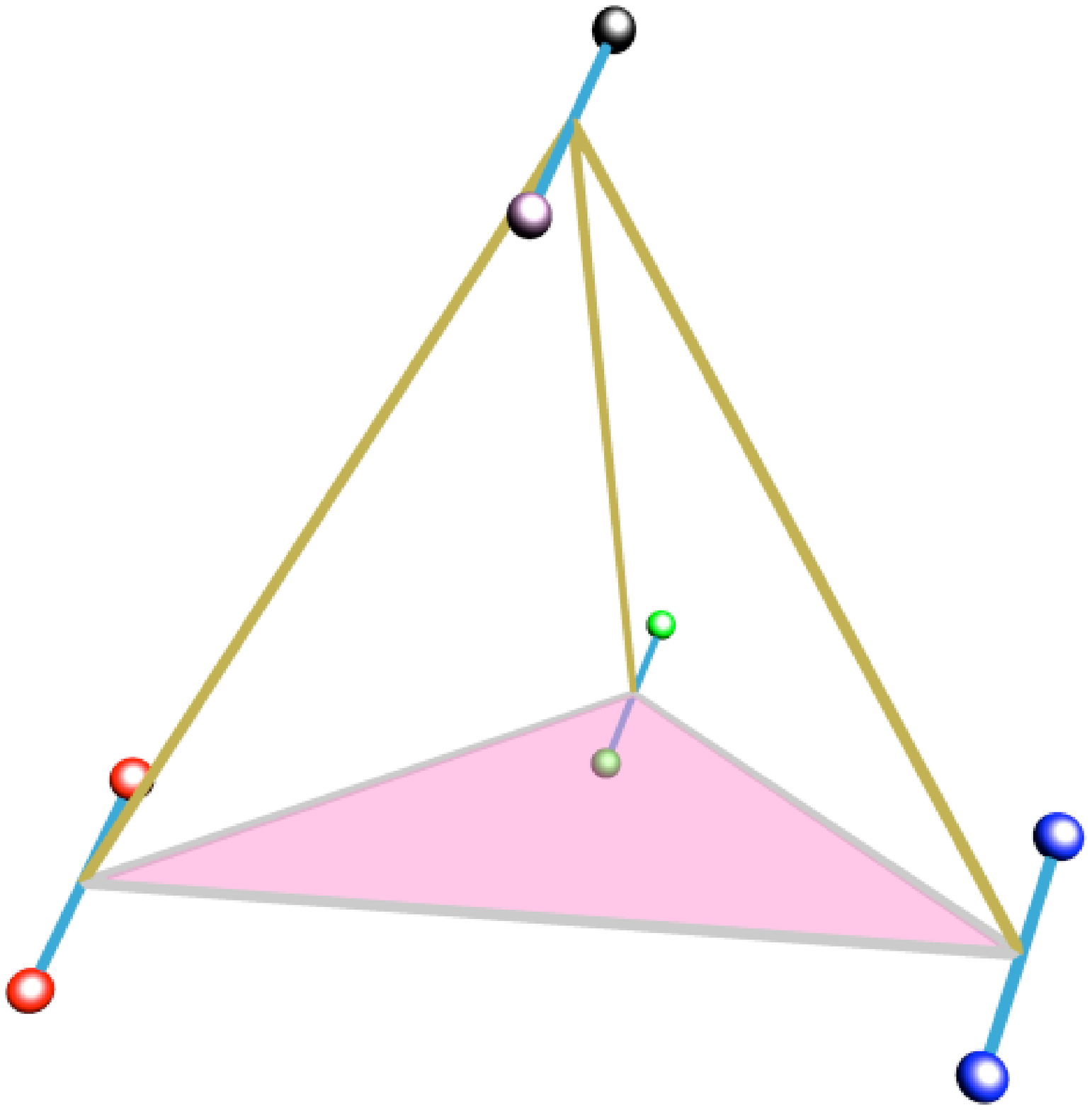}
\caption{The inverted tetrahedron, decorated with right-handed 
quark $(d_{\mathrm{R}},u_{\mathrm{R}})$ and lepton $(e_{\mathrm{R}}, \nu_{\mathrm{R}})$ pairs. The left 
panel depicts our current understanding, without right-handed charged 
currents; the right panel shows how a $W_{\mathrm{R}}$-boson could be 
added.\label{fig:RHdec}}
\end{center}
\end{figure}

What about the right-handed fermions? In quantum field theory, it is 
equivalent to talk about left-handed \textit{antifermions.} That 
observation motivates me to display the right-handed quarks and 
leptons as decorations on an inverted tetrahedron. The right-handed 
fermions are, by definition, singlets under the usual left-handed weak 
isospin, $\mathrm{SU(2)}_{\mathrm{L}}$, so I give the decorations a different 
orientation. We do not know whether the pairs of quarks and leptons 
carry a right-handed weak isospin, in other words, whether they make 
up $\mathrm{SU(2)}_{\mathrm{R}}$ doublets. We do know that we have---as 
yet---no experimental evidence for right-handed charged-current weak 
interactions. Accordingly, I will generally display the right-handed fermions 
without a connecting $W_{\mathrm{R}}$-boson, as shown in the left panel of 
Figure~\ref{fig:RHdec}.
Is there a right-handed charged-current interaction? If not, we come 
back to the question that shook our ancestors: what is the meaning of 
parity violation, and what does it tell us about the world? If we 
should discover---or wish to conjecture---a right-handed charged 
current, it can be added to our graphic, as shown in the right-panel 
of Figure~\ref{fig:RHdec}. If there is a right-handed charged-current 
interaction, restoring parity invariance at high energy scales, what 
makes that interaction so feeble that we haven't yet observed it?

Neutrino oscillations make us almost certain that a right-handed 
neutrino exists,\footnote{A purely left-handed Majorana mass term 
remains a logical, though not especially likely, possibility. For additional discussion of the sources 
of neutrino mass and the existence and nature of $\nu_{\mathrm{R}}$, see the 
lectures by Bel\'{e}n Gavela and Pilar Hern\'{a}ndez.} so I have placed a 
right-handed neutrino in Figure~\ref{fig:RHdec}. I have given it a 
different coloration from the established leptons as a reminder that we 
have not proved its existence, and we do not know its nature.

\begin{figure}[t!]
\begin{center}
\raisebox{10pt}{\includegraphics[width=5.2cm]{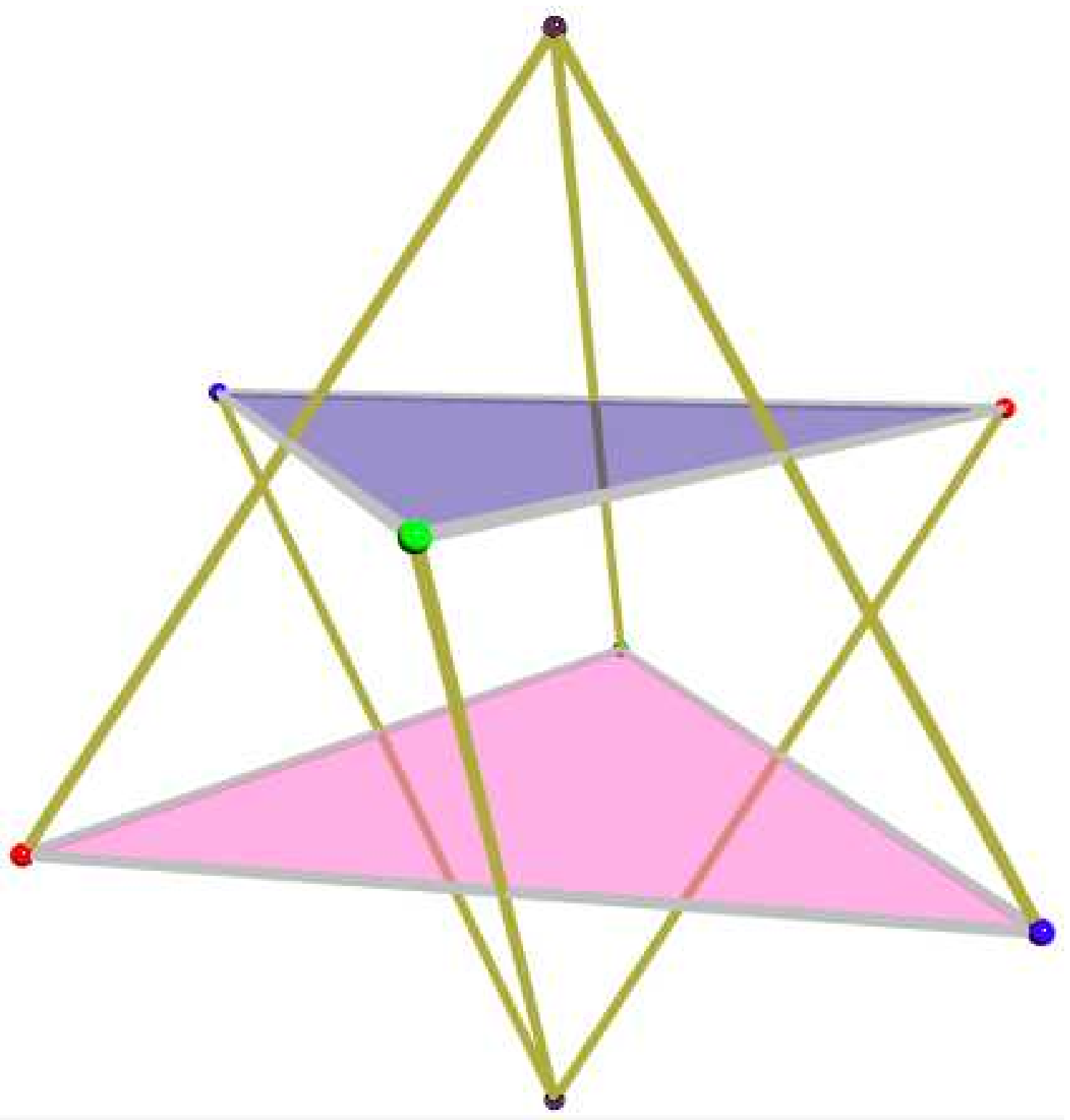}}
\qquad\qquad
\includegraphics[width=6cm]{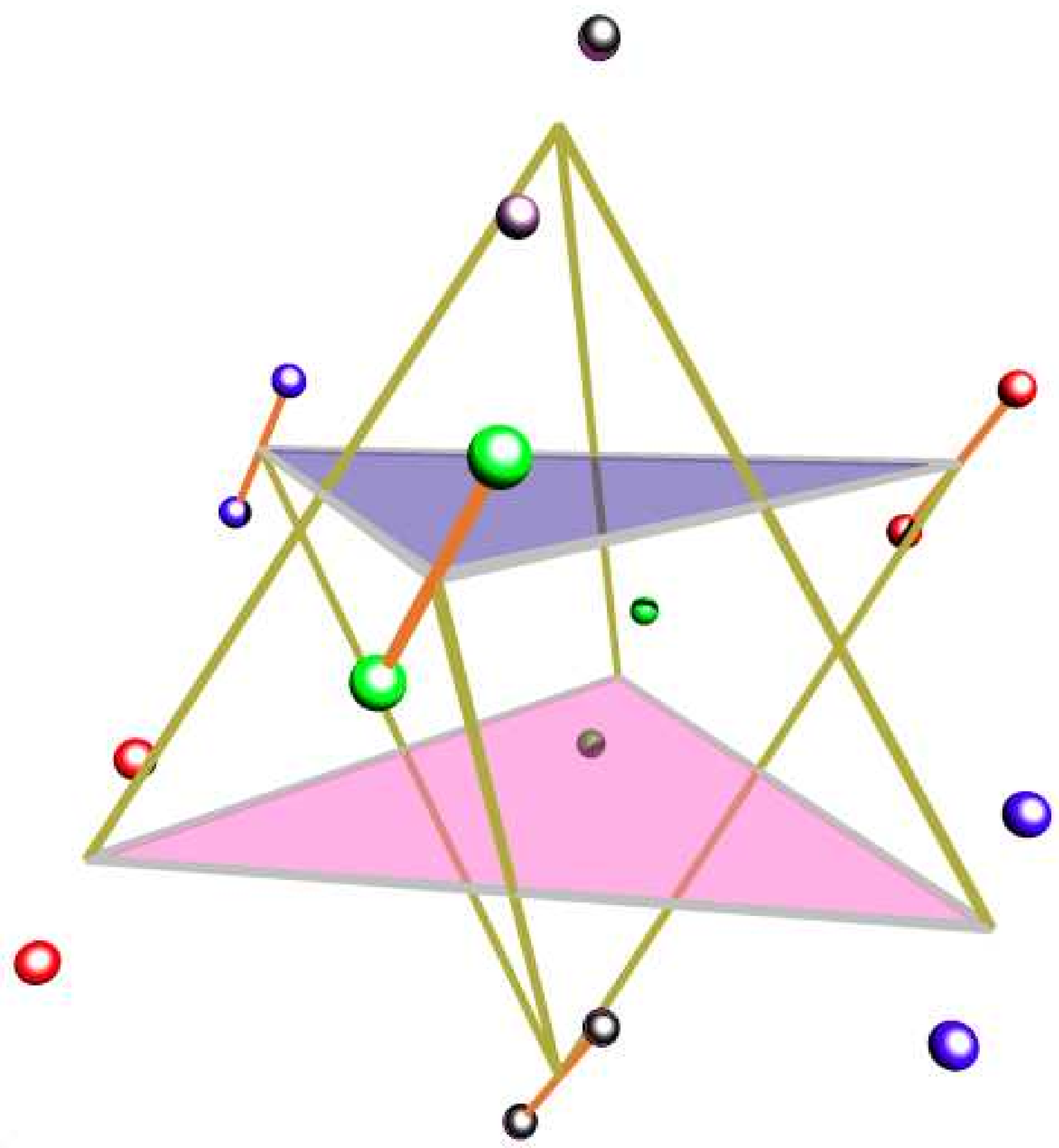}
\caption{The double simplex, undecorated (left panel) and decorated
with one generation of quarks and leptons (right panel).  \label{fig:onegenDS}}
\end{center}
\end{figure}

If parity violation in the weak interactions teaches us of an 
important asymmetry between left-handed and right-handed fermions, the 
nonvanishing masses of the quarks and leptons inform us that left and 
right cannot be entirely separate. Coupling the left-handed particle 
to its right-handed counterpart is what endows fermions with mass. 
For example, the mass term of the electron in the Lagrangian of 
quantum electrodynamics is 
\begin{equation}
    \mathcal{L}_{e} = -m_{e}\bar{e}e = 
-m_{e}\bar{e}\left[\cfrac{1}{2}(1-\gamma_{5}) + 
\cfrac{1}{2}(1+\gamma_{5})\right]e =
-m_{e}(\bar{e}_{\mathrm{R}}e_{\mathrm{L}} + 
\bar{e}_{\mathrm{L}}e_{\mathrm{R}})\;.
\label{eq:emass}
\end{equation}
How shall we combine left with right?  A suggestive structure is the 
pair of interpenetrating tetrahedra shown in 
Figure~\ref{fig:onegenDS}.
Mathematicians refer to a tetrahedron as a \textit{simplex} in 
three-dimensional space, so I call this construction the \textit{double 
simplex.}\footnote{My sketchbook, with interactive graphics and 
photographs of ball-and-stick models, is available for browsing at 
\url{http://lutece.fnal.gov/DoubleSimplex}.}

The structure of the double simplex is based on the
$\mathrm{{SU(4)}\otimes {SU(2)}\otimes {SU(2)}}$ decomposition of
$\mathrm{SO(10)}$.  A three-dimensional solid (tetrahedron) represents
the fundamental \textbf{4} representation of $\mathrm{SU(4)}$.  It is
decorated at the vertices with dumbbells representing the
{$\mathrm{SU(2)_{L}}$} and {$\mathrm{SU(2)_{R}}$} quantum numbers.  The
vertical coordinate of $\mathrm{SU(4)}$ can be read as $B-L$, the
difference of baryon number and lepton number.  The group
$\mathrm{SO(10)}$ is a useful classification symmetry, because its
16-dimensional fundamental representation contains an entire generation
of the known quarks and leptons.  Using $\mathrm{SO(10)}$ as a
coordinate system, if you like, carries no implication that it is the
symmetry of the world, or that it is the basis of a unified theory of
the strong, weak, and electromagnetic interactions.  The idea of the
double simplex is to represent what we know is true, what we hope might
be true, and what we don't know---in other terms, to show the
connections that are firmly established, those we believe must be
there, and the open issues.

Fermion masses tell us that the left-handed and right-handed fermions 
are linked, but we do not know what agent makes the connection. In the 
standard \ewgg\ electroweak theory, it is the Higgs boson---the avatar 
of electroweak symmetry breaking---that endows the fermions with mass. 
But this has not been proved by experiment, and it is certainly 
conceivable that some entirely different mechanism is the source of 
fermion mass.

\begin{figure}[t!]
\begin{center}
\includegraphics[width=6cm]{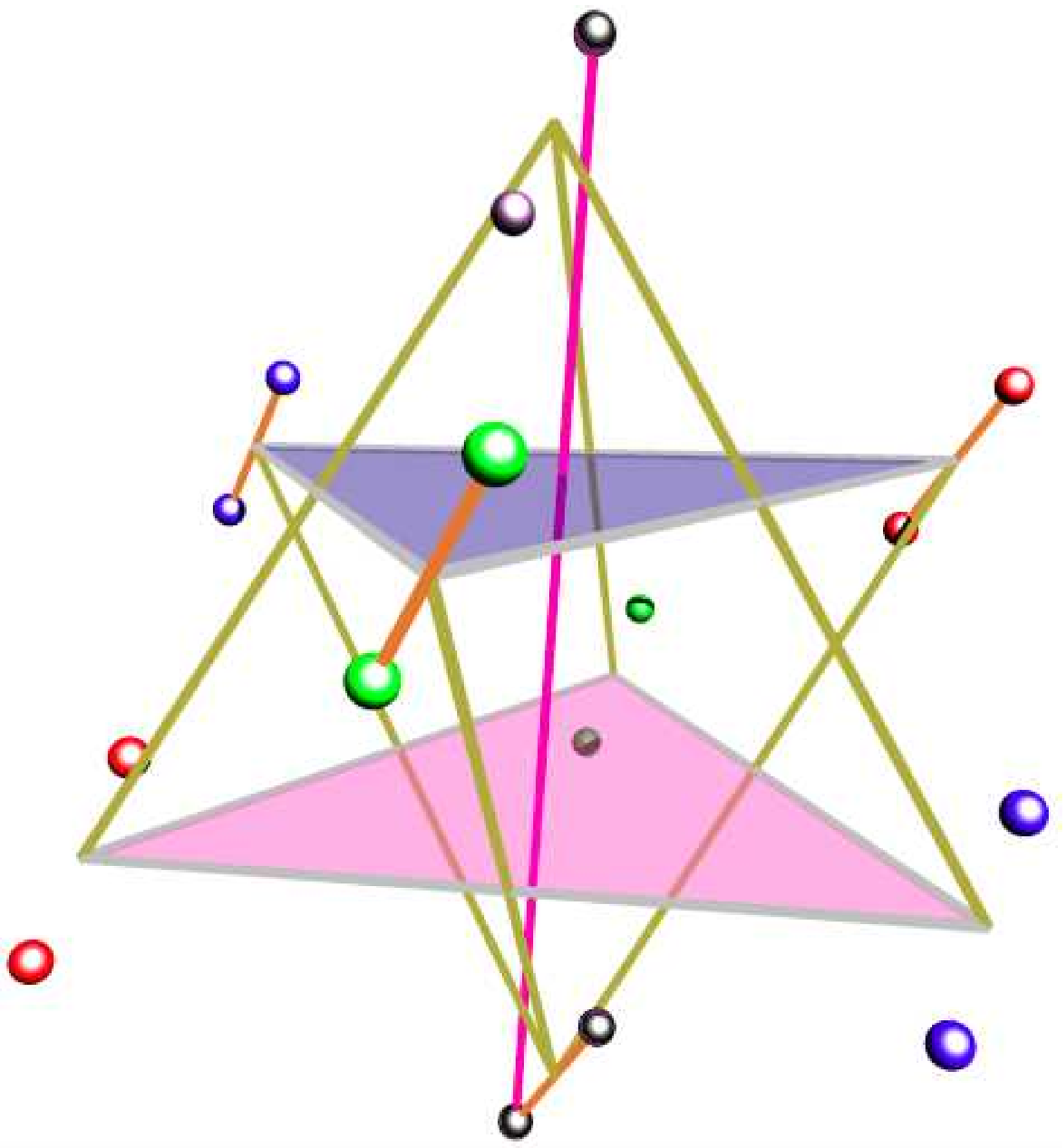}
\qquad\qquad
\includegraphics[width=6cm]{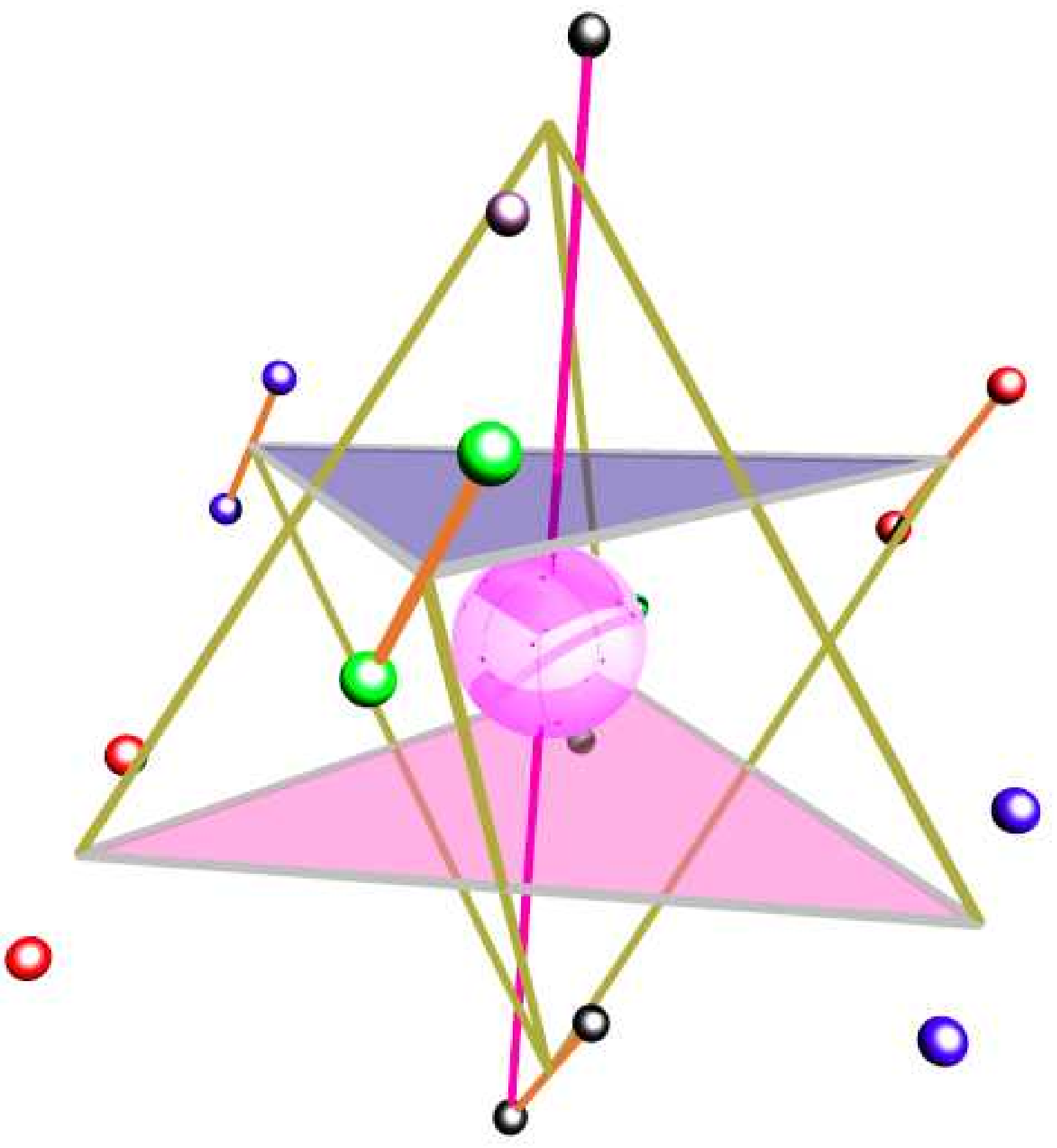}
\caption{The connection between $e_{\mathrm{R}}$ and $e_{\mathrm{L}}$
implied by the electron's nonzero mass.  \label{fig:onegene}}
\end{center}
\end{figure}

I draw the connection between the left-handed and right-handed
electrons in Figure~\ref{fig:onegene}.  
The left-hand panel shows the link between $e_{\mathrm{L}}$ and
$e_{\mathrm{R}}$.  In the right-hand panel, I show the connection
veiled within an opalescent globe that represents our ignorance of the
symmetry-hiding phase transition that links left and right. It is 
excellent to find that the central mystery of the standard model---the 
nature of electroweak symmetry breaking---appears at the center of the 
double simplex!

\begin{figure}[b!]
\begin{center}
\includegraphics[width=6cm]{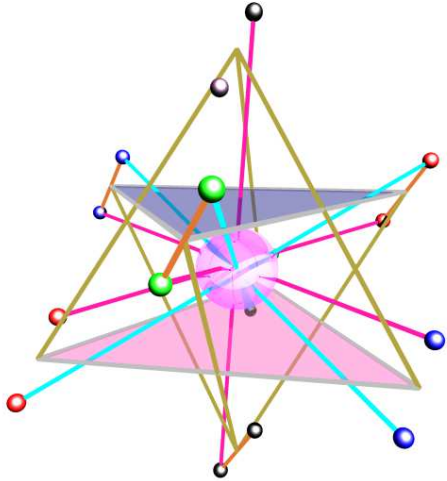}
\caption{The connections that give rise to mass for the quarks and 
leptons of the first generation. \label{fig:onegen}}
\end{center}
\end{figure}

Connecting all the left-handed fermions to their right-handed 
counterparts\footnote{I omit the neutrinos in this brief tour, because 
there are several possible origins for neutrino mass.} leads us to the 
representation given in Figure~\ref{fig:onegen}.
Does one agent give masses to all the quarks and leptons? (That is 
the standard-model solution.) If so, what distinguishes one fermion 
species from another? We do not know the answer, and for that reason I 
contend that \textit{fermion mass is evidence for physics beyond the standard 
model.} Let us illustrate the point in the standard-model context. The 
 mass of fermion $f$ is given by
\begin{equation}
    \mathcal{L}_{f} =
-\displaystyle{\frac{{}{\zeta_{f}}v}{\sqrt{2}}}(\bar{f}_{\mathrm{R}}f_{\mathrm{L}} +
\bar{f}_{\mathrm{L}}f_{\mathrm{R}}) =
-\displaystyle{\frac{{}{\zeta_{f}}v}{\sqrt{2}}}\bar{f}f\;,
\label{eq:fmassYuk}
\end{equation}
where
$v/\sqrt{2} = 
    (G_{\mathrm{F}}\sqrt{8})^{-1/2} \approx 174\gev$
is the vacuum expectation value of the Higgs field. The 
\textit{Yukawa coupling} $\zeta_{f}$ is not predicted by the 
electroweak theory, nor does the standard model relate different
 Yukawa couplings. In any 
event, we do not know whether one agent, or two, or three, will give rise 
to the electron, up-quark, and down-quark masses. 

Of course, the world we have discovered until now consists not only 
of one family of quarks and one family of leptons, but of the three 
pairs of quarks and three pairs of leptons enumerated in 
\eqn{eq:quarks} and \eqn{eq:leptons}. We do not know the 
meaning of the replicated generations, and indeed we have no 
experimental indication to tell us which pair of quarks is to be 
associated with which pair of leptons. 

\begin{figure}[t!]
\begin{center}
\includegraphics[width=6.3cm]{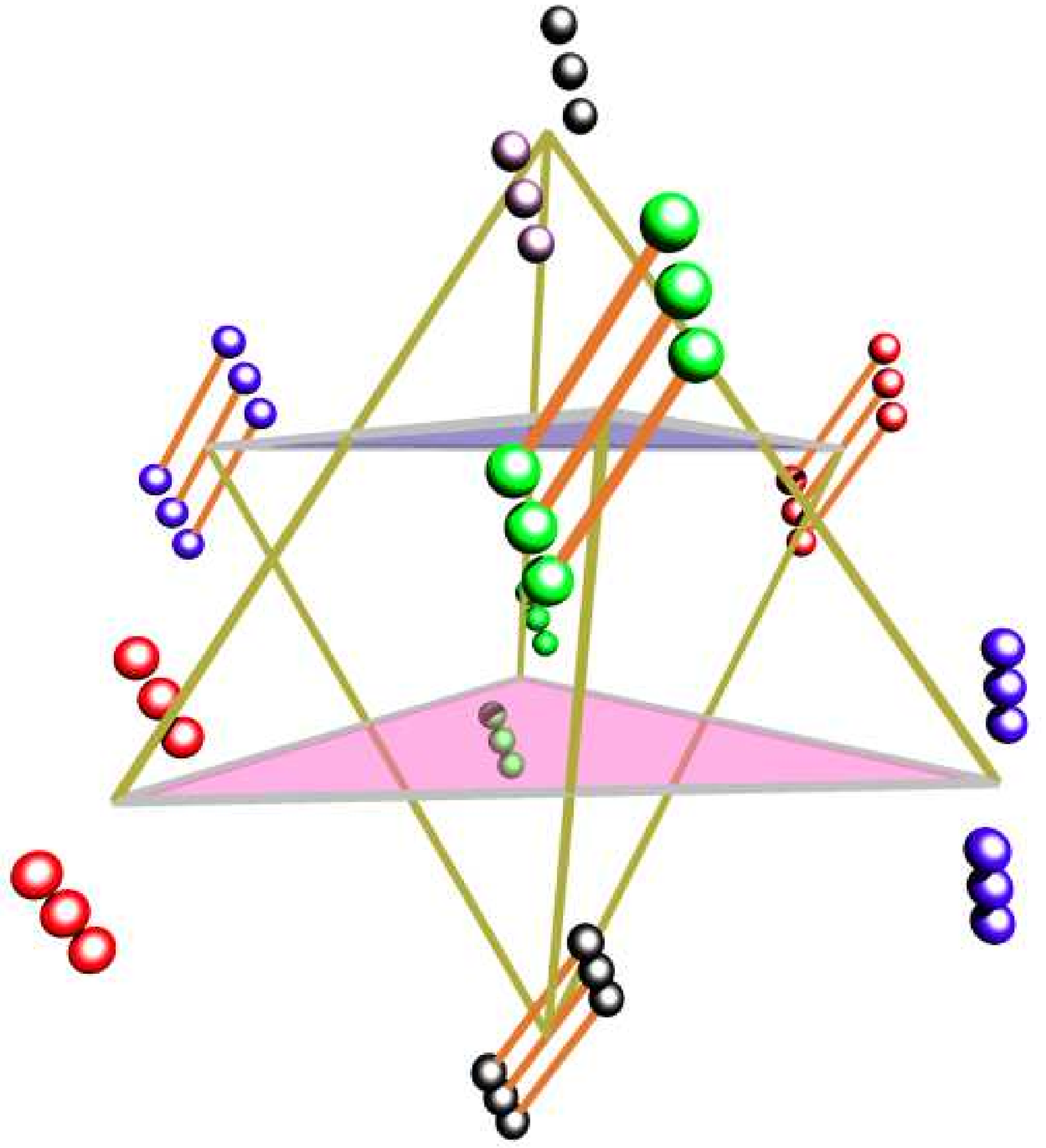}\qquad
\qquad\includegraphics[width=6.3cm]{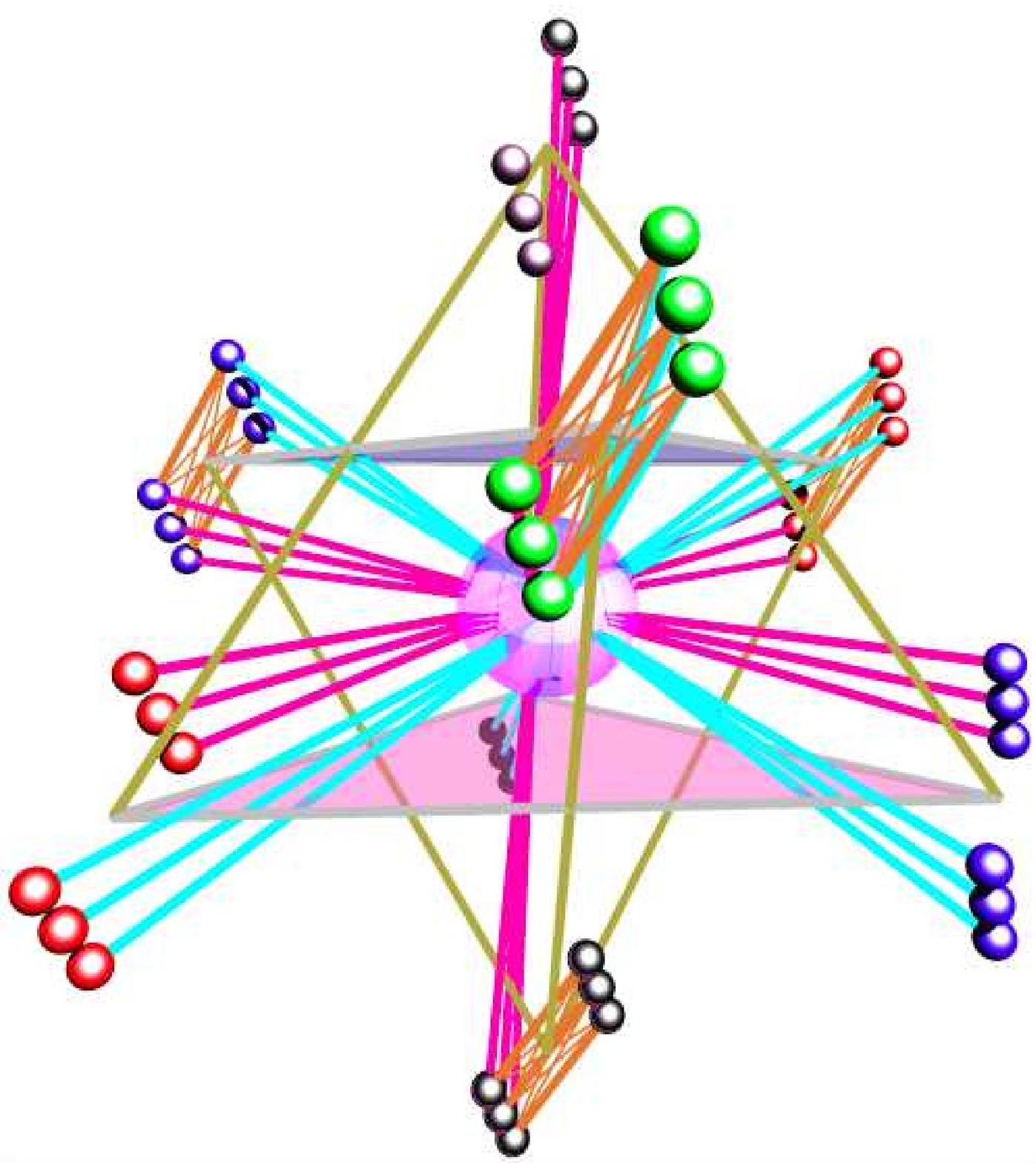}
\caption{Left panel: Three generations of quarks and leptons.  Right
panel: The connections that give rise to mass and mixing for three
generations of quarks and leptons.  \label{fig:threegen}}
\end{center}
\end{figure}

In the absence of any understanding of the relation of one generation 
to another, I depict the three generations in the double simplex 
simply by replicating the decorations to include three pairs of 
quarks and three pairs of leptons, as shown in the left panel of 
Figure~\ref{fig:threegen}.
The connections that generate the fermion masses are indicated in the
right panel of Figure~\ref{fig:threegen}.  The Yukawa couplings of the
charged leptons and quarks range from $\zeta_{e} \approx 3 \times
10^{-6}$ for the electron to $\zeta_{t} \approx 1$ for the top quark.
In the case of more than one generation, the connections that endow 
the fermions with mass also determine the mixing among generations, 
the suppressed transitions such as $ u \leftrightarrow s$ and $u 
\leftrightarrow b$. With three generations, the Yukawa couplings may 
have complex phases that give rise to $\mathcal{CP}$-violating
transitions. Although it is correct to say that the standard model 
describes the observed examples of $\mathcal{CP}$ violation, I would 
like to insist that because the standard model does not prescribe the 
Yukawa couplings, $\mathcal{CP}$ \textit{violation---like fermion mass---is 
evidence for physics beyond the standard model.}

Let us return to the point that the charge conjugate of a left-handed
field is right-handed.  If the field $\psi$ annihilates a particle,
then its charge-conjugate filed $\psi^{\mathrm{c}} \equiv
\mathcal{C}\bar{\psi}^{\mathrm{T}}$ annihilates the corresponding
antiparticle. In terms of Dirac matrices, the charge-conjugation 
operator is
\begin{equation}
    \mathcal{C} = i\gamma^{2}\gamma^{0} = 
    -\mathcal{C}^{-1} = -\mathcal{C}^{\dagger} = 
    -\mathcal{C}^{\mathrm{T}} \;.
\end{equation}
The left-handed component of the charge-conjugate field is
\begin{eqnarray}
{\psi^{\mathrm{c}}_{\mathrm{L}}} & = & \cfrac{1}{2}(1 - 
\gamma_{5})\psi^{\mathrm{c}} = \cfrac{1}{2}(1 - \gamma_{5})
\mathcal{C}\bar{\psi}^{\mathrm{T}}  \nonumber \\
 & = & \mathcal{C}\cfrac{1}{2}(1 - 
 \gamma_{5})\bar{\psi}^{\mathrm{T}} = 
 \mathcal{C}[\bar{\psi}\cfrac{1}{2}(1 - \gamma_{5})]^{\mathrm{T}} 
 \label{eq:CC}\\
  & = & \mathcal{C}(\bar{\psi}_{\mathrm{R}})^{\mathrm{T}} = 
  {(\psi_{\mathrm{R}})^{\mathrm{c}}} \nonumber \;, \nonumber
\end{eqnarray}
which is indeed the charge conjugate of the right-handed component of 
the Dirac field $\psi$.

\begin{figure}[t!]
\begin{center}
\raisebox{13pt}{\includegraphics[width=5.2cm]{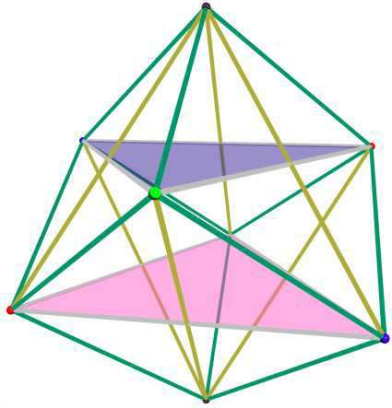}}
\caption{The double simplex, with additional interactions suggested by 
the shape of the figure indicated as green links. \label{fig:cube}}
\end{center}
\end{figure}

With this connection in mind, we can now think of the double simplex 
as composed of left-handed particles and left-handed antiparticles. 
When we combine the two sets of particles into one representation, we 
are invited to consider the possibility of new transformations that 
take any member of the extended family into any other. The agents of 
change will be new gauge bosons, since gauge-boson interactions preserve 
chirality. I connect the hitherto unconnect vertices of the 
(undecorated) double simplex in Figure~\ref{fig:cube}.
The hypothetical new interactions are easy to visualize, because the 
double simplex can be inscribed in a cube. Do some of these 
interactions exist? If so, why are they so weak that we have not yet 
observed them?

The object of our double-simplex construction project has been to 
identify important topical questions for particle physics without 
plunging into formalism. As a theoretical physicist, I have deep 
respect for the power of mathematics to serve as a refiner's fire for 
our ideas. But I hope this exercise has helped you to see the power 
and scope of physical reasoning and the insights that can come from 
building and looking at a physical object with an inquiring 
spirit---even if the physical object inhabits an abstract space!

In the spirit of providing homework for life, here are some of the 
questions we have encountered in this first lecture:

\begin{questions}{First}
    \item  Are quarks and leptons elementary?
	
    \item  What is the relationship of quarks to leptons?
	
    \item  Are there right-handed weak interactions?
	
    \item  Are there new quarks and leptons?
	
    \item  Are there new gauge interactions linking quarks and leptons?
	
    \item  What is the relationship of left-handed \& right-handed 
    particles?
	
    \item  What is the nature of the right-handed neutrino?
	
    \item  What is the nature of the mysterious new force that hides 
    electroweak symmetry?
	
    \item  Are there different kinds of matter?
	
    \item  Are there new forces of a novel kind?
    
    \item  What do generations mean? Is there a family symmetry?
	
    \item  What makes a top quark a top quark, and an electron an 
    electron?
	
    \item  What is the (grand) unifying symmetry?

\end{questions}

\section{The Electroweak Theory \protect\footnote{Much more detail can
be found in my 2002 European School of High-Energy Physics (Pylos, 
Greece) lectures,
\protect\url{http://lutece.fnal.gov/Talks/CQPylos.pdf}, and in my 
2000 TASI lectures, Ref.~\cite{Quigg:2002td}.}\label{sec:electroweak}}
To provide us with a common starting point for our investigation of
theories that extend the standard model, we devote this lecture to a
survey of the electroweak theory.  As I have emphasized
elsewhere~\cite{Quigg:2002td}, the theory of the strong interactions,
quantum chromodynamics, is an essential element of the standard model,
but it is by contemplating the electroweak theory that we are led most
quickly to see the shortcomings of the standard model.

We shall begin by recalling the idea of gauge theories, and then use 
the strategy we uncover there to construct the electroweak theory. 
Applying the theory to quarks, we come upon the need to inhibit 
flavor-changing neutral currents that motivated the 
introduction of the charmed quark. Then we swiftly review the tests of 
the electroweak theory that have led us, over the past decade, to 
elevate it to the status of a (provisional!) law of nature. A 
profound puzzle raised by the electroweak theory, as we shall see, is 
why empty space---the vacuum---is so nearly massless. We will recall 
bounds on the mass of the Higgs boson and then conclude our little 
tour by looking at the electroweak scale and beyond.

\subsection{How Symmetries Lead to Interactions \label{subsec:symint}}
Suppose that we knew the Schr\"{o}dinger equation, but not the laws of 
electrodynamics.  Would it be possible to derive---in other words, to 
guess---Maxwell's equations from a gauge principle.  The answer is yes!  
it is worthwhile to trace the steps in the argument in detail.

A quantum-mechanical state is described by a complex Schr\"{o}dinger wave 
function $\psi(x)$.  Quantum-mechanical observables involve inner 
products of the form
\begin{equation}
	\langle \mathcal{O} \rangle = \int d^n\!x\: \psi^*{\mathcal{O}}\psi ,
	\label{observ}
\end{equation} which are unchanged under a global phase rotation:
\begin{equation}
	\psi(x) \rightarrow e^{i\theta} \psi(x) \qquad
	\psi^{*}(x) \rightarrow e^{-i\theta} \psi^{*}(x).
	\label{phaserot}
\end{equation}  In other words, the absolute phase of the wave function 
cannot be measured and is a matter of convention.  \emph{Relative} phases 
between wave functions, as measured in interference experiments, are 
unaffected by such a global rotation.

This raises the question: Are we free to choose one phase convention in 
San Miguel Regla and another in Geneva?  Differently stated, can quantum mechanics 
be formulated to be invariant under local (position-dependent) phase rotations
\begin{equation}
	\psi(x) \rightarrow \psi^\prime(x) = e^{i\alpha(x)} \psi(x)\; ?
	\label{locphase}
\end{equation} We shall see that this can be accomplished, but at the 
price of introducing an interaction that we will  construct to be 
electromagnetism.  

The quantum-mechanical equations of motion, such as the Schr\"{o}dinger 
equation, always involve derivatives of the wave function $\psi$, as do 
many observables.  Under local phase rotations, these transform as
\begin{equation}
	\partial_\mu \psi(x) \rightarrow \partial_\mu \psi^\prime =
	e^{i\alpha(x)}[\partial_\mu \psi(x) + i(\partial_\mu\alpha(x))\psi(x)],
	\label{gradtrans}
\end{equation} which involves more than a mere phase change.  The 
additional gradient-of-phase term spoils local phase invariance.  Local 
phase invariance may be achieved, however, if the equations of motion and 
the observables involving derivatives are modified by the introduction of 
the electromagnetic field $A_\mu(x)$.  If the gradient $\partial_\mu$ is 
everwhere replaced by the \emph{gauge-covariant derivative}
\begin{equation}
	\mathcal{D}_\mu \equiv \partial_\mu + ieA_\mu ,
	\label{gcder}
\end{equation} where $e$ is the charge in natural units of the particle 
described by $\psi(x)$ and the field $A_\mu(x)$ transforms under phase 
rotations \eqn{locphase} as
\begin{equation}
	A_\mu(x) \rightarrow A_\mu^\prime(x) \equiv A_\mu(x) - 
	(1/e)\partial_\mu\alpha(x), 
	\label{Atrans}
\end{equation}it is easily verified that under local phase transformations
\begin{equation}
	\mathcal{D}_\mu\psi(x) \rightarrow e^{i\alpha(x)}\mathcal{D}_\mu\psi(x) .
	\label{covphinv}
\end{equation}  Consequently quantities such as $\psi^*\mathcal{D}_\mu\psi$ 
are invariant under local phase transformations.  The required 
transformation law \eqn{Atrans} for the four-vector potential $A_\mu$ is 
precisely the form of a gauge transformation in 
electrodynamics.  Moreover, the covariant derivative defined in 
\eqn{gcder} corresponds to the familiar replacement $p\rightarrow p -eA$. 
Thus the form of the coupling 
$(\mathcal{D}_\mu\psi)$ between the electromagnetic field and matter is 
suggested, if not uniquely dictated, by local phase invariance.

A photon mass term would have the form
\begin{equation}
	\mathcal{L}_\gamma = \half m^2 A^\mu A_\mu ,
	\label{photmass}
\end{equation} which obviously violates local gauge invariance because
\begin{equation}
	A^\mu A_\mu \rightarrow 
	(A^\mu - \partial^\mu \alpha)(A_\mu - \partial_\mu\alpha)
	\ne A^\mu A_\mu .
	\label{AArot}
\end{equation} Thus we find that local gauge invariance has led us to the 
existence of a massless photon. 

This example has shown the possibility of using local gauge invariance as 
a dynamical principle. We have derived the content of Maxwell's 
equations from a symmetry principle. We can think of quantum 
electrodynamics as the gauge theory based on $\mathrm{U(1)}$ phase 
symmetry.

We can abstract from this discussion a general procedure. First, 
recognize a symmetry of Nature, perhaps by observing a conservation 
law, and build it into the laws of physics.\footnote{Recall that 
Noether's theorem correlates a conservation law with every continuous 
symmetry transformation under which the Lagrangian is invariant in 
form.} Then impose the symmetry 
in a stricter \textit{local} form. By a generalization of the 
arithmetic we have just recited, the local gauge symmetry leads to 
new interactions, mediated by massless vector fields, the gauge bosons. 
As we have seen, the interaction of the gauge fields with matter is 
given by ``minimal coupling'' to the conserved current that 
corresponds to the symmetry. If the symmetry is non-Abelian, imposing 
the symmetry also leads to interactions among the gauge bosons, since 
they carry the conserved charge.

Posed as a problem in mathematics, construction of a gauge theory is 
always possible, at the level of a classical Lagrangian. Formulating a 
consistent quantum theory may require additional vigilance. The 
formalism offers no guarantee that the gauge symmetry was chosen 
wisely; that verdict is left to experiment!

\subsection{Hiding a Gauge Symmetry \label{subsec:cache}}
The gauge-theory paradigm is constraining and it is predictive, but 
there is an obstacle to surmount if we want to apply it to all the 
interactions. As we have just seen, local gauge invariance is 
incompatible with a massive gauge boson. Yet we have known since the 
1930s that the (charged-current) weak interaction has a very short 
range, on the order of $10^{-15}\cm$, so must be mediated by a 
massive $\mathcal{O}(100\gev)$ force carrier. Happily, 
condensed-matter physics provides us with an example of a physical 
system in which the photon of QED acquires a mass inside a medium, 
as a consequence of a symmetry-reducing phase transition: 
\textit{superconductivity.}

Superconducting materials display two kinds of miraculous behavior: 
they carry an electric current without resistance, and they expel 
magnetic fields.  In the Ginzburg-Landau 
description~\cite{Ginzburg:1950sr} of the superconducting phase transition, a 
superconducting material is regarded as a collection of two kinds of 
charge carriers: normal, resistive carriers, and superconducting, 
resistanceless carriers.

In the absence of a magnetic field, the free energy of the superconductor 
is related to the free energy in the normal state through
\begin{equation}
G_{\rm super}(0) = G_{\rm normal}(0) + \alpha \abs{\psi}^2 + \beta 
\abs{\psi}^4\;\;,
\end{equation}
where $\alpha$ and $\beta$ are phenomenological parameters and 
$\abs{\psi}^2$ is an order parameter that measures the density of 
superconducting charge carriers.  The parameter $\beta$ is non-negative, 
so that the free energy is bounded from below.

\begin{figure}[t!]
    \begin{center}
\includegraphics[width=10cm]{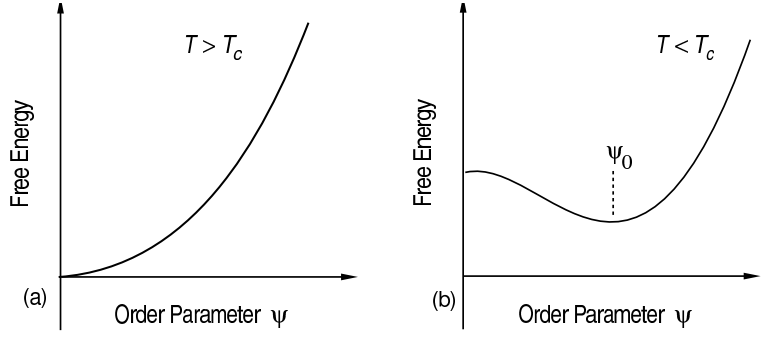}
\caption{Ginzburg-Landau description of the superconducting phase 
transition.}
\protect\label{fig:GL}
\end{center}
\end{figure}

Above the critical temperature for the onset of superconductivity, the 
parameter $\alpha$ is positive and the free energy of the substance is 
supposed to be an increasing function of the density of 
superconducting carriers, as shown in Figure \ref{fig:GL}(a).  The state 
of minimum energy, the vacuum state, then corresponds to a purely 
resistive flow, with no superconducting carriers active.  Below the 
critical temperature, the parameter $\alpha$ becomes negative and the 
free energy is minimized when $\vev{|\psi|^{2}} = \psi^{2}_0 \ne 0$, as illustrated in 
Figure \ref{fig:GL}(b).

This is a nice cartoon description of the superconducting phase 
transition, but there is more.  In an applied magnetic field $\vec{H}$, 
the free energy is
\begin{equation}
G_{\rm super}(\vec{H}) = G_{\rm super}(0) + \frac{\vec{H}^2}{8\pi} + 
\frac{1}{2m^\star}|-i\hbar\nabla\psi-(e^\star/c)\vec{A}\psi|^2 
\;\;,
\end{equation}
where $e^\star$ and $m^\star$ are the charge ($-2$ units) and effective 
mass of the superconducting carriers.  In a weak, slowly varying field 
$\vec{H} \approx 0$, when we can approximate $\psi\approx\psi_0$ and 
$\nabla\psi\approx 0$, the usual variational analysis leads to the 
equation of motion,
\begin{equation}
\nabla^2\vec{A}-\frac{4\pi e^\star}{m^\star c^2}\abs{\psi_0}^2\vec{A} = 
0\;\;,
\end{equation}
the wave equation of a massive photon.  In other words, the photon 
acquires a mass within the superconductor.  This is the origin of the 
Meissner effect, the exclusion of a magnetic field from a 
superconductor.  More to the point for our purposes, it shows how a 
symmetry-hiding phase transition can lead to a massive gauge boson.

\subsection{Constructing the Electroweak Theory {\protect{\label{sub:cache}}}}
Let us review the essential elements of the \ewgg\ electroweak 
theory \cite{GT}.
The electroweak theory takes three crucial clues from experiment:
\begin{itemize}
    \item  The existence of left-handed weak-isospin doublets,
    \begin{displaymath}
	\left( 
		\begin{array}{c}
	    \nu_{e}  \\
	    e
	\end{array}
	\right)_{\mathrm{L}} \qquad
		\left( 
		\begin{array}{c}
	    \nu_{\mu}  \\
	    \mu
	\end{array}
	\right)_{\mathrm{L}} \qquad
	\left( 
		\begin{array}{c}
	    \nu_{\tau}  \\
	    \tau
	\end{array}
	\right)_{\mathrm{L}}
    \end{displaymath}
    and
    \begin{displaymath}
	\left( 
		\begin{array}{c}
	    u  \\
	    d^{\prime}
	\end{array}
	\right)_{\mathrm{L}} \qquad
		\left( 
		\begin{array}{c}
	    c  \\
	    s^{\prime}
	\end{array}
	\right)_{\mathrm{L}} \qquad
	\left( 
		\begin{array}{c}
	    t  \\
	    b^{\prime}
	\end{array}
	\right)_{\mathrm{L}}\; ; 
    \end{displaymath}

    \item  The universal strength of the weak interactions;

    \item  The idealization that neutrinos are massless.
\end{itemize}

To save writing, we shall construct the electroweak theory as it 
applies to a single generation of leptons.  In this form, it is 
neither complete nor consistent: anomaly cancellation requires that a 
doublet of color-triplet quarks accompany each doublet of 
color-singlet leptons.  However, the needed generalizations are simple 
enough to make that we need not write them out.

To incorporate electromagnetism into a theory of the weak 
interactions, we add to the $\mathrm{SU(2)_{L}}$ family symmetry suggested by 
the first two experimental clues a $\mathrm{U(1)}_{Y}$ weak-hypercharge phase 
symmetry.  We begin by specifying the fermions: a left-handed weak 
isospin doublet
\begin{equation}
{{\sf L}} = \left(\begin{array}{c} \nu_e \\ e 
\end{array}\right)_{\mathrm{L}}
\end{equation}
with weak hypercharge $Y_{\mathrm{L}}=-1$, and a right-handed weak isospin singlet
\begin{equation}
      {{\sf R}}\equiv e_{\mathrm{R}}
\end{equation}
with weak hypercharge $Y_{\mathrm{R}}=-2$.

The electroweak gauge group, \ewgg, implies two sets of gauge fields:
a weak isovector $\vec{b}_\mu$, with coupling constant $g$, and a
weak isoscalar
${{\mathcal A}}_\mu$, with coupling constant $g^\prime$. Corresponding
to these gauge fields are the field-strength tensors 
\begin{equation}
    F^{\ell}_{\mu\nu} = \partial_{\nu}b^{\ell}_{\mu} - 
    \partial_{\mu}b^{\ell}_{\nu} + 
    g\varepsilon_{jk\ell}b^{j}_{\mu}b^{k}_{\nu}\; ,
    \label{eq:Fmunu}
\end{equation}
for the weak-isospin symmetry, and 
\begin{equation}
    f_{\mu\nu} = \partial_{\nu}{{\mathcal A}}_\mu - \partial_{\mu}{{\mathcal 
    A}}_\nu \; , 
    \label{eq:fmunu}
\end{equation}
for the weak-hypercharge symmetry.  We may summarize the interactions 
by the Lagrangian
\begin{equation}
\mathcal{L} = \mathcal{L}_{\rm gauge} + \mathcal{L}_{\rm leptons} \ ,                           
\end{equation}             
with
\begin{equation}
\mathcal{L}_{\rm gauge}=-\cfrac{1}{4}F_{\mu\nu}^\ell F^{\ell\mu\nu}
-\cfrac{1}{4}f_{\mu\nu}f^{\mu\nu},
\end{equation}
and
\begin{eqnarray}     
\mathcal{L}_{\rm leptons} & = & \overline{{\sf R}}\:i\gamma^\mu\!\left(\partial_\mu
+i\frac{g^\prime}{2}{\cal A}_\mu Y\right)\!{\sf R} 
\label{eq:matiere} \\ 
& + & \overline{{\sf
L}}\:i\gamma^\mu\!\left(\partial_\mu 
+i\frac{g^\prime}{2}{\cal
A}_\mu Y+i\frac{g}{2}\vec{\tau}\cdot\vec{b}_\mu\right)\!{\sf L}. \nonumber
\end{eqnarray}
The \ewgg\ gauge symmetry forbids a mass term for the electron in the 
matter piece \eqn{eq:matiere}.  Moreover, the theory we have described 
contains four massless electroweak gauge bosons, namely ${{\mathcal A}}_\mu$, 
$b^{1}_{\mu}$, $b^{2}_{\mu}$, and $b^{3}_{\mu}$, whereas Nature has 
but one: the photon.  To give masses to the gauge bosons and 
constituent fermions, we must hide the electroweak symmetry.

To endow the intermediate bosons of the weak interaction with mass, we 
take advantage of a relativistic generalization of the Ginzburg-Landau 
phase transition known as the Higgs mechanism \cite{Higgs:1964ia}.  We introduce 
a complex doublet of scalar fields
\begin{equation}
\phi\equiv \left(\begin{array}{c} \phi^+ \\ \phi^0 \end{array}\right)
\end{equation}
with weak hypercharge $Y_\phi=+1$.  Next, we add to the Lagrangian new 
(gauge-invariant) terms for the interaction and propagation of the 
scalars,
\begin{equation}
      \mathcal{L}_{\rm scalar} = (\D^\mu\phi)^\dagger(\D_\mu\phi) - V(\phi^\dagger \phi),
\end{equation}
where the gauge-covariant derivative is
\begin{equation}
      \D_\mu=\partial_\mu 
+i\frac{g^\prime}{2}{\cal A}_\mu
Y+i\frac{g}{2}\vec{\tau}\cdot\vec{b}_\mu \; ,
\label{eq:GcD}
\end{equation}
and the potential interaction has the form
\begin{equation}
      V(\phi^\dagger \phi) = \mu^2(\phi^\dagger \phi) +
\abs{\lambda}(\phi^\dagger \phi)^2 .
\label{SSBpot}
\end{equation}
We are also free to add a Yukawa interaction between the scalar fields
and the leptons,
\begin{equation}
      \mathcal{L}_{\rm Yukawa} = -\zeta_e\left[\overline{{\sf R}}(\phi^\dagger{\sf
L}) + (\overline{{\sf L}}\phi){\sf R}\right].
\label{eq:Yukterm}
\end{equation}

We then arrange 
the scalar self-interactions so that the vacuum state corresponds to a 
broken-symmetry solution.  The electroweak symmetry is spontaneously broken if the parameter
$\mu^2<0$. The minimum energy, or vacuum state, may then be chosen
to correspond to the vacuum expectation value
\begin{equation}
\vev{\phi} = \left(\begin{array}{c} 0 \\ v/\sqrt{2} \end{array}
\right)\;,
\label{eq:vevis}
\end{equation}
where $v = \sqrt{-\mu^2/\abs{\lambda}}$.
Let us verify that the vacuum \eqn{eq:vevis} indeed breaks the gauge 
symmetry.  The vacuum state $\vev{\phi}$ is invariant under a symmetry 
operation $\exp{(i \alpha {\mathcal G})}$ corresponding to the 
generator ${\mathcal G}$ provided that $\exp{(i \alpha {\mathcal 
G})}\vev{\phi} = \vev{\phi}$, \ie, if ${\mathcal G}\vev{\phi} = 0$.  
We easily compute that 
\begin{eqnarray}
    \tau_{1}\vev{\phi} & = \left( 
    \begin{array}{cc}
	0 & 1  \\
	1 & 0
    \end{array}
    \right) \left( 
    \begin{array}{c}
	0  \\
	v/\sqrt{2}
    \end{array}
    \right) & = \left( 
    \begin{array}{c}
	v/\sqrt{2}  \\
	0
    \end{array}
    \right) \neq 0 \quad\hbox{broken!}
    \nonumber  \\
    \tau_{2}\vev{\phi} & = \left( 
    \begin{array}{cc}
	0 & -i  \\
	i & 0
    \end{array}
    \right) \left( 
    \begin{array}{c}
	0  \\
	v/\sqrt{2}
    \end{array}
    \right) & = \left( 
    \begin{array}{c}
	-iv/\sqrt{2}  \\
	0
    \end{array}
    \right) \neq 0 \quad\hbox{broken!}
    \nonumber  \\
    \tau_{3}\vev{\phi} & = \left( 
    \begin{array}{cc}
	1 & 0  \\
	0 & -1
    \end{array}
    \right) \left( 
    \begin{array}{c}
	0  \\
	v/\sqrt{2}
    \end{array}
    \right) & = \left( 
    \begin{array}{c}
	0  \\
	-v/\sqrt{2}
    \end{array}
    \right) \neq 0 \quad\hbox{broken!}
    \nonumber  \\
    Y\vev{\phi} & = Y_{\phi}\vev{\phi} = +1 \vev{\phi} = & \left( 
    \begin{array}{c}
	0  \\
	v/\sqrt{2}
    \end{array}
    \right) \neq 0 \quad\hbox{broken!}
    \label{eq:brisure}
\end{eqnarray}
However, if we examine the effect of the electric charge operator $Q$ 
on the (electrically neutral) vacuum state, we find that
\begin{eqnarray}
    Q \vev{\phi} & = \cfrac{1}{2}(\tau_{3} + Y)\vev{\phi} \qquad\quad& = 
    \cfrac{1}{2} \left( 
    \begin{array}{cc}
	Y_{\phi}+1 & 0  \\
	0 & Y_{\phi}- 1
    \end{array}
    \right) \vev{\phi}
    \nonumber  \\
     & = \left( 
     \begin{array}{cc}
	 1 & 0  \\
	 0 & 0
     \end{array}
      \right) \left(     \begin{array}{c}
	0  \\
	v/\sqrt{2}
    \end{array}
\right) & = \left( 
     \begin{array}{c}
	 0  \\
	 0
     \end{array}
     \right) \quad\hbox{\textit{unbroken!}}
    \label{eq:Qok}
\end{eqnarray}
The original four generators are all broken, but electric charge is
not.  It appears that we have accomplished our goal of breaking
$\mathrm{SU(2)_{L}}\otimes \mathrm{U(1)}_{Y} \to \mathrm{U(1)}_{\mathrm{em}}$.  We expect the
photon to remain massless, and expect the gauge bosons that correspond
to the generators $\tau_{1}$, $\tau_{2}$, and $\kappa \equiv
\cfrac{1}{2}(\tau_{3} - Y)$ to acquire masses.

To establish the particle content of the theory, we expand about the 
vacuum state, letting
\begin{equation}
    \phi = \displaystyle{\left(\begin{array}{c} 0 \\ (v+\eta)/\sqrt{2} \end{array}
    \right)}
    \label{eq:expvac}
\end{equation}
in unitary gauge. The Lagrangian for the scalars becomes
\begin{eqnarray}
	\mathcal{L}_{\mathrm{scalar}} & = &
	\cfrac{1}{2}(\partial^{\mu}\eta)(\partial_{\mu}\eta) - 
	\mu^{2}\eta^{2} \nonumber\\
	 & & + \frac{v^{2}}{8}[g^{2}\abs{b_{1} - ib_{2}}^{2} + 
	 (g^{\prime}\mathcal{A}_{\mu} - g b_{\mu}^{3})^{2}] \\
	 & & + \mbox{ interaction terms} . \nonumber \label{eq:scallag}
\end{eqnarray}
The Higgs boson $\eta$ has acquired a $(\mbox{mass})^{2}$
$M_{H}^{2} = -2\mu^{2} > 0$. Now let us expand the terms proportional 
to $v^{2}/8$. Identifying $W^{\pm} = \cfrac{1}{\sqrt{2}}(b_{1} \mp ib_{2})$, we find
\begin{equation}
	\frac{g^{2}v^{2}}{8}(\abs{W^{+}_{\mu}}^{2} + \abs{W^{-}_{\mu}}^{2}) \;,
    \end{equation}
which implies $M_{W^{\pm}} = gv/2$. Next, we define the orthogonal 
combinations
\begin{equation}
Z_{\mu} = \frac{-g^{\prime}\mathcal{A}_{\mu} + g
b_{\mu}^{3}}{\sqrt{g^{2} + g^{\prime 2}}}\;,
	\quad\quad\qquad
A_{\mu} = \frac{g\mathcal{A}_{\mu} + g^{\prime}
b_{\mu}^{3}}{\sqrt{g^{2} + g^{\prime 2}}}\;,
\end{equation}
and conclude that $M_{Z^{0}} = \sqrt{g^{2} + g^{\prime 2}}\;v/2 =
M_{W}\sqrt{1 + g^{\prime 2}/g^{2}}$ and $M_{A} = 0$.
In the broken-symmetry situation, the Yukawa term becomes
\begin{eqnarray}
\mathcal{L}_{\mathrm{Yukawa}} & = & -\zeta_{e}\frac{(v + \eta)}{\sqrt{2}}
(\bar{e}_{\mathsf{R}}e_{\mathsf{L}} +
\bar{e}_{\mathsf{L}}e_{\mathsf{R}}) \nonumber\\
 & = & - \frac{\zeta_{e}v}{\sqrt{2}}\bar{e}e -
 \frac{\zeta_{e}\eta}{\sqrt{2}}\bar{e}e\;, \label{eq:eYuk}
\end{eqnarray}
so that the electron acquires a mass $m_{e} = \zeta_{e}v/\sqrt{2}$ 
and the Higgs-boson coupling to electrons is $m_{e}/v \propto 
\hbox{fermion mass}$.

Let us summarize. As a result of spontaneous symmetry breaking, the weak bosons acquire 
masses, as auxiliary scalars assume the role of the third 
(longitudinal) degrees of freedom of what had been massless gauge 
bosons.  Specifically, the mediator of the charged-current weak 
interaction, $W^{\pm} = (b_{1} \mp ib_{2})/\sqrt{2}$, acquires a 
mass characterized by 
$M_W^2=\pi\alpha/G_{\mathrm{F}}\sqrt{2}\sin^2{\theta_W}$, where 
$\sin^{2}\theta_W = g^{\prime2}/(g^{2}+g^{\prime2})$ is the 
weak mixing parameter. The mediator of the neutral-current weak 
interaction, $Z = b_{3}\cos{\theta_{W}} - \mathcal{A}\sin{\theta_{W}}$, 
acquires a mass characterized by 
$M_Z^2=M_W^2/\cos^2{\theta_W}$.  After spontaneous symmetry breaking, 
there remains an unbroken $\mathrm{U(1)}_{\mathrm{em}}$ phase symmetry, so that 
electromagnetism is mediated by a massless photon, $A = 
\mathcal{A}\cos{\theta_{W}} + b_{3}\sin{\theta_{W}}$, coupled to the 
electric charge $e = gg^{\prime}/\sqrt{g^{2} + g^{\prime 2}}$.  As a vestige 
of the spontaneous breaking of the symmetry, there remains a massive, 
spin-zero particle, the Higgs boson.  The mass of the Higgs scalar is 
given symbolically as $M_{H}^{2} = -2\mu^{2} > 0$, but we have no 
prediction for its value.  Though what we take to be the work of the 
Higgs boson is all around us, the Higgs particle itself has not yet 
been observed.
The fermions (the electron in our abbreviated treatment) acquire 
masses as well; these are determined not only by the scale of 
electroweak symmetry breaking, $v$, but also by  their Yukawa interactions with
the scalars.  

To determine the values of the coupling constants and the electroweak 
scale---hence the masses of $W^{\pm}$ and $Z^{0}$---we now examine 
the interactions terms we wrote symbolically in \eqn{eq:scallag}.

\subsubsection{Charged-current interactions}

The interactions of the $W$-boson with the leptons are given by 
\begin{equation}
    \mathcal{L}_{W\mathrm{-lep}} = \frac{-g}{2\sqrt{2}}\left[ 
    \bar{\nu}_{e}\gamma^{\mu}(1 - \gamma_{5})eW^{+}_{\mu} +
    \bar{e}\gamma^{\mu}(1 - \gamma_{5})\nu_{e}W^{-}_{\mu}
    \right], \mathrm{etc.},
    \label{eq:Wleplag}
\end{equation}
so the Feynman rule for the $\nu_{e}eW$ vertex is
\begin{center}
\begin{picture}(160,100)(0,0)
	\ArrowLine(30,50)(10,90)
	\Text(5,85)[]{$e$}
	\ArrowLine(10,10)(30,50)
	\Text(5,15)[]{$\nu$}
	\ZigZag(30,50)(90,50){3}{6}
	\Text(95,58)[]{$\lambda$}
	\Text(110,50)[l]{${\displaystyle\frac{-ig}{2\sqrt{2}}}\gamma_{\lambda}(1 - \gamma_{5})$}
\end{picture}
\end{center}
The $W$-boson propagator (in unitary gauge) is     
\qquad\begin{picture}(100,15)(0,0)
	\ZigZag(0,0)(60,0){3}{4}
	\Text(75,0)[l]{$= \displaystyle{\frac{-i(g_{\mu\nu} -
		k_{\mu}k_{\nu}/M_{W}^{2})}{k^{2} - M_{W}^{2}}}\; .$}
\end{picture}\\[12pt]

Let us compute the cross section for inverse muon decay in the 
electroweak 
theory.   We find
\begin{equation}
    \sigma(\nu_{\mu}e \to \mu\nu_{e}) = 
    \frac{g^{4}m_{e}E_{\nu}}{16\pi M_{W}^{4}} \; \frac{\left[ 1 - 
    (m_{\mu}^{2} - m_{e}^{2})/2m_{e}E_{\nu}\right]^{2}}{(1 + 
    2m_{e}E_{\nu}/M_{W}^{2})}\; ,
    \label{eq:invmudkW}
\end{equation}
which coincides with the familiar four-fermion result at low
energies, provided we identify
\begin{equation}
    \frac{g^{4}}{16M_{W}^{2}} = 2 G_{\mathrm{F}}^{2}\; ,
    \label{eq:gidGF}
\end{equation}
(where $G_{\mathrm{F}} =  1.16639 \times 10^{-5}\gev^{-2}$ is the Fermi constant) which implies that
\begin{equation}
    \frac{g}{2\sqrt{2}} = 
    \left(\frac{G_{\mathrm{F}}M_{W}^{2}}{\sqrt{2}}\right)^{\cfrac{1}{2}}\; .
    \label{eq:gidGF3}
\end{equation}
With the aid of our result for the $W$-boson mass, $M_{W^{\pm}} = 
gv/2$, we 
determine the electroweak scale,
\begin{equation}
    v = \left(G_{\mathrm{F}}\sqrt{2}\right)^{-\cfrac{1}{2}} \approx 246\gev\; ,
    \label{eq:vevval}
\end{equation}
which implies that $\vev{\phi^{0}} = 
    (G_{\mathrm{F}}\sqrt{8})^{-\cfrac{1}{2}} \approx 174\gev$.

Let us now investigate 
the properties of the $W$-boson in terms of its mass, $M_{W}$.  
Consider first the leptonic disintegration of the $W^{-}$, with decay 
kinematics specified thus:
    \begin{center} \begin{picture}(200,100)(0,0)
	\ArrowLine(20,50)(20,90)
	\Text(28,80)[l]{$e(p)$\qquad$\displaystyle{p\approx\left(\frac{M_{W}}{2};
	    \frac{M_{W}\sin\theta}{2}, 0, \frac{M_{W}\cos\theta}{2}\right)}$}
	\ArrowLine(20,50)(20,10)
	\Text(28,20)[l]{$\bar{\nu}_{e}(q)$\qquad$\displaystyle{q\approx\left(\frac{M_{W}}{2};
	    -\,\frac{M_{W}\sin\theta}{2}, 0, -\,\frac{M_{W}\cos\theta}{2}\right)}$}
	\Vertex(20,50){3}
	\Text(12,50)[r]{$W^{-}$}
    \end{picture}   \end{center}
The Feynman amplitude for the decay is
\begin{equation}
    \M = -i \left( \frac{G_{\mathrm{F}}M_{W}^{2}}{\sqrt{2}}\right)^{\cfrac{1}{2}}
    \bar{u}(e,p)\gamma_{\mu}(1 - \gamma_{5})v(\nu,q)\, 
    \varepsilon^{\mu}\; ,
    \label{eq:Wdkamp}
\end{equation}
where $\varepsilon^{\mu}= (0; \hat{\varepsilon})$ is the polarization 
vector of the $W$-boson in its rest frame.  The square of the 
amplitude is
\begin{eqnarray}
    \abs{\M}^{2} & = & \frac{G_{\mathrm{F}}M_{W}^{2}}{\sqrt{2}}
    \tr{\left[ \slashii{\varepsilon}(1-\gamma_{5})\slashii{q}(1+\gamma_{5})
    \slashii{\varepsilon}^{*}\slashiv{p}\right]}  
    \label{eq:Wdk2} \\
     & = & \displaystyle{\frac{8G_{\mathrm{F}}M_{W}^{2}}{\sqrt{2}}}\left[ 
     \varepsilon\cdot q \: \varepsilon^{*}\cdot p -
     \varepsilon \cdot \varepsilon^{*} \: q\cdot p +
     \varepsilon \cdot p \: \varepsilon^{*}\cdot q +
     i\epsilon_{\mu\nu\rho\sigma}\varepsilon^{\mu}q^{\nu}\varepsilon^{*\rho}p^{\sigma}
     \right] \nonumber \; .
\end{eqnarray}
The \textit{decay rate} is independent of the $W$ polarization, so 
let us look first at the case of longitudinal polarization 
$\varepsilon^{\mu}=(0;0,0,1)=\varepsilon^{*\mu}$, to eliminate the 
last term.  For this case, we find
\begin{equation}
     \abs{\M}^{2} = \frac{4G_{\mathrm{F}}M_{W}^{4}}{\sqrt{2}}\sin^{2}\theta\; ,
    \label{eq:Wdk3}
\end{equation}
so the differential decay rate is
\begin{equation}
    \frac{d\Gamma_{0}}{d\Omega} =  \frac{\abs{\M}^{2}}{64\pi^{2}} \: 
    \frac{{\mathcal S}_{12}}{M_{W}^{3}} \; ,
    \label{eq:Wdk4}
\end{equation}
where ${\mathcal S}_{12} = \sqrt{[M_{W}^{2}-(m_{e}+m_{\nu})^{2}]
[M_{W}^{2}-(m_{e}-m_{\nu})^{2}]} = M_{W}^{2}$, so that
\begin{equation}
    \frac{d\Gamma_{0}}{d\Omega} = 
    \frac{G_{\mathrm{F}}M_{W}^{3}}{16\pi^{2}\sqrt{2}}\sin^{2}\theta \; ,
    \label{eq:Wdk5}
\end{equation}
and 
\begin{equation}
    \Gamma(W \to e\nu) = \frac{G_{\mathrm{F}}M_{W}^{3}}{6\pi\sqrt{2}} \; .
    \label{eq:Wdktot}
\end{equation}

\subsubsection{Neutral Currents}
The interactions of the $Z$-boson with leptons are given by 
\begin{equation}
    \mathcal{L}_{Z\mathrm{-}\nu} = \frac{-g}{4\cos\theta_{W}} 
    \bar{\nu}\gamma^{\mu}(1 - \gamma_{5})\nu\:Z_{\mu}
    \label{eq:Znulag}
\end{equation}
and
\begin{equation}
    \mathcal{L}_{Z\mathrm{-}e} = \frac{-g}{4\cos\theta_{W}} 
    \bar{e}\left[L_{e}\gamma^{\mu}(1 - \gamma_{5}) +
    R_{e}\gamma^{\mu}(1 + \gamma_{5})\right]e\:Z_{\mu}\; ,
    \label{eq:Zelag}
\end{equation}
where the chiral couplings are
\begin{eqnarray}
    L_{e} & = & 2 \sin^{2}\theta_{W} - 1 = 2x_{W} + \tau_{3} \; ,
    \nonumber  \\
    R_{e} & = & 2 \sin^{2}\theta_{W}\; .
    \label{eq:lepchicoup}
\end{eqnarray}
By analogy with the calculation of the $W$-boson total width 
\eqn{eq:Wdktot}, we easily compute that
\begin{eqnarray}
    \Gamma(Z \to \nu\bar{\nu}) & = & \frac{G_{\mathrm{F}}M_{Z}^{3}}{12\pi\sqrt{2}}\;,
    \nonumber \\
    \Gamma(Z \to e^{+}e^{-}) & = & \Gamma(Z \to 
    \nu\bar{\nu})\left[L_{e}^{2} + R_{e}^{2}\right]\; .
    \label{eq:Zwidths}
\end{eqnarray}

The neutral weak current mediates a reaction that did not arise in 
the $V-A$ theory, $\nu_{\mu}e \to \nu_{\mu}e$, which proceeds 
entirely by $Z$-boson exchange:
\begin{center} \begin{picture}(130,100)(0,0)
	\ArrowLine(30,50)(10,90)
	\Text(5,85)[]{$\nu_{\mu}$}
	\ArrowLine(10,10)(30,50)
	\Text(5,15)[]{$\nu_{\mu}$}
	\ZigZag(30,50)(100,50){4}{5}
	\ArrowLine(100,50)(120,90)
	\Text(125,85)[]{$e$}
	\ArrowLine(120,10)(100,50)
	\Text(125,5)[]{$e$}
   \end{picture}   \end{center}
This was, in fact, the reaction in which the first evidence for the 
weak neutral current was seen by the Gargamelle collaboration in 
1973~\cite{Hasert:1973ff} 
\begin{figure}[t!]
\begin{center}
\includegraphics[angle=90, height=6cm]{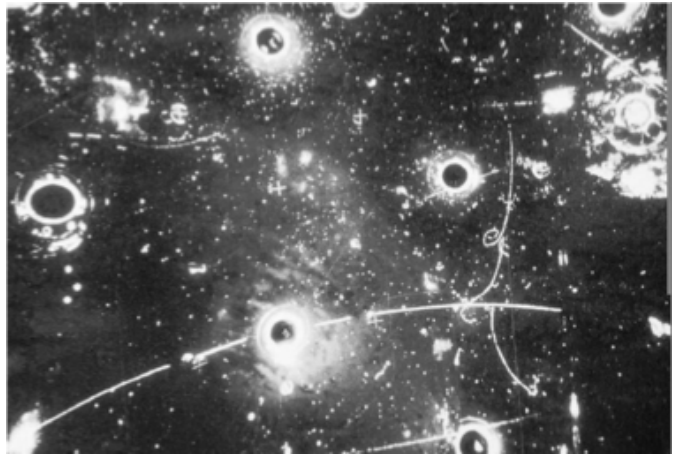}
\caption{First $\nu_{\mu}e$ elastic scattering event observed by the 
Gargamelle Collaboration~\cite{Hasert:1973ff} at CERN. Muon neutrinos 
enter the Freon (CF$_{3}$Br) bubble chamber from the right. A 
recoiling electron appears near the 
center of the image and travels toward the left, 
initiating a shower of curling branches. \label{fig:GGM}}
\end{center}
\end{figure}
(see Figure~\ref{fig:GGM}).

To exercise your calculational muscles, please do
\begin{problem}
It's an easy exercise to compute all the cross 
sections for neutrino-electron elastic scattering.  Show that
\begin{eqnarray}
    \sigma(\nu_{\mu}e \to \nu_{\mu}e) & = &
    \displaystyle{\frac{G_{\mathrm{F}}^{2}m_{e}E_{\nu}}{2\pi}} \left[L_{e}^{2} + 
    R_{e}^{2}/3\right]  \; ,
    \nonumber  \\
    \sigma(\bar{\nu}_{\mu}e \to \bar{\nu}_{\mu}e) & = & 
    \displaystyle{\frac{G_{\mathrm{F}}^{2}m_{e}E_{\nu}}{2\pi}} \left[L_{e}^{2}/3 + 
    R_{e}^{2}\right]  \; ,
    \nonumber  \\
    \sigma(\nu_{e}e \to \nu_{e}e) & = &
    \displaystyle{\frac{G_{\mathrm{F}}^{2}m_{e}E_{\nu}}{2\pi}} \left[(L_{e}+2)^{2} + 
    R_{e}^{2}/3\right]  \; ,
    \nonumber  \\
    \sigma(\bar{\nu}_{e}e \to \bar{\nu}_{e}e) & = & 
    \displaystyle{\frac{G_{\mathrm{F}}^{2}m_{e}E_{\nu}}{2\pi}} \left[(L_{e}+2)^{2}/3 + 
    R_{e}^{2}\right]  \; .
    \label{eq:signueel}
\end{eqnarray}
By measuring all the cross sections, one may undertake a 
``model-independent'' determination of the chiral couplings $L_{e}$ 
and $R_{e}$, or the traditional vector and axial-vector couplings $v$ 
and $a$, which are related through
\begin{equation}
    \begin{array}{ccc}
	a = \cfrac{1}{2}(L_{e}-R_{e}) & \quad & v = 
	\cfrac{1}{2}(L_{e}-R_{e})  \\[3pt]
	L_{e} = v+a & \quad & R_{e} = v-a
    \end{array}\; .
    \label{eq:chiva}
\end{equation}
By inspecting \eqn{eq:signueel}, you can see that even after measuring 
all four cross sections, there remains a two-fold ambiguity: the same 
cross sections result if we interchange $R_{e} \leftrightarrow 
-R_{e}$, or, equivalently, $v \leftrightarrow a$.  The ambiguity is 
resolved by measuring the forward-backward asymmetry in a reaction 
like $e^{+}e^{-} \to \mu^{+}\mu^{-}$ at energies well below the 
$Z^{0}$ mass.  The asymmetry is proportional to 
$(L_{e}-R_{e})(L_{\mu}-R_{\mu})$, or to $a_{e}a_{\mu}$, and so 
resolves the sign ambiguity for $R_{e}$, or the $v$-$a$ ambiguity.
\end{problem}

\subsubsection{Electroweak Interactions of Quarks \label{subsubsec:EWq}}
To extend our theory to include the electroweak interactions of 
quarks, we observe that each generation consists of a left-handed 
doublet
\begin{equation}
    \begin{array}{cccc}
	 & I_{3} & Q & Y = 2(Q - I_{3})  \\[6pt]
	\mathsf{L}_{q}= \left( 
	\begin{array}{c}
	    u  \\[3pt]
	    d
	\end{array}
	\right)_{\mathrm{L}}\quad & 
	\begin{array}{c}
	    + \cfrac{1}{2}  \\[3pt]
	    - \cfrac{1}{2}
	\end{array}
	 & 
	 \begin{array}{c}
	     +\cfrac{2}{3}  \\[3pt]
	     -\cfrac{1}{3}
	 \end{array}
	  & \cfrac{1}{3}\; ,
    \end{array}
    \label{eq:LHq}
\end{equation}
and two right-handed singlets, 
\begin{equation}
    \begin{array}{cccc}
	 & I_{3} & Q & Y = 2(Q - I_{3})  \\[6pt]
	 
	\begin{array}{c}
	    \mathsf{R}_{u} = u_{\mathrm{R}}  \\[3pt]
	    \mathsf{R}_{d} = d_{\mathrm{R}}
	\end{array}
	\quad & 
	\begin{array}{c}
	    0  \\[3pt]
	    0
	\end{array}
	 & 
	 \begin{array}{c}
	     +\cfrac{2}{3}  \\[3pt]
	     -\cfrac{1}{3}
	 \end{array}
	  &          \begin{array}{c}
	     +\cfrac{4}{3}  \\[3pt]
	     -\cfrac{2}{3}
	 \end{array}\; ,

    \end{array}
    \label{eq:RHq}
\end{equation}
Proceeding as before, we find the Lagrangian terms for the $W$-quark 
charged-current interaction,
\begin{equation}
    \mathcal{L}_{W\mathrm{-quark}} = \frac{-g}{2\sqrt{2}}\left[ 
    \bar{u}_{e}\gamma^{\mu}(1 - \gamma_{5})d \:W^{+}_{\mu} +
    \bar{d}\gamma^{\mu}(1 - \gamma_{5})u \:W^{-}_{\mu}
    \right]\; , 
    \label{eq:Wqlag}
\end{equation}
which is identical in form to the leptonic charged-current 
interaction \eqn{eq:Wleplag}.  Universality is ensured by the fact 
that the charged-current interaction is determined by the weak 
isospin of the fermions, and that both quarks and leptons come in 
doublets.

The neutral-current interaction is also equivalent in form to its 
leptonic counterpart, \eqn{eq:Znulag} and \eqn{eq:Zelag}.  We may write it compactly as
\begin{equation}
    \mathcal{L}_{Z\mathrm{-quark}} = \frac{-g}{4\cos\theta_{W}} 
    \sum_{i=u,d} \bar{q}_{i}\gamma^{\mu}\left[ L_{i}(1 - 
    \gamma_{5}) + R_{i}(1 + \gamma_{5})\right]q_{i}\:Z_{\mu}\; ,
    \label{eq:Zqlag}
\end{equation}
where the chiral couplings are
\begin{eqnarray}
    L_{i} & = & \tau_{3} - 2Q_{i} \sin^{2}\theta_{W} \; ,
    \nonumber  \\
    R_{i} & = & -2Q_{i} \sin^{2}\theta_{W}\; .
    \label{eq:qchicoup}
\end{eqnarray}
Again we find a quark-lepton universality in the form---but not the 
values---of the chiral couplings.

\subsubsection{Trouble in Paradise \label{subsubsec:paradis}}
Until now, we have based our construction on the idealization that 
the $u \leftrightarrow d$ transition is of universal strength.  The 
unmixed doublet
\begin{displaymath}
    \left( 
    \begin{array}{c}
	u  \\
	d
    \end{array}
    \right)_{\mathrm{L}}
\end{displaymath}
does not quite describe our world.  We attain a better description by 
replacing
\begin{displaymath}
    \left( 
    \begin{array}{c}
	u  \\
	d
    \end{array}
    \right)_{\mathrm{L}} \to
    \left( 
    \begin{array}{c}
	u  \\
	d_{\theta}
    \end{array}
    \right)_{\mathrm{L}} \; ,   
\end{displaymath}
where 
\begin{equation}
    d_{\theta} \equiv d\,\cos\theta_{C} + s\,\sin\theta_{C}\; ,
    \label{eq:cabrot}
\end{equation}
with $\cos\theta_{C} = 0.9736 \pm 0.0010$.\footnote{The arbitrary 
Yukawa couplings that give masses to the quarks can easily be chosen 
to yield this result.}  The change to the ``Cabibbo-rotated'' doublet 
perfects the charged-current interaction---at least up to small 
third-generation effects that we could easily incorporate---but leads 
to serious trouble in the neutral-current sector, for which the 
interaction now becomes
\begin{eqnarray}
    \mathcal{L}_{Z\mathrm{-quark}} & = & \frac{-g}{4\cos\theta_{W}} \:Z_{\mu}
     \left\{\bar{u}\gamma^{\mu}\left[ L_{u}(1 - 
    \gamma_{5}) + R_{u}(1 + \gamma_{5})\right]u \right. \nonumber \\
      & & + \bar{d}\gamma^{\mu}\left[ L_{d}(1 - 
    \gamma_{5}) + R_{d}(1 + 
    \gamma_{5})\right]d\,\cos^{2}\theta_{C}  \nonumber \\
      & &   + \bar{s}\gamma^{\mu}\left[ L_{d}(1 - 
    \gamma_{5}) + R_{d}(1 + \gamma_{5})\right]s\,\sin^{2}\theta_{C} 
    \nonumber \\
      & &  +  \bar{d}\gamma^{\mu}\left[ L_{d}(1 - 
    \gamma_{5}) + R_{d}(1 + 
    \gamma_{5})\right]s\,\sin\theta_{C}\cos\theta_{C} 
    \nonumber \\
      & &   + \left. \bar{s}\gamma^{\mu}\left[ L_{d}(1 - 
    \gamma_{5}) + R_{d}(1 + 
    \gamma_{5})\right]d\,\sin\theta_{C}\cos\theta_{C} \right\}
    \; ,
    \label{eq:Zqrotlag}
\end{eqnarray}
Until the discovery and systematic study of the weak neutral current, 
culminating in the heroic measurements made at LEP and the SLC, there 
was not enough knowledge to challenge the first three terms.  The last 
two \textit{strangeness-changing} terms were known to be poisonous, 
because many of the early experimental searches for neutral currents 
were fruitless searches for precisely this sort of interaction.  
Strangeness-changing neutral-current interactions are not seen at an 
impressively low level.\footnote{For more on rare kaon decays, see 
the TASI 2000 lectures by Tony Barker~\cite{Barker:2000fi} and Gerhard 
Buchalla~\cite{Buchalla:2001ux}.}

Only recently have Brookhaven Experiments 787~\cite{Adler:2004hp} and 
939~\cite{Artamonov:2004hr}  detected three 
 candidates for the decay $K^{+} \to \pi^{+}\nu\bar{\nu}$,
    \begin{center} \begin{picture}(100,80)(0,0)
	\ArrowLine(50,30)(20,30)
	\ArrowLine(80,30)(50,30)
	\Vertex(50,30){2}
	\ArrowLine(20,20)(80,20)
	\Text(17,25)[r]{$K^{+}$}
	\Text(85,25)[bl]{$\pi^{+}$}
	\Text(25,35)[b]{$\bar{s}$}
	\Text(75,35)[b]{$\bar{d}$}
	\Text(25,15)[t]{$u$}
	\ZigZag(50,30)(65,55){2}{5}
	\ArrowLine(60,70)(65,55)
	\Text(60,76)[]{$\bar{\nu}$}
	\Text(85,65)[]{$\nu$}
	\ArrowLine(65,55)(80,60)
    \end{picture}   \end{center}
and inferred a branching ratio ${\mathcal B}(K^{+} \to 
\pi^{+}\nu\bar{\nu}) = 1.47^{+1.30}_{-0.89}\times 10^{-10}$.

The good 
agreement between the standard-model prediction, ${\mathcal B}(K_{L} 
\to \mu^{+}\mu^{-}) = 0.77 \pm 0.11 \times 10^{-10}$ (through the 
process $K_{L} \to \gamma\gamma \to \mu^{+}\mu^{-}$), and 
experiment~\cite{Ambrose:2000gj} 
leaves little room for a strangeness-changing neutral-current 
contribution:  
    \begin{center} \begin{picture}(150,50)(0,0)
	\ArrowLine(70,30)(20,30)
	\ArrowLine(20,20)(70,20)
	\CArc(70,25)(5,-90,90)
	\Vertex(75,25){2}
	\ZigZag(75,25)(115,25){2}{5}
	\ArrowLine(125,50)(115,25)
	\ArrowLine(115,25)(125,0)
	
	\Text(17,25)[r]{$K^{+}$}
	\Text(130,45)[l]{$\mu^{+}$}
	\Text(130,5)[l]{$\mu^{-}$}
	\Text(25,35)[b]{$\bar{s}$}
	\Text(25,15)[t]{$d$}
	\Text(149,25)[r]{,}
    \end{picture}   \end{center}
that is easily normalized to the normal charged-current leptonic decay 
of the $K^{+}$:
    \begin{center} \begin{picture}(150,50)(0,0)
	\ArrowLine(70,30)(20,30)
	\ArrowLine(20,20)(70,20)
	\CArc(70,25)(5,-90,90)
	\Vertex(75,25){2}
	\ZigZag(75,25)(115,25){2}{5}
	\ArrowLine(125,50)(115,25)
	\ArrowLine(115,25)(125,0)
	
	\Text(17,25)[r]{$K^{+}$}
	\Text(130,45)[l]{$\mu^{+}$}
	\Text(130,5)[l]{$\nu$}
	\Text(25,35)[b]{$\bar{s}$}
	\Text(25,15)[t]{$u$}
	\Text(149,25)[r]{.}
    \end{picture}   \end{center}

The cure for this fatal disease was put forward by Glashow, 
Iliopoulos, and Maiani~\cite{Glashow:1970gm}.  Expand the model of quarks to 
include two left-handed doublets,
\begin{equation}
		\left(
		\begin{array}{c}
			\nu_{e}  \\
			e^{-}
		\end{array}
		 \right)_{\mathrm{L}} \;\;
		\left(
		\begin{array}{c}
			\nu_{\mu}  \\
			\mu^{-}
		\end{array}
		 \right)_{\mathrm{L}} 
  \;\;\;\;\;\;
		\left(
		\begin{array}{c}
			u  \\
			d_{\theta}
		\end{array}
		 \right)_{\mathrm{L}} \;\;
		\left(
		\begin{array}{c}
			c  \\
			s_{\theta}
		\end{array}
		 \right)_{\mathrm{L}}  \; ,  \label{eq:gim}
\end{equation}
where 
\begin{equation}
    s_{\theta} = s\,\cos\theta_{C} - d\,\sin\theta_{C}\; ,
    \label{eq:sthdef}
\end{equation}
plus the corresponding right-handed singlets, $e_{\mathrm{R}}$, $\mu_{\mathrm{R}}$, 
$u_{\mathrm{R}}$, $d_{\mathrm{R}}$, $c_{\mathrm{R}}$, and $s_{\mathrm{R}}$.  This required the 
introduction of the charmed quark, $c$, which had not yet been 
observed.  By the addition of the second quark generation, the 
flavor-changing cross terms vanish in the $Z$-quark interaction, and 
we are left with:    \begin{center} \begin{picture}(280,100)(0,0)
	\ArrowLine(30,50)(10,90)
	\Text(5,85)[]{$q_{i}$}
	\ArrowLine(10,10)(30,50)
	\Text(5,15)[]{$q_{i}$}
	\ZigZag(30,50)(90,50){3}{6}
	\Text(95,58)[]{$\lambda$}
	\Text(110,50)[l]{${\displaystyle\frac{-ig}{4\cos\theta_{W}}}\gamma_{\lambda}[(1 - 
	\gamma_{5})L_{i} + (1 + \gamma_{5})R_{i}]$\quad ,}

    \end{picture}   \end{center}
which is flavor diagonal!

The generalization to $n$ quark doublets is straightforward.  Let the 
charged-current interaction be
\begin{equation}
    \mathcal{L}_{W\mathrm{-quark}} = \frac{-g}{2\sqrt{2}}\left[\bar{\Psi}\gamma^{\mu}
    (1 - \gamma_{5}){\mathcal O}\Psi\:W^{+}_{\mu} + \mathrm{h.c.} \right]\; ,
    \label{eq:compWq}
\end{equation}
where the composite quark spinor is 
\begin{equation}
    \Psi = \left( 
    \begin{array}{c}
	u  \\
	c  \\
	\vdots  \\
	   \\
	d  \\
	s  \\
	\vdots
    \end{array}
    \right)
    \label{eq:compspin}
\end{equation}
and the flavor structure is contained in
\begin{equation}
    {\mathcal O} = \left( 
    \begin{array}{cc}
	0 & U  \\
	0 & 0
    \end{array}
    \right)\; ,
    \label{eq:flav}
\end{equation}
where $U$ is the unitary quark-mixing matrix.  The weak-isospin 
contribution to the neutral-current interaction has the form
\begin{equation}
    \mathcal{L}_{Z\mathrm{-quark}}^{\mathrm{iso}} = \frac{-g}{4\cos\theta_{W}}
    \bar{\Psi}\gamma^{\mu}(1 - \gamma_{5})\left[{\mathcal O},{\mathcal 
    O}^{\dagger}\right]\Psi\; .
    \label{eq:isoZq}
\end{equation}
Since the commutator
\begin{equation}
    \left[{\mathcal O},{\mathcal O}^{\dagger}\right] =
    \left(
    \begin{array}{cc}
	I & 0  \\
	0 & -I
    \end{array}
    \right)
    \label{eq:commut}
\end{equation}
the neutral-current interaction is flavor diagonal, and the 
weak-isospin piece is, as expected, proportional to $\tau_{3}$.

In general, the $n \times n$ quark-mixing matrix $U$ can be 
parametrized in terms of $n(n-1)/2$ real mixing angles and 
$(n-1)(n-2)/2$ complex phases, after exhausting the freedom to 
redefine the phases of quark fields.  The $3\times 3$ case of three 
mixing angles and one phase, often called the 
Cabibbo--Kobayashi-Maskawa matrix, presaged the discovery of the third 
generation of quarks and leptons~\cite{Kobayashi:1973fv}.

\subsection{Precision Tests of the Electroweak 
Theory \label{subsec:prete}}
In its 
simplest form, with the electroweak gauge symmetry broken by the Higgs 
mechanism, the \ewgg\ theory has scored many 
qualitative successes: the prediction of neutral-current interactions, 
the necessity of charm, the prediction of the existence and properties 
of the weak bosons $W^{\pm}$ and $Z^{0}$.  Over the past ten years, in 
great measure due to the beautiful experiments carried out at the $Z$ 
factories at CERN and SLAC, precision measurements have tested the
electroweak theory as a quantum field theory~\cite{Swartz:1999xv, Altarelli:2004fq}, at
the one-per-mille level, as indicated in Table \ref{tbl:Zmeas}.  
\begin{table}[h!]
    \centering
    \caption{Precision measurements at the $Z^{0}$ pole. (For 
    sources of the data, see 
    {\protect\cite{PDBook}} and  \cite{LEPEWWG}.)}
    \vspace*{6pt}
    \begin{tabular}{cc}
	    \hline\\[-6pt]
	    $M_{Z}$ & $91\,187.6 \pm 2.1\mev$ \\
	    $\Gamma_{Z}$ & $2\,495.2 \pm 2.3\mev$  \\
	    $\sigma^{0}_{\mathrm{hadronic}}$ & $41.540 \pm 0.037\nb$  \\
	    $\Gamma_{\mathrm{hadronic}}$ & $1744.4 \pm 2.0\mev$  \\
	    $\Gamma_{\mathrm{leptonic}}$ & $83.984 \pm 0.086\mev$  \\
	    $\Gamma_{\mathrm{invisible}}$ & $499.0 \pm 1.5\mev$  \\[4pt]
	    \hline
    \end{tabular}
    \label{tbl:Zmeas}
\end{table}

\begin{figure}[t!]
\begin{center}
\includegraphics[width=8.1cm]{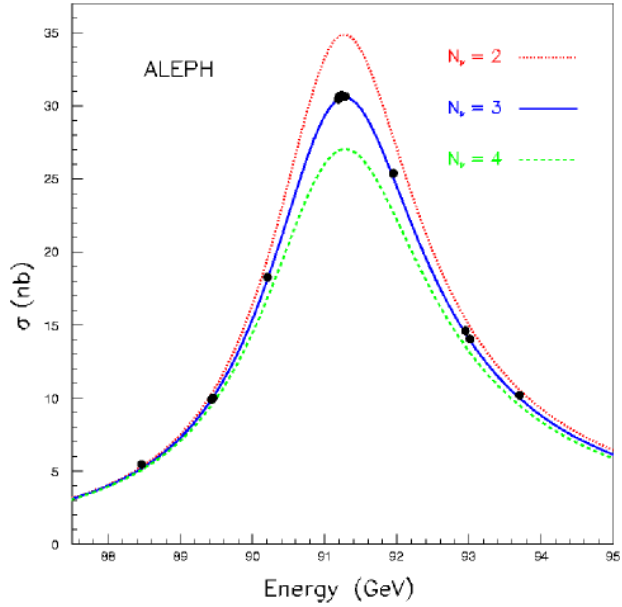}
\vspace*{-6pt}
\caption{Comparison of the $Z^{0}$ line shape with predictions based 
on two, three, and four light neutrino species~\cite{alephZ}.}
	\label{fig:zshape}
\end{center}
\end{figure}

A classic achievement of the $Z$ factories is the determination of the
number of light neutrino species.  If we define the invisible width of
the $Z^{0}$ as
\begin{equation}
    \Gamma_{\mathrm{invisible}} = \Gamma_{Z} - 
    \Gamma_{\mathrm{hadronic}} - 
    3\Gamma_{\mathrm{leptonic}}\; ,
    \label{eq:gaminv}
\end{equation}
then we can compute the number of light neutrino species as
\begin{equation}
    N_{\nu} = \Gamma_{\mathrm{invisible}}/\Gamma^{\mathrm{SM}}(Z \to 
    \nu_{i}\bar{\nu}_{i})\;.
    \label{eq:nnudef}
\end{equation}
A typical current value is $N_{\nu} = 2.994 \pm 0.012$, in excellent 
agreement with the observation of light $\nu_{e}$, $\nu_{\mu}$, and 
$\nu_{\tau}$. A graphical indication that only three neutrino species 
are accessible as $Z^{0}$ decay products is given in 
Figure~\ref{fig:zshape}. 

\begin{figure}[t!]
\begin{center}
\includegraphics[width=14cm]{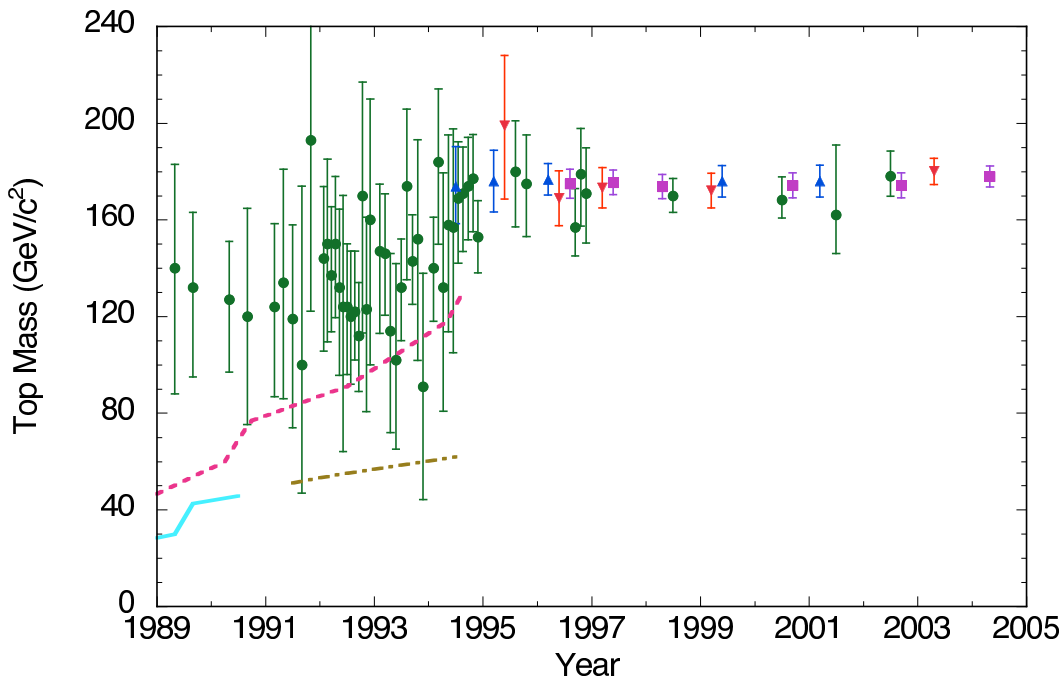}
\caption{Indirect determinations of the top-quark mass from fits to 
electroweak observables (open circles) and 95\% confidence-level lower 
bounds on the top-quark mass inferred from direct searches in 
$e^{+}e^{-}$ annihilations (solid line) and in $\bar{p}p$ collisions, 
assuming that standard decay modes dominate (broken line).  An 
indirect lower bound, derived from the $W$-boson width inferred from 
$\bar{p}p \rightarrow (W\hbox{ or }Z)+\hbox{ anything}$, is shown as 
the dot-dashed line.  Direct measurements of $m_{t}$ by the CDF 
(triangles) and D\O\ (inverted triangles) Collaborations are shown at 
the time of initial evidence, discovery claim, and at the conclusion 
of Run 1.  The world averages from direct observations are shown as 
squares.  For sources of data, see Ref.  {\protect\cite{PDBook}}. 
(From Ref.\
{\protect\cite{Quigg:1997uh}}.)}
	\label{fig:EWtop}
\end{center}
\end{figure}

As an example of the insights precision 
measurements have brought us (one that mightily impressed the Royal 
Swedish Academy of Sciences in 1999), I show in Figure \ref{fig:EWtop} the time 
evolution of the top-quark mass favored by simultaneous fits to many 
electroweak observables.  Higher-order processes involving virtual top 
quarks are an important element in quantum corrections to the 
predictions the electroweak theory makes for many observables.  
A new world-average top mass has been reported by the Tevatron 
Collider experiments~\cite{Group:2004rc}: $m_t = 178.0 \pm 4.3\gev$.

\begin{figure}[t!]
\begin{center}
\includegraphics[width=9cm]{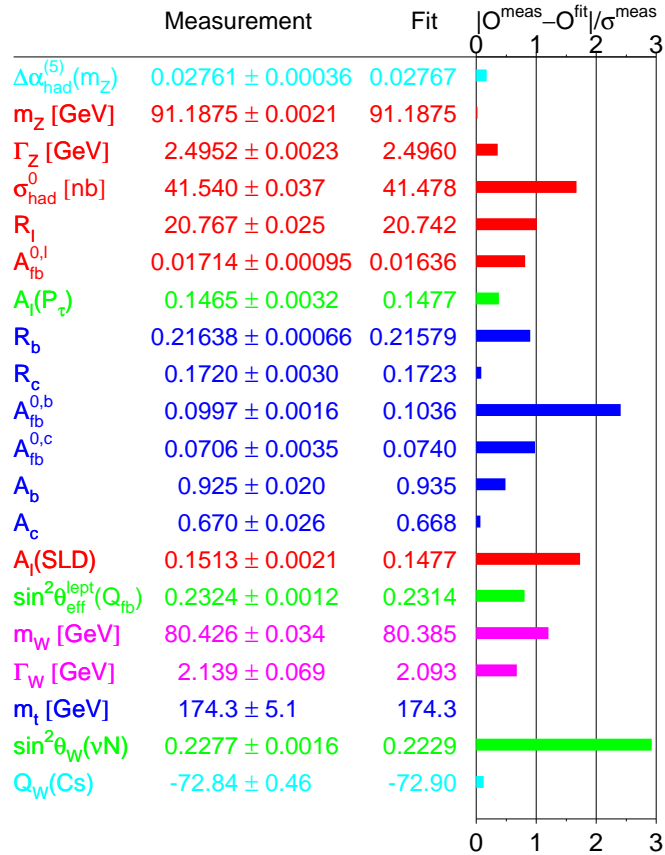}
\vspace*{-36pt}
\caption{Precision electroweak measurements and the pulls they exert 
on a global fit to the standard model, from Ref.\ 
{\protect\cite{LEPEWWG}}.}
\label{fig:pulls}
\end{center}
\end{figure}

The comparison between the electroweak theory and a considerable 
universe of data is shown in Figure \ref{fig:pulls}
where the pull, or difference between the global fit and measured 
value in units of standard deviations, is shown for some twenty 
observables~\cite{LEPEWWG}.
The distribution of pulls for this fit, due to the LEP Electroweak 
Working Group, is not noticeably different from a normal 
distribution, and only a couple of observables differ from the fit by 
as much as about two standard deviations.  This is the case for any 
of the recent fits.  From fits of the kind represented here, 
we learn that the standard-model interpretation of the data favors a 
light Higgs boson.  We will revisit this conclusion in 
\S\ref{Hclues}.

The beautiful agreement between the electroweak theory and a vast 
array of data from neutrino interactions, hadron collisions, and 
electron-positron annihilations at the $Z^{0}$ pole and beyond means 
that electroweak studies have become a modern arena in which we can 
look for new physics ``in the sixth place of 
decimals.''

\subsection{Why the Higgs boson must exist}
\begin{figure}[t!]
\begin{center}
\includegraphics[width=10cm]{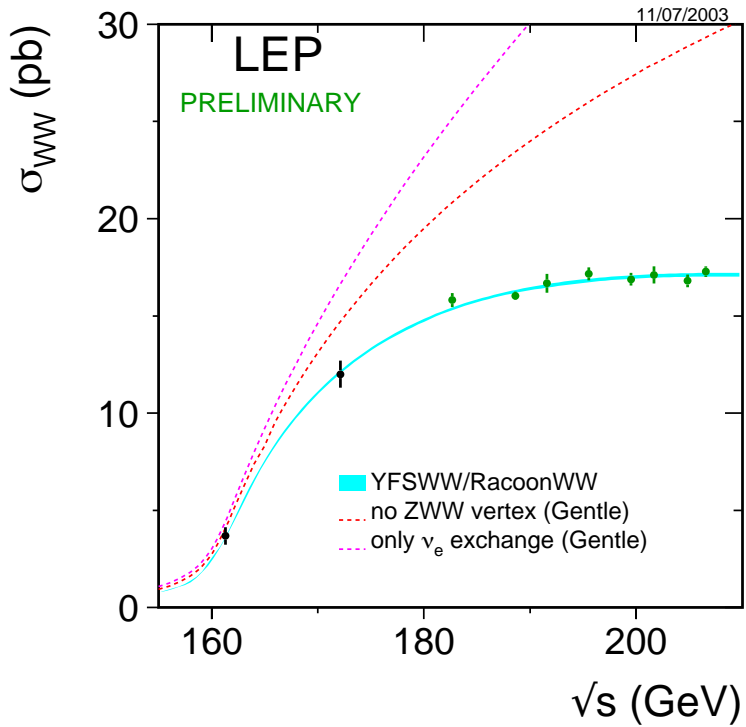}
\caption{Cross section for the reaction $e^{+}e^{-} \to W^{+}W^{-}$
measured by the four LEP experiments, together with the full
electroweak-theory simulation and the cross sections that would result
from $\nu$-exchange alone and from
$(\nu+\gamma)$-exchange~\cite{LEPEWWG}.  \label{fig:LEPgc}.}
\end{center}
\end{figure}

How can we be sure that a Higgs boson, or something very like it, will be 
found? One 
path to the \emph{theoretical} discovery of the Higgs boson
involves its role in the cancellation of 
high-energy divergences. An illuminating example is provided by the 
reaction
\begin{figure}[b!]
    \begin{center}
	\includegraphics[height=6cm]{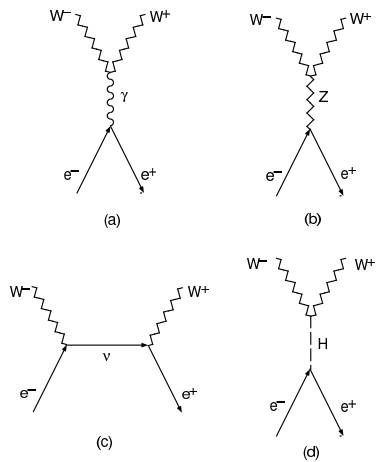}
	\caption{Lowest-order contributions to the $e^+e^- \rightarrow 
	W^{+}W^{-}$ scattering amplitude.}
	\protect\label{fig:eeWW}
    \end{center}
\end{figure}
$e^+e^- \to W^+W^-$,
which is described in lowest order by the four 
Feynman graphs in Figure \ref{fig:eeWW}. The contributions of the direct-channel 
$\gamma$- and $Z^0$-exchange diagrams 
of Figs.~\ref{fig:eeWW}(a) and (b) cancel the leading divergence in the $J=1$ 
partial-wave amplitude of 
the neutrino-exchange diagram in Figure~\ref{fig:eeWW}(c).  This is 
the famous ``gauge cancellation'' observed in experiments at LEP~2 
and the Tevatron.  The LEP measurements in Figure~\ref{fig:LEPgc}
agree well with the predictions of electroweak-theory Monte Carlo
generators, which predict a benign high-energy behavior.  If the 
$Z$-exchange contribution is omitted (middle dashed line) or if both the 
$\gamma$- and $Z$-exchange contributions are omitted (upper dashed 
line), the calculated cross section grows unacceptably with 
energy---and disagrees with the measurements.  The gauge cancellation 
in the $J=1$ partial-wave amplitude is thus observed.

However, this is not the end of the high-energy story: the $J=0$
partial-wave amplitude, which exists in this case because the
electrons are massive and may therefore be found in the ``wrong''
helicity state, grows as $s^{1/2}$ for the production of
longitudinally polarized gauge bosons.  The resulting divergence is
precisely cancelled by the Higgs boson graph of
Figure~\ref{fig:eeWW}(d).  If the Higgs boson did not exist, something
else would have to play this role.  From the point of view of
$S$-matrix analysis, the Higgs-electron-electron coupling must be
proportional to the electron mass, because ``wrong-helicity''
amplitudes are always proportional to the fermion mass.

Let us underline this result.
If the gauge symmetry were unbroken, there would be 
no Higgs boson, no longitudinal gauge bosons, and no extreme divergence 
difficulties. But there would be no viable low-energy phenomenology
 of the 
weak interactions. The most severe divergences of individual diagrams 
are eliminated by the gauge 
structure of the couplings among gauge bosons and leptons. A lesser, but 
still potentially fatal, divergence arises because the electron has 
acquired mass---because of the Higgs mechanism. Spontaneous symmetry 
breaking provides its own cure by supplying a Higgs boson to remove the 
last divergence. A similar interplay and compensation must exist in any 
satisfactory theory.

\subsection{The vacuum energy problem} 
I want to spend a moment to revisit a 
longstanding, but usually unspoken, challenge to the completeness of 
the electroweak theory as we have defined it: the vacuum energy 
problem~\cite{Linde:1974at, Veltman:1975au}.
I do so not only for its intrinsic interest, but also to 
raise the question, ``Which problems of completeness and 
consistency do we worry about at a given moment?''  It is perfectly 
acceptable science---indeed, it is often essential---to put certain 
problems aside, in the expectation that we will return to them at the 
right moment.  What is important is never to forget that the problems 
are there, even if we do not allow them to paralyze us.  

\begin{figure}[t!]
\begin{center}
\includegraphics[width=9cm]{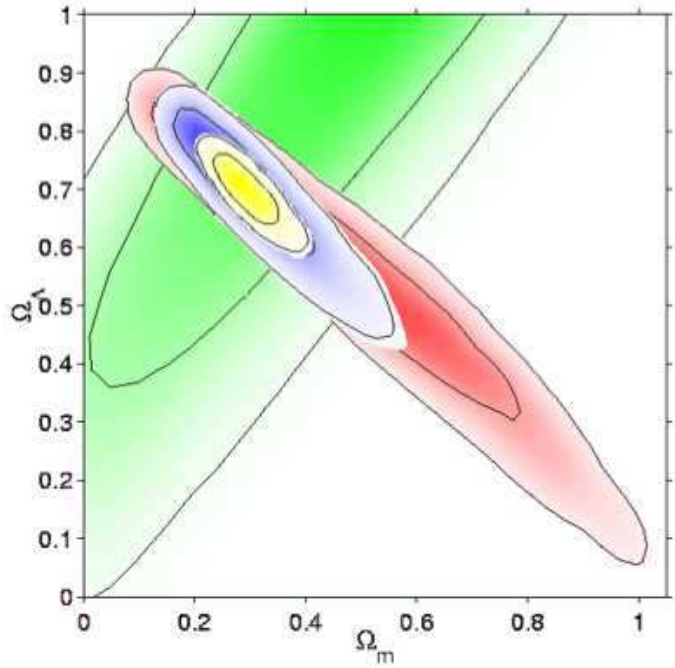}
\caption{Constraints on the mass ($\Omega_{m}$) and dark-energy 
($\Omega_{\Lambda}$) densities of the universe, at
68\% and 95\% confidence levels.  Bottom layer (green): supernova
constraints; next layer up (red): CMB data alone; next (blue): CMB data
plus HST Key Project prior; top layer (yellow): all data combined
\cite{Lewis:2002ah}\label{fig:CosConc}.}
\end{center}
\end{figure}

For the usual Higgs potential, 
$V(\phi^{\dagger}\phi) = \mu^{2}(\phi^{\dagger}\phi) + 
\abs{\lambda}(\phi^{\dagger}\phi)^{2}$, the value of 
the potential at the minimum is
\begin{equation}
    V(\vev{\phi^{\dagger}\phi}) = \frac{\mu^{2}v^{2}}{4} = 
    - \frac{\abs{\lambda}v^{4}}{4} < 0.
    \label{minpot}
\end{equation}
Identifying $M_{H}^{2} = -2\mu^{2}$, we see that the Higgs potential 
contributes a field-independent constant term,
\begin{equation}
    \varrho_{H} \equiv \frac{M_{H}^{2}v^{2}}{8}.
    \label{eq:rhoH}
\end{equation}
I have chosen the notation $\varrho_{H}$ because the constant term in the 
Lagrangian plays the role of a vacuum energy density.  When we 
consider gravitation, adding a vacuum energy density 
$\varrho_{\mathrm{vac}}$ is equivalent to adding a cosmological constant 
term to Einstein's equation.  Although recent
observations~\footnote{For a cogent summary of current knowledge of the
cosmological parameters, including evidence for a cosmological
constant, see Ref.\ \cite{Carroll:2003qq}. For a useful summary of gravitational
theory, see the essay by T. d'Amour in \S14 of the 2000 \textit{Review
of Particle Physics,} Ref.\ \cite{Groom:2000in}.} raise the intriguing possibility
that the cosmological constant may be different from zero (see 
Figure~\ref{fig:CosConc}), the
essential observational fact is that the vacuum energy density must be
very tiny indeed,
\begin{equation}
    \varrho_{\mathrm{vac}} \ltap 10^{-46}\gev^{4}\; .
    \label{eq:rhovaclim}
\end{equation}
Therein lies the puzzle: if we take
$v = (G_{\mathrm{F}}\sqrt{2})^{-\frac{1}{2}}  \approx 246\gev$  
and insert the current experimental lower bound~\cite{Barate:2003sz} 
$M_{H} \gtap 114.4\gev$ into \eqn{eq:rhoH}, we find that the 
contribution of the Higgs field to the vacuum energy density is
\begin{equation}
    \varrho_{H} \gtap  10^{8}\gev^{4},
    \label{eq:rhoHval}
\end{equation}
some 54 orders of magnitude larger than the upper bound inferred from 
the cosmological constant.

What are we to make of this mismatch?  The fact that $\varrho_{H} \gg 
\varrho_{\mathrm{vac}}$ means that the smallness of the cosmological 
constant needs to be explained.  In a unified theory of the strong, 
weak, and electromagnetic interactions, other (heavy!) Higgs fields 
have nonzero vacuum expectation values that may give rise to still 
greater mismatches.  At a fundamental level, we can therefore conclude 
that a spontaneously broken gauge theory of the strong, weak, and 
electromagnetic interactions---or merely of the electroweak 
interactions---cannot be complete.  Either we must find a separate 
principle to zero the vacuum energy density of the Higgs field, or 
we may suppose that a proper quantum theory of gravity, in combination 
with the other interactions, will resolve the puzzle of the 
cosmological constant.  The vacuum energy problem must be an important 
clue.  But to what?

\subsection{Bounds on $M_{H}$}
The Standard Model does not give a precise 
prediction for the mass of the Higgs boson. We can, however, use arguments 
of self-consistency to place plausible lower and upper bounds on the mass of 
the Higgs particle in the minimal model. Unitarity arguments~\cite{Lee:1977eg} lead to a conditional upper bound on the Higgs 
boson mass. It is straightforward to compute the 
amplitudes ${\cal M}$ for gauge boson scattering at high energies, and to make
a partial-wave decomposition, according to
\begin{equation}
      {\cal M}(s,t)=16\pi\sum_J(2J+1)a_J(s)P_J(\cos{\theta}) \; .
\end{equation}
 Most channels ``decouple,'' in the sense 
that partial-wave amplitudes are small at all energies (except very
near the particle poles, or at exponentially large energies), for
any value of the Higgs boson mass $M_H$. Four channels are interesting:
\begin{equation}
\begin{array}{cccc}
W_L^+W_L^- & Z_L^0Z_L^0/\sqrt{2} & HH/\sqrt{2} & HZ_L^0 \; ,
\end{array}
\end{equation}
where the subscript $L$ denotes the longitudinal polarization
states, and the factors of $\sqrt{2}$ account for identical particle
statistics. For these channels, the $s$-wave amplitudes are all asymptotically
constant (\ie, well-behaved) and  
proportional to $G_{\mathrm{F}}M_H^2.$ In the high-energy 
limit,\footnote{It is convenient to calculate these amplitudes by 
means of the Goldstone-boson equivalence theorem, which 
reduces the dynamics of longitudinally polarized gauge bosons to a 
scalar field theory with interaction Lagrangian given by 
$\mathcal{L}_{\mathrm{int}} = -\lambda v h 
(2w^{+}w^{-}+z^{2}+h^{2}) - 
(\lambda/4)(2w^{+}w^{-}+z^{2}+h^{2})^{2}$, with $1/v^{2} = 
G_{\mathrm{F}}\sqrt{2}$ and $\lambda = G_{\mathrm{F}}M_{H}^{2}/\sqrt{2}$. {In the
high-energy limit, an amplitude for longitudinal gauge-boson
interactions may be replaced by a corresponding amplitude for the
scattering of massless Goldstone bosons: $\mathcal{M}(W_{L},Z_{L}) =
\mathcal{M}(w,z) + \mathcal{O}(M_{W}/\sqrt{s})$. The 
equivalence theorem can be traced to the work of Cornwall, 
  Levin, and  Tiktopoulos~\cite{Cornwall:1974km}.  It was applied to this problem by Lee, Quigg, and 
Thacker~\cite{Lee:1977eg}, and developed extensively by Chanowitz and 
Gaillard~\cite{Chanowitz:1985hj}, and others.}}
\begin{equation}
\lim_{s\gg M_H^2}(a_0)\to\frac{-G_{\mathrm{F}} M_H^2}{4\pi\sqrt{2}}\cdot \left[
\begin{array}{cccc} 1 & 1/\sqrt{8} & 1/\sqrt{8} & 0 \\
      1/\sqrt{8} & 3/4 & 1/4 & 0 \\
      1/\sqrt{8} & 1/4 & 3/4 & 0 \\
      0 & 0 & 0 & 1/2 \end{array} \right] \; .
\end{equation} 
Requiring that the largest eigenvalue respect the 
partial-wave unitarity condition $\abs{a_0}\le 1$ yields
\begin{equation}
	M_H \le \left(\frac{8\pi\sqrt{2}}{3G_{\mathrm{F}}}\right)^{1/2} =1\tev
\end{equation}
as a condition for perturbative unitarity.

If the bound is respected, weak interactions remain weak at all
energies, and perturbation theory is everywhere reliable. If the
bound is violated, perturbation theory breaks down, and weak
interactions among $W^\pm$, $Z$, and $H$ become strong on the \onetev.
This means that the features of strong interactions at GeV energies
will come to characterize electroweak gauge boson interactions at
TeV energies. We interpret this to mean that new phenomena are to
be found in the electroweak interactions at energies not much larger
than 1~TeV.

Lower bounds on the Higgs mass that follow from the 
requirement that electroweak symmetry be broken in the vacuum, even 
in  the presence of quantum corrections, date from the work of 
Linde~\cite{Linde:1976sw} and Weinberg~\cite{Weinberg:1976pe}.
The effects of heavy fermions---important for the top quark---are explored in \cite{Hung:1979dn, 
Altarelli:1994rb, Espinosa:1995se}.

\subsection{The electroweak scale and beyond \label{subsec:hierarchy}}
We have seen that the scale of electroweak symmetry breaking, $v =
(G_{\mathrm{F}}\sqrt{2})^{-\frac{1}{2}} \approx 246\gev$, sets the values of
the $W$- and $Z$-boson masses.  But the electroweak scale is not the
only scale of physical interest.  It seems certain that we must also
consider the Planck scale, derived from the strength of Newton's
constant, and it is also probable that we must take account of the
\smgg\ unification scale around
$10^{15\mathrm{-}16}\gev$.  There may well be a distinct flavor scale. 
The existence of other significant energy scales is behind the famous
problem of the Higgs scalar mass: how to keep the distant scales from
mixing in the face of quantum corrections, or how to stabilize the
mass of the Higgs boson on the electroweak scale, or why is the 
electroweak scale small? We call this puzzle the hierarchy problem.

The \ewgg\ electroweak theory does not explain how the 
scale of electroweak symmetry breaking is maintained in the presence 
of quantum corrections.  The problem of the scalar sector can be 
summarized neatly as follows~\cite{Veltman:1981mj, LlewellynSmith:1984zn}. The Higgs potential is
\begin{equation}
      V(\phi^\dagger \phi) = \mu^2(\phi^\dagger \phi) +
\abs{\lambda}(\phi^\dagger \phi)^2 \;.
\end{equation}
With $\mu^2$ chosen to be less than zero, the electroweak symmetry is 
spontaneously broken down to the $\mathrm{U(1)}$ of electromagnetism, as the 
scalar field acquires a vacuum expectation value that is fixed by the low-energy
phenomenology, 
\begin{equation}
	\vev{\phi} = \sqrt{-\mu^2/2|\lambda|} \equiv (G_{\mathrm{F}}\sqrt 8)^{-1/2}
		\approx 174 {\rm \;GeV}\;.
		\label{hvev}
\end{equation}

Beyond the classical approximation, scalar mass parameters receive 
quantum corrections from loops that contain particles of spins 
$J=1, 1/2$, and $0$:
\begin{equation}
\raisebox{-24pt}{\includegraphics[width=12cm]{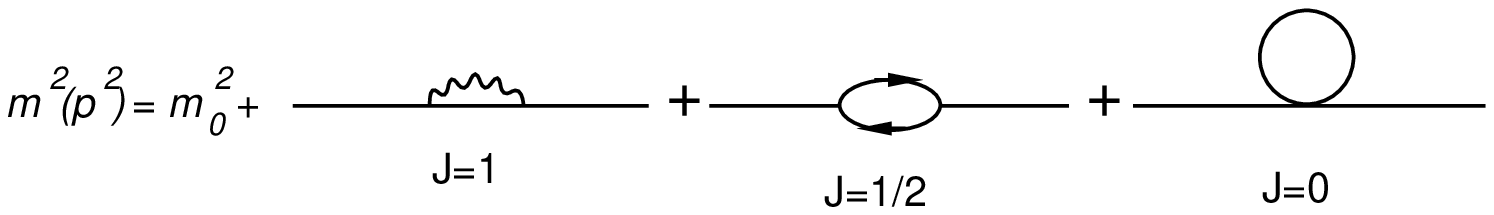}}
\label{loup}
\end{equation}
The loop integrals are potentially divergent.  Symbolically, we may 
summarize the content of \eqn{loup} as
\begin{equation}
	m^2(p^2) = m^2(\Lambda^2) + Cg^2\int^{\Lambda^2}_{p^2}dk^2 
	+ \cdots \;,
	\label{longint}
\end{equation}
where $\Lambda$ defines a reference scale at which the value of 
$m^2$ is known, $g$ is the coupling constant of the theory, and the 
coefficient $C$ is calculable in any particular theory.  
Instead of dealing with the relationship between observables and 
parameters of the Lagrangian, we choose to describe the variation of 
an observable with the momentum scale.  In order for the mass shifts 
induced by radiative corrections to remain under control (\ie , not to 
greatly exceed the value measured on the laboratory scale), either 
$\Lambda$ must be small, so the range of integration is not 
enormous, or new physics must intervene to cut off the integral.

If the fundamental interactions are described by an 
\smgg\ gauge symmetry, \ie, by quantum
chromodynamics and the electroweak theory, then the 
natural reference scale is the Planck mass,\footnote{It is because
$M_{\mathrm{Planck}}$ is so large (or because $G_{\mathrm{Newton}}$ is
so small) that we normally consider gravitation irrelevant for
particle physics.  The graviton-quark-antiquark coupling is
generically $\sim E/M_{\mathrm{Planck}}$, so it is easy to make a
dimensional estimate of the branching fraction for a gravitationally
mediated rare kaon decay: $B(K_{L} \to \pi^{0}G) \sim
(M_{K}/M_{\mathrm{Planck}})^{2} \sim 10^{-38}$, which is truly
negligible!}

\begin{equation}
	\Lambda \sim M_{\rm Planck}  = 
	\left(\frac{\hbar c}{G_{\mathrm{Newton}}}\right)^{1/2} \approx 1.22 
	\times 10^{19} {\rm \; GeV}\;.
\end{equation}
In a unified theory of the strong, weak, and electromagnetic 
interactions, the natural scale is the unification scale,
\begin{equation}
	\Lambda \sim U \approx 10^{15}\hbox{-}10^{16} {\rm \; GeV}\;.
\end{equation}
Both estimates are very large compared to the scale of electroweak 
symmetry breaking \eqn{hvev}.  We are therefore assured that new physics must 
intervene at an energy of approximately 1~TeV, in order that the 
shifts in $m^2$ not be much larger than \eqn{hvev}.

\subsection{Clues to the Higgs-boson mass \label{Hclues}}
We have seen in our discussion of Figure~\ref{fig:EWtop} that the sensitivity of 
electroweak observables to the (long unknown) mass of the top quark 
gave early indications for a very massive top.  For example, the 
quantum corrections to the standard-model predictions given below \eqn{eq:eYuk} for 
$M_{W}$ and  $M_{Z}$ arise from different quark loops:
\begin{center} \begin{picture}(280,80)(0,0)
	 \ZigZag(10,40)(50,40){2}{5}
	 \ZigZag(75,40)(115,40){2}{5}
	 \ArrowArc(62.5,40)(12.5,0,180)
	 \ArrowArc(62.5,40)(12.5,180,360)
	 \Text(62.5,58)[b]{{\large $\bar{b}$}}
	 \Text(62.5,22)[t]{{\large $t$}}
	 \Text(5,40)[r]{{\large $W^{+}$}}
	 \Text(120,40)[l]{{\large $W^{+}$}}
	 
	 \ZigZag(165,40)(205,40){2}{5}
	 \ZigZag(230,40)(270,40){2}{5}
	 \ArrowArc(217.5,40)(12.5,0,180)
	 \ArrowArc(217.5,40)(12.5,180,360)
	 \Text(217.5,58)[b]{{\large $\bar{t}$}}
	 \Text(217.5,22)[t]{{\large $t$}}
	 \Text(160,40)[r]{{\large $Z^{0}$}}
	 \Text(275,40)[l]{{\large $Z^{0}$,}}
    \end{picture}   \end{center}
$t\bar{b}$ for $M_{W}$, and $t\bar{t}$ (or $b\bar{b}$) for $M_{Z}$.  
These quantum corrections alter the link  between the $W$- and 
$Z$-boson masses, so that
\begin{equation}
    M_{W}^{2} = M_{Z}^{2}\left(1 - \sin^{2}\theta_{W}\right)
    \left(1 + \Delta\rho\right)\; ,
    \label{eq:MWqc}
\end{equation}
where
\begin{equation}
    \Delta\rho \approx \Delta\rho^{(\mathrm{quarks})} = 
    \frac{3G_{\mathrm{F}}m_{t}^{2}}{8\pi^{2}\sqrt{2}} \; .
    \label{eq:drhoq}
\end{equation}
The strong dependence on $m_{t}^{2}$ is characteristic, and it 
accounts for the precision of the top-quark mass estimates derived 
from electroweak observables.

Now that $m_{t}$ is known to about 2.5\% from direct observations at the
Tevatron, it becomes profitable to look beyond the quark loops to the
next most important quantum corrections, which arise from Higgs-boson
effects.  The Higgs-boson quantum corrections are typically smaller
than the top-quark corrections, and exhibit a more subtle dependence
on $M_{H}$ than the $m_{t}^{2}$ dependence of the top-quark
corrections.  For the case at hand,
\begin{equation}
    \Delta\rho^{\mathrm{(Higgs)}} = \mathsf{C} \cdot \ln \left(\frac{M_{H}}{v}\right) 
    \; ,
    \label{eq:drhoh}
\end{equation}
where I have arbitrarily chosen to define the coefficient
$\mathsf{C}$ at the electroweak scale $v$.  

\begin{figure}[t!]
\begin{center}
    \includegraphics[width=9.5cm]{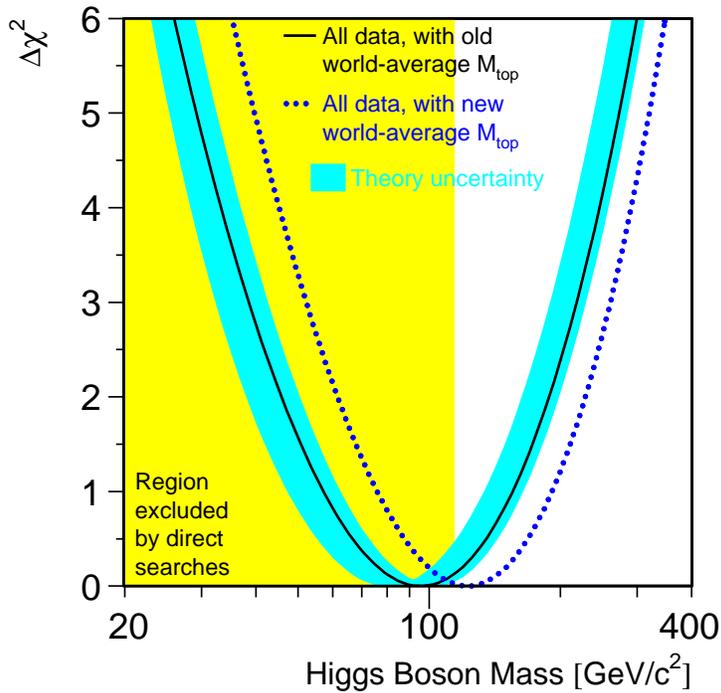}
    \caption{$\Delta\chi^{2}=\chi^2-\chi^2_{\mathrm{min}}$ from a
    global fit to precision data {\it vs.} the Higgs-boson mass,
    $M_{H}$.  The solid line is the result of the fit with $m_{t} =
    174.3\gev$; the band represents an estimate of the theoretical
    error due to missing higher order corrections.  The yellow-shaded region
    shows the 95\% CL exclusion limit on $M_{H}$ from the direct search
    at LEP. The dashed curve shows the change brought about by the new
    Tevatron average top mass, $m_{t} = 178.0\gev$.  (From Ref.~{\protect\cite{LEPEWWG}}.)
    \label{fig:TeVBB}.}
\end{center}
\end{figure}

Figure \ref{fig:TeVBB} shows how the goodness of the 
LEP Electroweak Working Group's 
global fit depends upon the Higgs-boson mass.  Within the standard model, they 
deduce a 95\% CL upper limit, $M_{H} \ltap 219 (251)\gev$, for 
$m_{t} = 174.3 \pm 5.1 (178.0 \pm 4.3)\gev$.  
The recent increase in the world-average top mass changes the 
best-fit Higgs-boson mass from $96^{+60}_{-38}\gev$ to 
$117^{+67}_{-45}\gev$.
The direct searches at LEP have concluded that $M_{H}>
114.4\gev$~\cite{Barate:2003sz}, excluding much of the favored region. Even with the 
additional breathing space afforded by a higher top mass, either the Higgs
boson is just around the corner, or the standard-model analysis is
misleading.  Things will soon be popping!

We will begin to explore the new physics that may lie beyond the 
standard model in Lecture 3, where we take up the possibility of 
unified theories of the strong, weak, and electromagnetic interactions. 
Let us conclude today's rapid survey of the electroweak theory by 
summarizing some of the questions we have encountered:

\begin{questions}{Second}
    \addtocounter{bean}{13}
    \item What contrives a Higgs potential that hides electroweak symmetry?

  \item  What separates the electroweak scale from higher scales?
  
  \item What \textit{are} the distinct scales of physical interest?
 
  \item  Why is empty space so nearly weightless?
  
  \item What determines the gauge symmetries?
  
  \item What accounts for the  range of fermion masses?
  
  \item Why is (strong-interaction) isospin a good symmetry? What does 
  it mean?    
\end{questions}
To prepare for our discussion of unified theories, please review the
elements of group theory and work out
\begin{problem}
    Examine the (standard-model) \smgg\ content of the \textbf{5},
    \textbf{10}, and \textbf{24} representations of
    $\mathrm{SU(5)}$.  Decompose the fundamental
    \textbf{16} and adjoint \textbf{45} representations of
    $\mathrm{SO(10)}$ into $\mathrm{{SU(5)} \otimes U(1)}$; into
    $\mathrm{{SU(4)} \otimes {SU(2)} \otimes {SU(2)}}$.
\end{problem}
\section{Unified Theories \label{sec:unified}}
\vspace*{6pt}
\centerline{REDES 239: \textit{Los Ladrillos del Universo} $\cdot$ 
19.5.2002 $\cdot$ TVE}
\vspace*{-9pt}
\begin{tabularx}{\linewidth}{X X}
    \begin{quotation}
\noindent\textit{Eduardo Punset:} Una teor\'{i}a acerca de todo?

\vspace*{6pt}
\noindent\textit{Chris Quigg:} 
Bueno, no me gusta la expresi\'{o}n de teor\'{i}a acerca del todo, porque
incluso despu\'{e}s de conocer todas las reglas todav\'{i}a queda por saber
c\'{o}mo aplicar esas reglas a este maravilloso mundo tan diverso y
complejo que nos rodea.  Por tanto, creo que deber\'{i}amos tener un poco
m\'{a}s de humildad cuando utilizamos expresiones como esa de ``teor\'{i}a
acerca del todo'', pero es una teor\'{i}a de ``mucho''.
    \end{quotation}
    &
    \begin{quotation}
\noindent\textit{Eduardo Punset:} A theory of everything?

\vspace*{6pt}
\noindent\textit{Chris Quigg:} 
I don't like the expression, ``a theory of everything,'' because even 
if we should ever know all the rules, we still must learn how to apply 
those rules to this marvelous world of diversity and change that 
surrounds us. For that reason, I believe we should display a little 
more humility when we use expressions like ``theory of everything.'' 
Nevertheless, it is a theory of quite a lot!
    \end{quotation}
\end{tabularx}

\vspace*{-18pt}
\subsection{Why Unify? \label{subsec:why}}
The standard model based on \smgg\ gauge symmetry encapsulates much 
of what we know and describes many observations, but it leaves many 
things unexplained. Both the success and the incompleteness of the 
standard model encourage us to look beyond it to a more comprehensive 
understanding. One attractive way to proceed is by \textit{enlarging 
the gauge group,} which we may attempt either by accreting F{9} new 
symmetries or by unifying the symmetries we have already recognized.

Left-right symmetric models, such as those based on the gauge 
symmetry $$\mathrm{SU(3)_{c} \otimes SU(2)_{L}\otimes SU(2)_{R} \otimes 
U(1)}_{B-L}\;,$$ follow the first path. Such models attribute the 
observed parity violation in the weak interactions to spontaneous 
symmetry breaking---the $\mathrm{SU(2)_{R}}$ symmetry is broken at a 
higher scale than the $\mathrm{SU(2)_{L}}$---and naturally 
accommodate Majorana neutrinos. We saw in Lecture~1 that they can be 
represented readily in the double simplex. Left-right symmetric 
theories also open new possibilities, including transitions that 
induce $n \leftrightarrow \bar{n}$ oscillations and a mechanism for 
spontaneous $\mathcal{CP}$ violation. More generally, enlarging the 
gauge group by accretion seeks to add a missing element or to explain 
additional observations.

Unified theories, on the other hand, seek to find a symmetry group 
$${\mathcal{G}} \supset \smgg$$
(usually a simple group, to maximize the predictive power) that 
contains the known interactions. This approach is motivated by the 
desire to unify quarks and leptons and to reduce the number of 
independent coupling constants, the better to understand the relative 
strengths of the strong, weak, and electromagnetic interactions at 
laboratory energies. Supersymmetric unified theories, which we will 
investigate briefly in Lecture~4, bring the added ambitions of 
incorporating gravity and joining constituents and forces.

Two very potent ideas are at play here. The first is the idea of 
unification itself: what Feynman calls \textit{amalgamation,} which is 
the central notion of \textit{generalization and synthesis} that scientific 
explanation represents. Examples from the history of physics include 
Maxwell's joining of electricity and magnetism and light; the atomic 
hypothesis, which places thermodynamics and statistical mechanics 
within the realm of Newtonian mechanics; and the links among atomic 
structure, chemistry, and quantum mechanics.

The second is the notion that the human scale of space and time is not
privileged for understanding Nature, and may even be disadvantaged.
Not only in physics, but throughout science, this has been a growing
recognition since the quantum-mechanical revolution of the 1920s.  To
understand why a rock is solid, or why a metal gleams, we must
understand its structure on a scale a billion  times smaller $(10^{-9})$
than the human scale, and we must understand the rules that prevail
there.  It may well be that certain scales are privileged for
understanding certain globally important aspects of the Universe: why,
for example, the fine structure constant $\alpha \approx 1/137$, and why
the strong coupling, measured at the $B$ factories is $\alpha_{s}\approx 
1/5$;  or why fermion masses have the (seemingly unintelligible) pattern 
they do. 

I believe that the discovery that \textit{the human scale is not 
preferred} is as important as the discoveries that 
the human location is not privileged (Copernicus) and that there is no 
preferred inertial frame (Einstein), and will prove to be as influential.

Let us examine the motivation for constructing a unified theory  in greater detail.
$\Box$ Quarks and leptons are structureless, spin-$\cfrac{1}{2}$
particles.  (How) are they related?
$\Box$ What is the meaning of electroweak universality, embodied in the
matching left-handed doublets of quarks and leptons?
$\Box$ Anomaly cancellation requires quarks \textit{and} leptons.
$\Box$ Can the three distinct coupling parameters of the standard model
($\alpha_{s}, \alpha_{\mathrm{em}}, \sin^{2}\theta_{W}$ or $g_{s}, g,
g^{\prime}$) be reduced to two or one?
$\Box$ $\alpha_{\mathrm{em}}$ increases with $Q^{2}$; $\alpha_{s}$ decreases.
Is there a unification point where all (suitably defined) couplings
coincide?
$\Box$ Why is charge quantized?  [$Q_{d}=\cfrac{1}{3}Q_{e}$,
$Q_{p}+Q_{e}=0$, $Q_{\nu}-Q_{e}=Q_{u}-Q_{d}$,
$Q_{\nu}+Q_{e}+3Q_{u}+3Q_{d}=0$.]

These questions lead us toward a more complete electroweak 
unification, which is to say a simple $\mathcal{G} \supset \ewgg$; a 
quark-lepton connection; a ``grand'' unification of the strong, weak, 
and electromagnetic interactions, based on a simple group 
$\mathcal{G} \supset \smgg$. If we choose the task of grand 
unification, we must find a group that contains the known interactions 
and that can accommodate the known fermions---either as one generation 
plus replicas, or as all three generations at once. The unifying group 
will surely contain interactions beyond the established ones, and we 
should be open to the possibility that the fermion representations 
require the existence of particles yet undiscovered.

\subsection{Toward a Unified Theory \label{subsec:toward}}
It is convenient to express all the fermions in terms of left-handed 
fields, for ease in counting degrees of freedom.\footnote{We 
established the correspondence between right-handed particles and 
left-handed antiparticles in our discussion of \eqn{eq:CC}.} 
Denoting the quantum numbers as $(\mathrm{SU(3)_{c}, 
SU(2)_{L}})_{Y}$, we can enumerate the fermions of the first 
generation as follows:
\begin{equation}
\begin{array}{rl}
u_{\mathrm{L}}, d_{\mathrm{L}}: & \repr{3}{2}{1/3} \\
d^{c}_{\mathrm{L}}: & \repr{3^{*}}{1}{2/3} \\
u^{c}_{\mathrm{L}}: & \repr{3^{*}}{1}{-4/3} \\
\nu_{\mathrm{L}}, e_{\mathrm{L}}: & \repr{1}{2}{-1} \\
e^{c}_{\mathrm{L}}: & \repr{1}{1}{2} \\
{\nu^{c}_{\mathrm{L}}:} &
{\repr{1}{1}{0}}\; .
\end{array}  \label{eq:fermreps}
\end{equation}
This collection of particles is not identical to its conjugate, so 
$\mathcal{G}$ must admit complex representations. The smallest 
appropriate group is $\mathrm{SU(5)}$, and we shall choose it to 
illustrate the strategy of unified theories.

Let us examine the low-dimensional representations of
$\mathrm{SU(5)}$.\footnote{For a quick review of Young tableaux,
see~\cite{WohlYT}.} The fundamental representation is

\centerline{{$\yng(1) \quad \mathbf{5} = 
(\mathbf{3},\mathbf{1})_{-2/3} \oplus (\mathbf{1},\mathbf{2})_{1}$},}
\noindent and its conjugate is

\centerline{{$\yng(1,1,1,1) \quad \mathbf{5^{*}} = 
(\mathbf{3^{*}},\mathbf{1})_{2/3} \oplus (\mathbf{1},\mathbf{2})_{-1}$}.}
\noindent To generate larger representations, we consider products of 
the fundamental, for example

\centerline{	 $\mathbf{5}\otimes\mathbf{5}$ : 
	$\yng(1)\otimes\yng(1) = \yng(1,1)\oplus\yng(2) = \mathbf{10} \oplus 
	\mathbf{15}$,}
\noindent where
\[
	\mathbf{15} = \repr{6}{1}{-4/3} \oplus \repr{3}{2}{1/3} \oplus 
	\repr{1}{3}{2}\;,
\]
and
\[
	\mathbf{10} = \repr{3^{*}}{1}{-4/3} \oplus \repr{3}{2}{1/3} \oplus 
	\repr{1}{1}{2}\;.
\]
The product of the fundamental with its conjugate is

\centerline{$\mathbf{5} \otimes \mathbf{5^{*}}$ : $\yng(1) \otimes 
	\yng(1,1,1,1) = \yng(1,1,1,1,1) \oplus \yng(2,1,1,1) = \mathbf{1} 
	\oplus \mathbf{24}$,}
\noindent where
\[
	{\mathbf{1} = \repr{1}{1}{0}}\;,
\]
and
\begin{displaymath}
\mathbf{24} = \mathbf{24^{*}}  = 
	{\repr{8}{1}{0}} \,\oplus 
	{\repr{3}{2}{-5/3}} \,\oplus  
	{\repr{3^{*}}{2}{5/3}}\, \oplus 
		  {\repr{1}{3}{0}} \oplus {\repr{1}{1}{0}}\;.
\end{displaymath}
Finally, consider the product

\centerline{$\mathbf{5} \otimes \mathbf{10^{*}}$ : $\yng(1) 
\otimes \yng(1,1,1) = \yng(1,1,1,1) \oplus \yng(2,1,1) = 
\mathbf{5^{*}} \oplus \mathbf{45^{*}}$,}
\noindent where
\begin{displaymath}
	\mathbf{45^{*}}  =  
	 \repr{8}{2}{-1} \oplus \repr{6}{1}{2/3} \oplus 
	\repr{3^{*}}{3}{2/3} \oplus 
	\repr{3}{2}{7/3} \oplus  \repr{3^{*}}{1}{2/3} \oplus 
	\repr{3}{1}{-8/3} \oplus \repr{1}{2}{-1}\;.
\end{displaymath}

Now we look for a fit.  $\Box$ Does a generation (without a right-handed
neutrino) fit in the 15-dimensional representation? It does not: 
$\mathbf{15} = \repr{6}{1}{-4/3} \oplus \repr{3}{2}{1/3} \oplus
\repr{1}{3}{2}$ contains color-sextet quarks! $\Box$ Do three 
generations fit in the 45-dimensional representation? No, 
$\mathbf{45^{*}} = \repr{8}{2}{-1} \oplus \repr{6}{1}{2/3} \oplus
\repr{3^{*}}{3}{2/3} \oplus \repr{3}{2}{7/3} \oplus
\repr{3^{*}}{1}{2/3} \oplus \repr{3}{1}{-8/3} \oplus \repr{1}{2}{-1}$
contains color-octet and color-sextet fermions. $\Box$ If nothing fits 
like a glove, perhaps we should try a glove and a mitten, placing one 
generation in several representations, 
$\mathbf{5^{*}}\,\oplus\mathbf{10} \,\oplus\mathbf{1}$:
\[
	\begin{array}{rlc}
{u_{L}, d_{L}:} & 
{\repr{3}{2}{1/3}}  & {\mathbf{10}}\\
{d^{c}_{L}:} & 
{\repr{3^{*}}{1}{2/3}} & {\mathbf{5^{*}}} \\
{u^{c}_{L}:} & 
{\repr{3^{*}}{1}{-4/3}} & {\mathbf{10}} \\
{\nu_{L}, e_{L}:} & 
{\repr{1}{2}{-1}} & {\mathbf{5^{*}}}  \\
{e^{c}_{L}:} & {\repr{1}{1}{2}} & {\mathbf{10}}\\
{\nu^{c}_{L}:} & {\repr{1}{1}{0}} &
		{\mathbf{1}}
	\end{array}  
\]
The presence of both quarks and leptons in either the $\mathbf{5^{*}}$ 
or $\mathbf{10}$ means that we can expect quark-lepton transformations.

\begin{figure}[t!]
\begin{center}
\includegraphics[width=6cm]{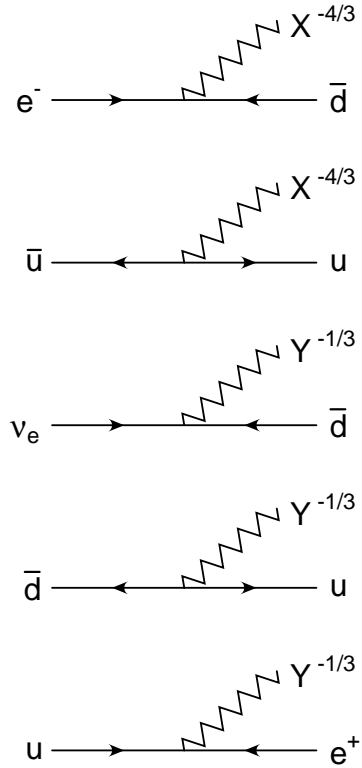}
\caption{Transitions mediated by the $X^{\pm4/3}$ and $Y^{\pm1/3}$ 
gauge bosons in the $\mathrm{SU(5)}$ unified theory. \label{fig:su5trans}}
\end{center}
\end{figure}

What about the gauge interactions? The twelve known gauge bosons fit 
in the adjoint $\mathbf{24}$ representation:
\[
	\begin{array}{cc}
		\repr{8}{1}{0} & \hbox{gluons}  \\
		\repr{1}{3}{0} & W^{\pm}, W_{3}  \\
		\repr{1}{1}{0} & \mathcal{A}\;.
	\end{array}  \hbox{ or }(b_{1}, b_{2}, b_{3})
\]
\noindent
The $\mathbf{24}$ also includes new fractionally charged
\textit{leptoquark} gauge bosons
\[
\begin{array}{ll}
	\repr{3}{2}{-5/3} & X^{-4/3}, Y^{-1/3} \\
	\repr{3^{*}}{2}{5/3} & X^{4/3}, Y^{1/3}
\end{array}
\]
that mediate the transitions illustrated in Figure~\ref{fig:su5trans}.
Recall that the price (or reward!) of the partial electroweak 
unification achieved in the \ewgg\ theory was a new interaction, the 
weak neutral current. Here again we find that new interactions are 
required to complete the symmetry---just as our geometrical 
discussion in Lecture~1 invited us to think.

The new vertices of Figure~\ref{fig:su5trans} can give rise to proton 
decay, which is known to be an exceedingly rare process. Accordingly, 
we must arrange that $X^{\pm4/3}$ and $Y^{\pm1/3}$ acquire very large 
masses, lest the proton decay rapidly. We hide the $\mathrm{SU(5)}$ 
symmetry in two steps. First, a $\mathbf{24}$ of auxiliary scalars breaks
$\mathrm{SU(5)} \to \smgg$ at a high scale, to give large masses to
$X^{\pm4/3}$ and $Y^{\pm1/3}$. The $\mathbf{24}$ does not occur in 
the $\bar{\mathsf{L}}\mathsf{R}$ products
\begin{eqnarray}
\mathbf{5}^{\ast}\otimes \mathbf{10} & = & \mathbf{5} \oplus \mathbf{45}
\label{su5fm}  \\
\mathbf{10} \otimes \mathbf{10} & = & \mathbf{5}^{\ast} \oplus 
\mathbf{45}^{\ast} \oplus \mathbf{50}^{\ast}
	\nonumber
\end{eqnarray}
that generate fermion masses, so the quarks and leptons escape large 
tree-level masses. At a second stage, a $\mathbf{5}$ of scalars 
containing the standard-model Higgs fields breaks
$\smgg \to \mathrm{SU(3)_{c}} \otimes \mathrm{U(1)}_{\mathrm{em}}$.

The \suf\ unification brings a pair of agreeable consequences. If 
built on a complete generation of quarks and leptons, the theory is 
anomaly free, which guarantees that the symmetries survive quantum 
corrections. The anomalies of the $\mathbf{5^{*}}$ and $\mathbf{10}$ 
representations are equal and opposite, $A(\mathbf{5^{*}}) = -1$, 
$A(\mathbf{10}) = +1$, while $A(\mathbf{1}) = 0$, so that 
$A(\mathbf{5^{*}}) + A(\mathbf{10}) = 0$. If this seems a little 
precarious, we can note that $\mathrm{SO(10)}$ representations are 
anomaly free, and that 
$\mathrm{SO(10)} \; \mathbf{16} = \mathbf{10} \oplus \mathbf{5^{*}}
\oplus \mathbf{1}\hbox{ of }\mathrm{SU(5)}$. In addition, the unified 
theory offers us an explanation of charge quantization. Because the 
charge operator $Q$ is a generator of \suf, the charges must sum to 
zero in any representation. Applied to the $\mathbf{5^{*}}$, we find 
that $Q(d^{c}) = -\cfrac{1}{3}Q(e)$, one of the quark-lepton 
``coincidences'' we wish to understand.

\subsection{The Interaction Lagrangian and Running Couplings\label{subsec:intlag}}
The \suf\ unified theory is based on a simple group, so the strength 
of all the gauge interactions is specified by a single coupling 
constant, $g_{5}$. The theory prescribes the relative normalization 
of the electroweak theory's independent couplings, $g$ and 
$g^{\prime}$, and predicts the weak mixing parameter
\begin{equation}
    \sin^{2}\theta_{W} = 
    \displaystyle{\frac{g^{\prime\,2}}{g^{2}+g^{\prime\,2}}}\;.
    \label{eq:xwdef}
\end{equation}
Let us see how this   coupling-constant concordance comes about. The 
\suf\ interaction Lagrangian is
\begin{eqnarray}
	\mathcal{L}_{\mathrm{int}} & = & -\frac{g_{5}}{2}G_{\mu}^{a} 
	\left(\bar{u}\gamma^{\mu}\lambda^{a}u + 
	\bar{d}\gamma^{\mu}\lambda^{a}d\right)  \nonumber\\
	 &  &  
	 -\frac{g_{5}}{2}W_{\mu}^{i}\left(\bar{\mathsf{L}}_{u}\gamma^{\mu}\tau^{i}\mathsf{L}_{u} + 
	 \bar{\mathsf{L}}_{e}\gamma^{\mu}\tau^{i}\mathsf{L}_{e} \right) 
	 \label{eq:su5lag}\\
	 &  & 
	 -\frac{g_{5}}{2}{\sqrt{\cfrac{3}{5}}}\mathcal{A}_{\mu}\sum_{\mathrm{fermions}} 
	 \bar{f}\gamma^{\mu}Yf  \nonumber\\
	 &  & + X\hbox{ and }Y\hbox{ pieces} \nonumber \;,
\end{eqnarray}
where, in a familiar notation,
\begin{equation}
    \mathsf{L}_{u} = \left(
    \begin{array}{c}
	    u  \\
	    d_{\theta}
    \end{array}
    \right)_{\mathrm{L}} \quad \mathsf{L}_{e} =
    \left(
    \begin{array}{c}
	    \nu_{e}  \\
	    e
    \end{array}
    \right)_{\mathrm{L}}\;.
\end{equation}

In the weak-hypercharge piece, the factor $\sqrt{\cfrac{3}{5}}$ 
arises from the form of the normalized generator in \suf,
\begin{equation}
    \cfrac{1}{\sqrt{30}}\left( 
    \begin{array}{ccccc}
	    -2 &  &  &  &   \\
	     & -2 &  & 0 &   \\
	     &  & -2 &  &   \\
	     & 0 &  & 3 &   \\
	     &  &  &  & 3
    \end{array}
    \right) \propto Y \;.\label{eq:Ynorm}
\end{equation}
In electroweak terms, we identify
\begin{equation}
    g^{\prime\,2} = \cfrac{3}{5} g^{2}\;,
\end{equation}
so that in \textit{unbroken} \suf, we predict
\begin{equation}
    \displaystyle{\sin^{2}\theta_{W} = \frac{g^{\prime\,2}}{g^{2}+g^{\prime\,2}} = 
    \frac{\cfrac{3}{5}}{1 + \cfrac{3}{5}} = \frac{3}{8}}\;. 
    \label{eq:xwsu5}
\end{equation}

We carry out experiments at low energies, whereas if \suf\ is the 
unifying symmetry it is unbroken at extremely high energies. To make 
the connection, we need to examine the evolution of coupling 
constants---the dependence on energy scale that occurs in quantum 
field theory. Below a possible unification scale, the \smgg\ coupling 
constants evolve separately. In leading-logarithmic approximation, we 
have
\begin{equation}
    \mathrm{SU(3)_{\mathrm{c}}:}\quad 1/\alpha_{3}(Q^{2}) = 1/\alpha_{3}(\mu^{2}) + 
    b_{3}\log(Q^{2}/\mu^{2})  \;, \label{eq:su3run}
\end{equation}
where $4\pi b_{3} = 11 - 4n_{\mathrm{gen}}/3$, so that 
$b_{3} = 7/4\pi$;

\begin{equation}
    \mathrm{SU(2)_{\mathrm{L}}:}\quad 1/\alpha_{2}(Q^{2}) = 
    1/\alpha_{2}(\mu^{2}) + b_{2}\log(Q^{2}/\mu^{2})  \;, \label{eq:su2run}
\end{equation}
where $4\pi b_{2} = (22-4n_{\mathrm{gen}})/3 -
n_{\mathrm{Higgs}}/6$, so that $b_{2} = 19/24\pi$; and 

\begin{equation}
    \mathrm{U(1)}_{Y}:\quad 1/\alpha_{1}(Q^{2}) =  1/\alpha_{Y}(\mu^{2}) + 
    b_{1}\log(Q^{2}/\mu^{2})  \;, \label{eq:u1run}
\end{equation}
with $4\pi b_{1} = -4n_{\mathrm{gen}}/3 - 
n_{\mathrm{Higgs}}/10$, so that $b_{1} = 
-41/40\pi$. In \eqn{eq:su3run}--\eqn{eq:u1run}, $Q^{2}$ is the scale 
of interest and $\mu^{2}$ is the reference scale.

\begin{figure}[t!]
\begin{center}
\includegraphics[width=13cm]{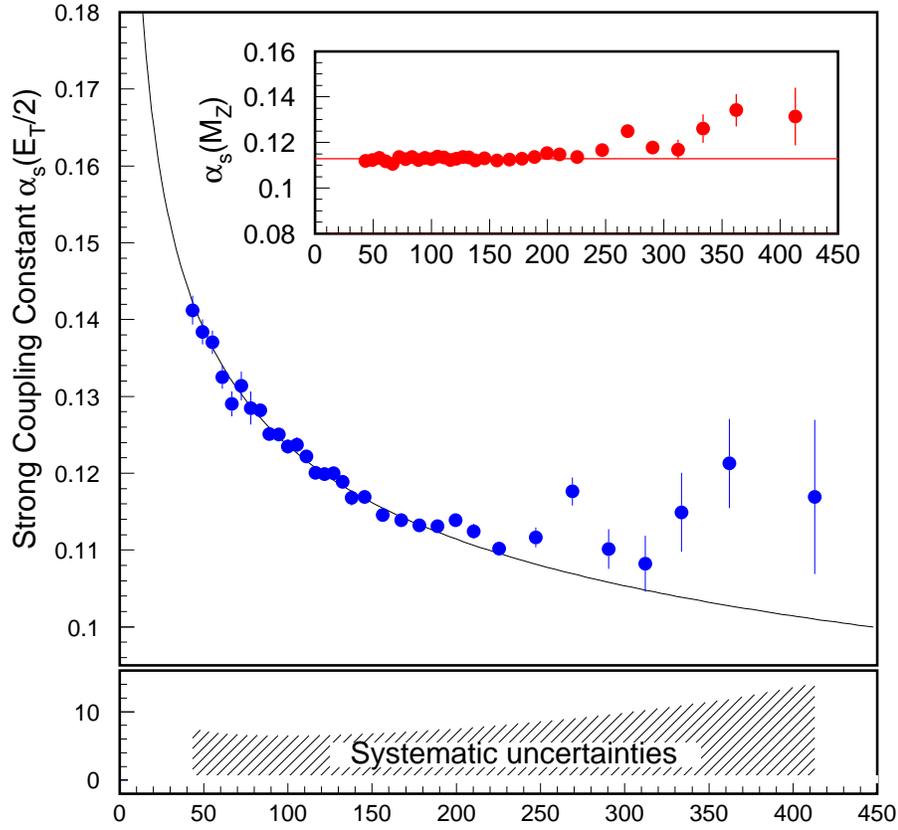}
\vspace*{-12pt}
\caption{Evolution of the strong coupling constant 
$\alpha_{s}(E_{T}/2)$ measured in jet studies at $1.8\tev$~\cite{Affolder:2001hn}.
The curve shows the expectations from QCD. The inset shows the values 
measured at different values of $E_{T}$ evolved to the $Z$-boson mass.
\label{fig:su3run}}
\end{center}
\end{figure}

\begin{figure}[t!]
\begin{center}
\includegraphics[width=10cm]{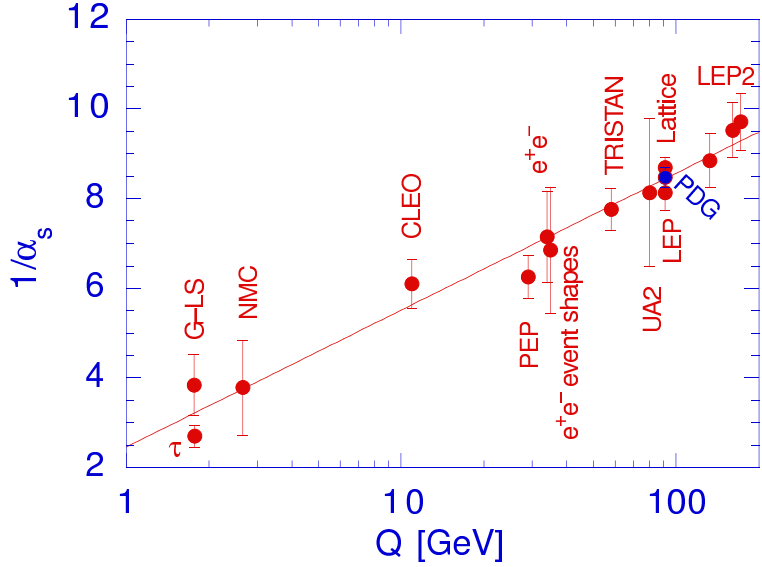}
\caption{Running of $1/\alpha_{s}(Q)$ determined in a variety of 
experiments at different energies. The point labeled PDG is the 
world-average value, presented at the $Z$-boson mass.
\label{fig:invalf}}
\end{center}
\end{figure}

Running couplings are not merely an artifact of quantum field theory, 
they are observed! Figure~\ref{fig:su3run} shows the evolution of the 
strong coupling constant as determined by the CDF collaboration from 
their study of $\bar{p}p \to \hbox{jets}$~\cite{Affolder:2001hn}.
A little easier to visualize is the compilation~\cite{PDBook} 
plotted in Figure~\ref{fig:invalf} as $1/\alpha_{s}$ over a wide range 
of energies. There the goodness of the form \eqn{eq:su3run} is readily 
apparent.

Having recalled the expectations for how coupling ``constants'' run, 
we can check the prediction of \suf\ unification. We recall that the 
electromagnetic coupling in the electroweak theory is a derived 
quantity that we can express in terms of the $\mathrm{SU(2)_{L}}$ and 
weak-hypercharge couplings as
\begin{equation}
    1/\alpha(Q^{2}) \equiv 1/\alpha_{Y}(Q^{2}) + 1/\alpha_{2}(Q^{2}) \;,
    \label{eq:emrun}
\end{equation}
where $\alpha_{Y}$ is proportional to $\alpha_{1}$. Relating the 
couplings to the \suf\ coupling $\alpha_{U}$ at the unification scale 
$U$, we have
\begin{equation}
    1/\alpha(Q^{2}) = \cfrac{8}{3}\cdot 1/\alpha_{U} + ({b_{Y}} + 
    {b_{2}})\log({Q^{2}}/{U^{2}}) \;, \label{eq:emrunu}
\end{equation}
which suggests that we form
\begin{eqnarray}
 \displaystyle{\frac{({8}/{3})}{\alpha_{s}(Q^{2})}} - 
\frac{1}{\alpha(Q^{2})} & = &
\underbrace{\left(\displaystyle{\frac{8{b_{3}}}{3}} - {b_{Y}} 
-{b_{2}}\right)}\log(\frac{Q^{2}}{U^{2}}) \label{eq:emrunu2}\\ & & 
\frac{22+n_{\mathrm{Higgs}}}{4\pi} \rightarrow \frac{67}{12\pi} \;. 
\nonumber
\end{eqnarray}
Using the measured values, $\alpha_{3}(M_{Z}^{2}) \approx 1/8.75$, 
$\alpha(M_{Z}^{2}) \approx 1/128.9$, and $M_{Z} \approx 91.19\gev$, we 
can use \eqn{eq:emrunu2} to estimate $U \approx  10^{15}\gev$ and 
$1/\alpha_{U} \approx 42$.

\begin{figure}[t!]
\begin{center}
\includegraphics[width=10cm]{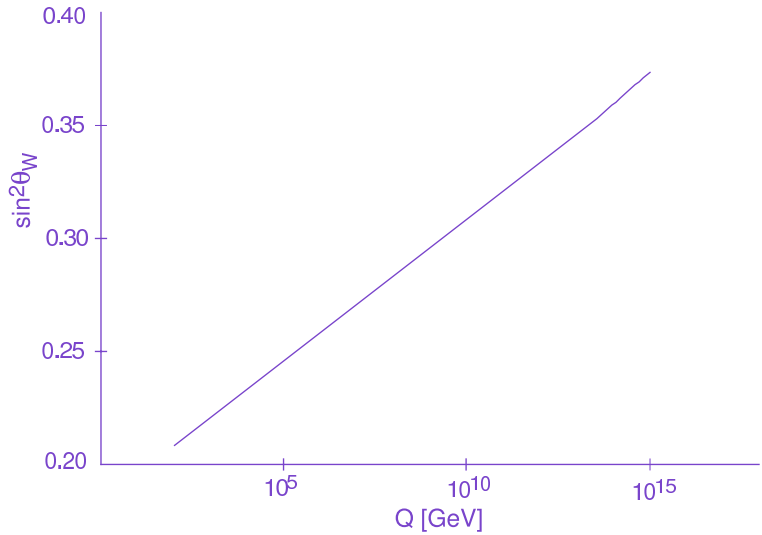}
\caption{Running of $\sin^{2}\theta_{W}$ in the \suf\ unified theory.
\label{fig:xwevol}}
\end{center}
\end{figure}

Now we are ready to test \suf\ unification using the weak mixing 
parameter
\begin{equation}
    x_{W} \equiv \sin^{2}\theta_{W} = \alpha/\alpha_{2} = 
    \frac{1/\alpha_{2}}{1/\alpha_{Y}+1/\alpha_{2}}\;. \label{eq:xwdef2}
\end{equation}
At the unification scale, the running couplings are simply related: 
\begin{equation}
\left.\begin{array}{l}
	1/\alpha_{2} = 1/\alpha_{U}  \\
	1/\alpha_{Y} = \cfrac{5}{3}\cdot 1/\alpha_{U}  \\
	1/\alpha = \cfrac{8}{3}\cdot 1/\alpha_{U}
\end{array}\right\} \;, \label{eq:unicoups}
\end{equation}
so that ${x_{W}(U^{2}) = \cfrac{3}{8}}$, as we have already noticed 
in \eqn{eq:xwsu5}. How does $x_{W}$ evolve? Putting together the 
pieces, we find that 
\begin{equation}
    x_{W}(Q^{2}) = \cfrac{3}{8} - 
    \underbrace{\cfrac{5}{8}(b_{1}-b_{2})}_{+109/96\pi}\alpha(Q^{2})\log\left(Q^{2}/U^{2}\right)\; ,
\end{equation}
which decreases as $Q$ decreases from the unification scale $U$, as 
sketched in Figure~\ref{fig:xwevol}.
At the $Z$-boson mass, we calculate
\begin{equation}
    \left.x_{W}(M_{Z}^{2})\right|_{\mathrm{SU(5)}} \approx 0.21 
    \;,\label{eq:xwsu5mz}
\end{equation}
to be compared with the measured value,
\begin{equation}
    \left.x_{W}(M_{Z}^{2})\right|_{\mathrm{exp}} = 0.2314 \pm 0.003
    \;.\label{eq:xwexp}
\end{equation}
So near, and yet so far!

\begin{figure}[b!]
\begin{center}
\includegraphics[width=10cm]{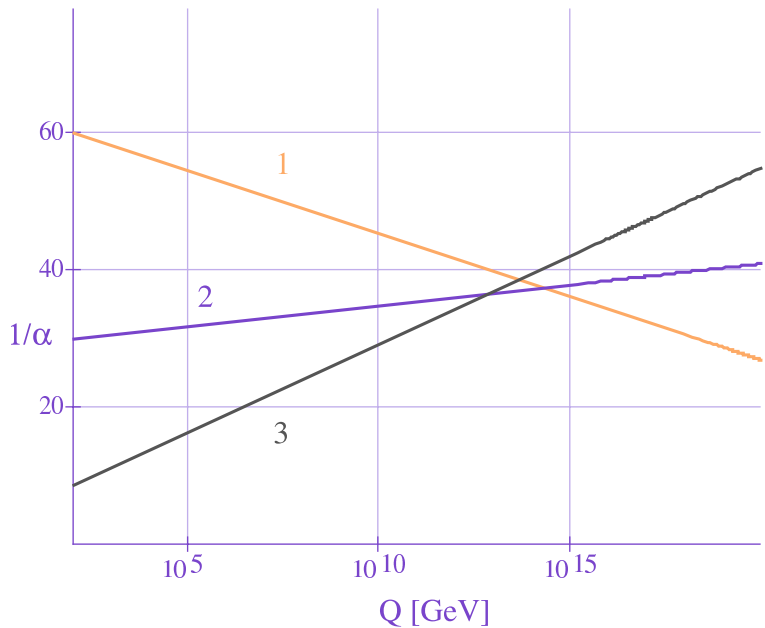}
\caption{Running of $1/\alpha_{1}, 1/\alpha_{2}, 1/\alpha_{3}$ in the
\suf\ unified theory.  \label{fig:alfevolve}}
\end{center}
\end{figure}

An equivalent way to display the same information is to combine the 
measured $\alpha_{1}(M_{Z}^{2}) \approx 1/60$ with 
$\alpha(M_{Z}^{2}) \approx 1/128.9$ to determine $\alpha_{2}(M_{Z}^{2}) 
\approx 1/30$, and then to evolve $1/\alpha_{1}, 1/\alpha_{2}, 
1/\alpha_{3}$ to high energies to see whether they meet. As we can 
anticipate from the near miss of $x_{W}$, the three couplings do not 
quite coincide at a single point at high energy, though they come 
close in the neighborhood of $10^{14\pm1}\gev$, as plotted in 
Figure~\ref{fig:alfevolve}. [With six Higgs doublets, they do 
coincide!]

\begin{problem}
    Suppose that a unified theory, $\mathrm{SU(5)}$ for definiteness,
    fixes the value of the unification scale, $U$, and the strength of
    the couplings, $1/\alpha_{U}$, at that scale.  The value of the
    coupling constants that we measure on a low scale have encrypted in
    them information about the spectrum of particles between our energy
    scale and $U$.  Assume that there are no particles in that range
    beyond those we know from the standard model.  How is the strong
    coupling constant $\alpha_{s}$ at low energies influenced by the
    mass of the top quark? What is the effect on the proton mass?
\end{problem}

{\small\noindent
This problem is a lovely example of the influence on the commonplace of
phenomena that we study far from the realm of everyday experience, so 
I will provide a brief answer. First, it is easy to see, by referring 
to \eqn{eq:su3run} for the evolution of $1/\alpha_{s}$ (which means 
$1/\alpha_{3}$) that the slope 
of $1/\alpha_{s}$ changes from $21/6\pi$ to $23/6\pi$ when we descend 
through top threshold, and decreased by another $2/6\pi$ at every 
succeeding threshold. Without doing any arithmetic,\footnote{I first 
did this analysis on a foggy shower door, but I am known to take very 
long showers!} we can sketch the 
evolution of $1/\alpha_{s}$ for two values of the top-quark mass, as 
I have done in Figure~\ref{fig:topinf}:
\begin{figure}[h!]
\begin{center}
\includegraphics[width=9.39cm]{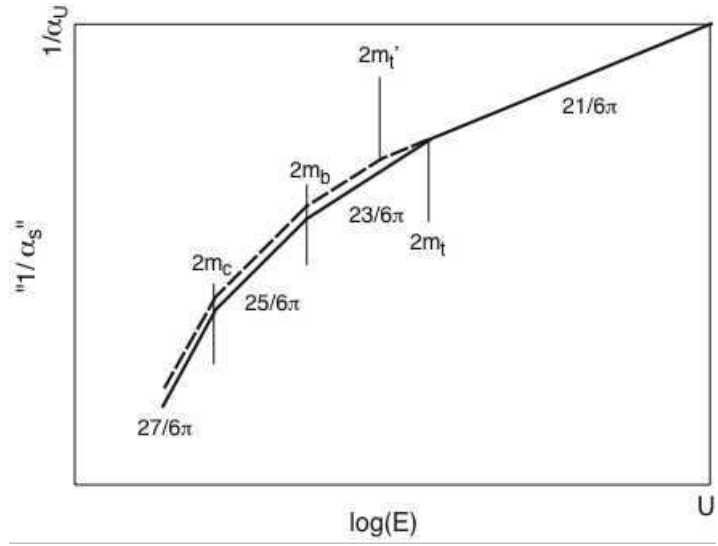}
\caption{Running of the strong coupling constant $1/\alpha_{s}$ for two values of $m_{t}$ in the
\suf\ unified theory.  \label{fig:topinf}}
\end{center}
\end{figure}
the smaller the value of $m_{t}$, the smaller the value of 
$\alpha_{s}$.

To determine the effect of varying the top-quark mass on the mass of 
the proton, we apply the lesson of (lattice) QCD that the mass of the 
proton is mostly determined by the energy stored up in the gluon 
field that confines three light quarks in a small volume. To good 
approximation, therefore, we can write the proton mass in terms of the 
QCD scale parameter as
\begin{equation}
    M_{\mathrm{proton}} \approx C \Lambda_{\mathrm{QCD}}\;,
\end{equation}
where the constant $C$ could be determined by lattice simulations. 
How, then, does $\Lambda_{\mathrm{QCD}}$ depend on $m_{t}$? We calculate 
$\alpha_{s}(2m_{t})$ evolving up from low energies and down from the 
unification scale, and match:
\begin{equation}
    \begin{array}{l}
    1/\alpha_U  +  {\displaystyle \frac{21}{6\pi}}\ln(2m_{t}/M_U) =   \\
     \quad 1/\alpha_s(2m_c) - {\displaystyle \frac{25}{6\pi}}\ln(m_c/m_b) 
     -{\displaystyle \frac{23}{6\pi}}\ln(m_b/m_{t}) \;.
     \end{array} \label{eq:runupdn}
\end{equation}
Using the convenient three-active-flavor definition
\begin{equation}
    1/\alpha_s(2m_c) \equiv {\displaystyle\frac{ 
     27}{6\pi}}\ln(2m_c/\Lambda_{\mathrm{QCD}})\;, 
     \label{eq:lambdef}  
\end{equation}
we solve for 
\begin{equation}
    \Lambda_{\mathrm{QCD}} = \hbox{constants}\cdot 
    \left(\frac{2m_{t}\cdot 2m_b\cdot 
    2m_c}{(1\gev)^{3}}\right)^{\!2/27}\gev \;. \label{eq:lambprop}
\end{equation}
The variation of $\Lambda_{\mathrm{QCD}}$ with the top-quark mass is 
shown in Figure~\ref{fig:lambtop}.
\begin{figure}[t!]
\begin{center}
\includegraphics[width=10cm]{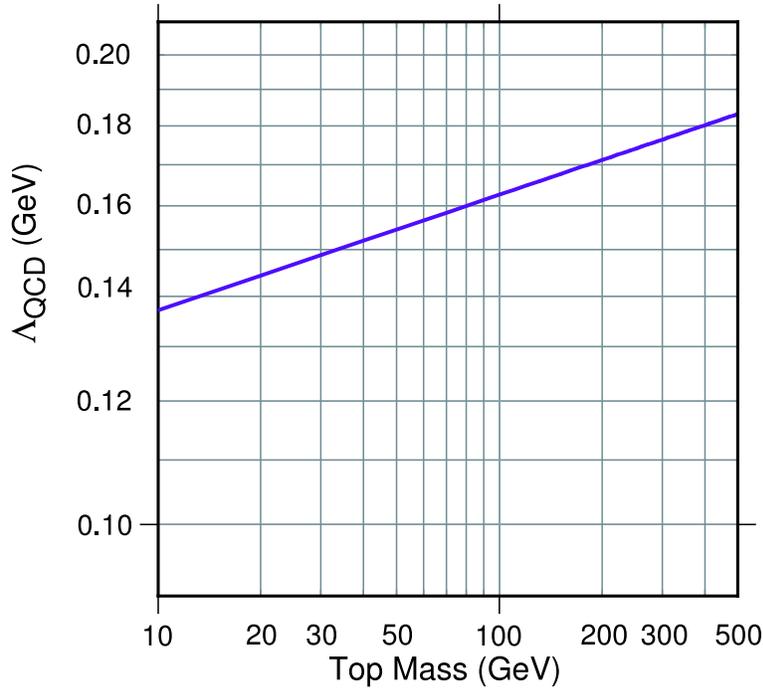}
\caption{Dependence of the QCD scale parameter $\Lambda_{\mathrm{QCD}}$ on $m_{t}$ in the
\suf\ unified theory.  \label{fig:lambtop}}
\end{center}
\end{figure}
With this, we have our answer. Although the population of top-antitop 
pairs within the proton is vanishingly small, because of the top 
quark's great mass, virtual effects of the top quark do affect the 
strong coupling constant we measure at low energies, within the 
framework of a unified theory.\footnote{Our use of \suf\ is not 
terribly restrictive here.} The proton mass is proportional to 
$(m_{t}/1\gev)^{2/27}$, for reasonable variations of $m_{t}$. This 
knowledge is of no conceivable technological value, but I find it 
utterly wonderful---the kind of below-the-surface connection that 
makes it such a delight to be a physicist!
}

\subsection{The Problem of Fermion Masses \label{subsec:massprob}}
\begin{figure}[t!]
\begin{center}
\includegraphics[width=10cm]{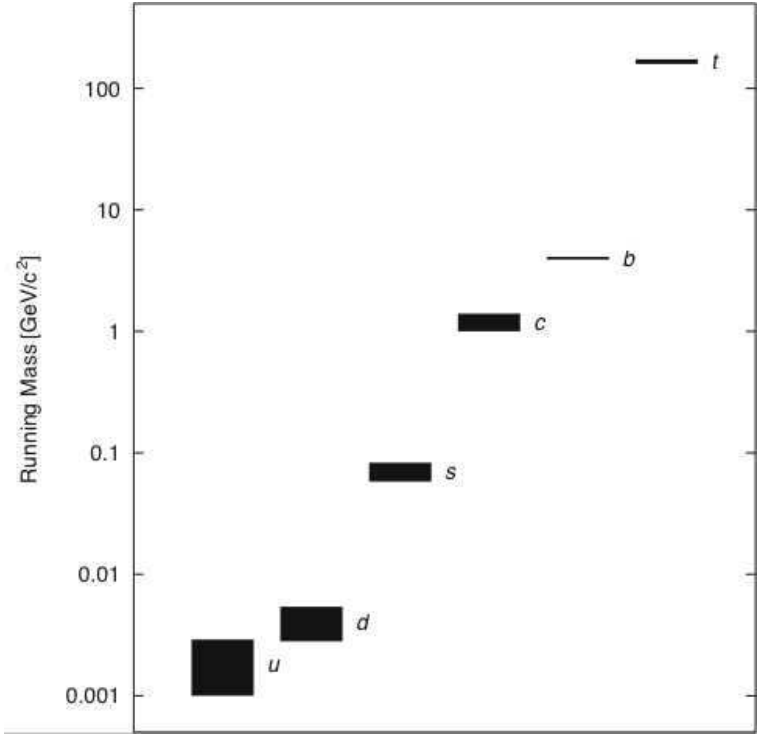}
\caption{Running ($\overline{\mathrm{MS}}$) masses of the quarks. The 
heavy-quark ($c,b,t$) masses are evaluated at the quark masses, 
$\overline{m}_{q}(m_{q})$, while the light-quark masses ($u,d,s$) are
evaluated at $1\gev$.   \label{fig:qmass}}
\end{center}
\end{figure}

Unraveling the origins of electroweak symmetry breaking will not 
necessarily give insight into the origin and pattern of fermion 
masses, because they are set by the Yukawa couplings $\zeta_{i},$ of 
unknown provenance,  that we first met in \eqn{eq:Yukterm}.
The puzzling pattern of quark masses is depicted in Figure~\ref{fig:qmass}.
The fact that masses---like coupling constants---are scale-dependent 
might encourage us to hope that what looks like an irrational pattern 
at low scales will reveal an underlying order at some other scale.

To illustrate the possibilities, let us adopt the specific framework 
of \suf\ unification, with the two-step spontaneously symmetry 
breaking we introduced in \S~\ref{subsec:toward}. At a high scale, a 
$\mathbf{24}$ of scalars breaks $\suf \to \smgg$, giving extremely large 
masses to the leptoquark gauge bosons $X^{\pm 4/3}$ and $Y^{\pm 
1/3}$. As we have already observed, the $\mathbf{24}$ does not occur in 
the $\bar{\mathsf{L}}\mathsf{R}$ products that generate fermion masses, 
so quarks and leptons escape large tree-level masses. At the 
electroweak scale, a $\mathbf{5}$ of scalars 
($\supset$ the standard-model Higgs fields) breaks $\smgg \to 
\mathrm{SU(3)}_{\mathrm{c}}\otimes U(1)_{\mathrm{em}}$, and endows fermions with 
mass. This approach relates quark and lepton masses at the 
unification scale,
\begin{equation}
	{\left.
	\begin{array}{c}
		m_{e} = m_{d}  \\
		m_{\mu} = m_{s}  \\
		m_{\tau} = m_{b}
	\end{array}
	\right\}\mbox{  at }U\; ;}
{\mbox{ separate parameters }\left\{
	\begin{array}{l}
		m_{u} \\
		m_{c} \\
		m_{t} \;,
	\end{array}
	\right. }
\end{equation}
with implications for the observed masses that we will now elaborate.

The fermion masses evolve from the unification scale $U$ to the 
experimental scale $\mu$:
\begin{eqnarray}
	\ln\left[m_{u,c,t}(\mu)\right] & \approx & \ln\left[m_{u,c,t}(U)\right]
	+ {\frac{12}{33-2n_{f}}\ln\left(\frac{\alpha_{3}(\mu)}{\alpha_{U}}\right)} 
	\label{eq:Uprun}\\
	 & & + 
	{\frac{27}{88-8n_{f}}\ln\left(\frac{\alpha_{2}(\mu)}{\alpha_{U}}\right)}
	- 
	{\frac{3}{10n_{f}}\ln\left(\frac{\alpha_{1}(\mu)}{\alpha_{U}}\right)} 
	\;, \nonumber
\end{eqnarray}
\begin{eqnarray}
	\ln\left[m_{d,s,b}(\mu)\right] & \approx & \ln\left[m_{d,s,b}({U})\right]
	+ {\frac{12}{33-2n_{f}}\ln\left(\frac{\alpha_{3}(\mu)}{\alpha_{U}}\right)} 
	\label{eq:Downrun} \\
	 & & + 
	{\frac{27}{88-8n_{f}}\ln\left(\frac{\alpha_{2}(\mu)}{\alpha_{U}}\right)}
	+ 
	{\frac{3}{20n_{f}}\ln\left(\frac{\alpha_{1}(\mu)}{\alpha_{U}}\right)} 
	\;,\nonumber
\end{eqnarray}
\begin{eqnarray}
	\ln\left[m_{e,\mu,\tau}(\mu)\right] & \approx & \ln\left[m_{e,\mu,\tau}({U})\right]
	\phantom{+ {\frac{12}{33-2n_{f}}\ln\left(\frac{\alpha_{3}(\mu)}{\alpha_{U}}\right)}}
	 \label{eq:Leprun}\\
	 & & + 
	{\frac{27}{88-8n_{f}}\ln\left(\frac{\alpha_{2}(\mu)}{\alpha_{U}}\right)}
	- 
	{\frac{27}{20n_{f}}\ln\left(\frac{\alpha_{1}(\mu)}{\alpha_{U}}\right)} 
	\;,\nonumber
\end{eqnarray}
where I have omitted a small Higgs-boson contribution to keep the 
formulas short. The classic success of \suf\ unification is the 
predicted relation between $m_{b}$ and $m_{\tau}$~\cite{Buras:1978yy}. Combining 
\eqn{eq:Downrun} and \eqn{eq:Leprun}, we have
\begin{equation}
\ln\left[\frac{m_{b}(\mu)}{m_{\tau}(\mu)}\right]  \approx 
{\ln\left[\frac{m_{b}({U})}{m_{\tau}({U})}\right]}
 + {\frac{12}{33-2n_{f}}\ln\left(\frac{\alpha_{3}(\mu)}{\alpha_{U}}\right) }
 -   {\frac{3}{2n_{f}}\ln\left(\frac{\alpha_{1}(\mu)}{\alpha_{U}}\right)}  \;,
\label{eq:btau}
\end{equation}
where the first term on the right-hand side vanishes. Choosing for 
illustration $n_{f}=6$, $1/\alpha_{U}=40$, $1/\alpha_{s}(\mu)=5$,
and $1/\alpha_{1}(\mu)=65$, we compute at a low scale 
\begin{equation}
    m_{b} = 2.91 m_{\tau} \approx 5.16\gev \;,
    \label{eq:btaueq}
\end{equation}
in suggestive agreement with experiment. The factor-of-three 
ratio arises because the quark masses, influenced by QCD, evolve more 
rapidly than the lepton masses.

The example of $b$-$\tau$ unification raises the hope that all 
fermion masses arise on high scales, and show simple patterns there. 
The other cases are not so pretty. You can see the situation yourself 
by working

\begin{problem}
    Choosing an observation scale $\mu \approx 1\gev$, compute 
    $m_{s}/m_{\mu}$ and $m_{d}/m_{e}$ and compare with experiment.
    A more elaborate symmetry breaking scheme that adds a $\mathbf{45}$ of 
    scalars can change the relation for $m_{e}/m_{d}$ at the unification 
    scale, and lead to a more agreeable result at low energies. Show 
    that the relations $m_{s} = \cfrac{1}{3}m_{\mu}$, $m_{d} = 3 
    m_{e}$ at the unification scale lead to the low-energy 
    predictions, $m_{s} \approx \cfrac{4}{3}m_{\mu}$ and $m_{d} \approx 
    12 m_{e}$.
\end{problem}

The prospect of finding order among the fermion masses has spawned a 
lively theoretical industry.~\footnote{For a recent review of unified models for 
fermion masses and mixings, with an emphasis on supersymmetric 
examples, see Ref.~\cite{Chen:2003zv}.} The essential strategy 
comprises four steps: $\Box$ Begin with supersymmetric \suf, which 
has advantages (as we shall see in Lecture~4) for 
$\sin^{2}\theta_{W}$, coupling-constant unification, and the proton 
lifetime, or with supersymmetric $\mathrm{SO(10)}$, which 
accommodates a massive neutrino gracefully. $\Box$ Find 
``textures''---simple patterns of Yukawa matrices---that lead to 
successful predictions for masses and mixing angles. $\Box$ Interpret 
the textures in terms of symmetry breaking patterns. $\Box$ Seek a 
derivation---or at least a motivation---for the winning entry.

{\vspace*{6pt}\small\noindent
\textit{Aside: varieties of neutrino mass.} We recall that the chiral decomposition 
of a Dirac spinor is
\begin{equation}
    \psi = \half(1-\gamma_{5})\psi + \half(1 + \gamma_{5})\psi \equiv 
        \psi_{\mathrm{L}} + \psi_{\mathrm{R}}\;,
    \label{eq:chiral}
\end{equation}
and that the charge conjugate of a right-handed field is left-handed,
$\psi_{\mathrm{L}}^{c} \equiv (\psi^{c})_{\mathrm{L}} =
(\psi_{\mathrm{R}})^{c}$. What are the possible forms for mass terms? 
The familiar Dirac mass term, as we have emphasized for the quarks 
and charged leptons, connects the left-handed and right-handed 
components of the same field,
\begin{equation}
    \mathcal{L}_{\mathrm{D}} = D(\bar{\psi}_{\mathrm{L}}\psi_{\mathrm{R}}+ 
    \bar{\psi}_{\mathrm{R}}\psi_{\mathrm{L}}) = D\bar{\psi}\psi\;,
    \label{eq:Dmassterm}
\end{equation}
(compare \eqn{eq:emass}) so that the mass eigenstate is $\psi = \psi_{\mathrm{L}} + 
\psi_{\mathrm{R}}$. This combination is invariant under 
global phase rotation $\nu \rightarrow e^{i\theta}\nu$, $\ell 
\rightarrow e^{i\theta}\ell$, so that lepton number is conserved. 

In contrast, Majorana mass terms connect the left-handed and 
right-handed components of conjugate fields,
\begin{eqnarray}
-\mathcal{L}_{\mathrm{MA}} & = & A(\bar{\psi}_{\mathrm{R}}^{c}\psi_{\mathrm{L}} + 
\bar{\psi}_{\mathrm{L}}\psi_{\mathrm{R}}^{c}) = A \bar{\chi}\chi
\nonumber  \\
-\mathcal{L}_{\mathrm{MB}} & = & 
B(\bar{\psi}^{c}_{\mathrm{L}}\psi_{\mathrm{R}} + 
\bar{\psi}_{\mathrm{R}}\psi^{c}_{\mathrm{L}}) = B\bar{\omega}\omega\;,
\end{eqnarray}
which is only possible for neutral fields.  In the Majorana case, the
mass eigenstates are
\begin{eqnarray}
\chi & \equiv & \psi_{\mathrm{L}} + \psi^{c}_{\mathrm{R}} = \chi^{c}
= \psi_{\mathrm{L}} + (\psi_{\mathrm{L}})^{c} 
\nonumber  \\
\omega & \equiv & \psi_{\mathrm{R}} + \psi^{c}_{\mathrm{L}} = \omega^{c}
= \psi_{\mathrm{R}} + (\psi_{\mathrm{R}})^{c} \;.
\end{eqnarray}
The mixing of particle and antiparticle fields means that the Majorana 
mass terms correspond to processes that violate lepton number by two 
units. Accordingly, a Majorana neutrino can mediate neutrinoless 
double beta decay, $(Z,A) \rightarrow (Z+2,A) + e^{-} + e^{-}$. 
Detecting neutrinoless double beta decay would offer decisive 
evidence for the Majorana nature of the neutrino.\footnote{For an 
excellent recent review, see Ref.~\cite{Elliott:2002xe}.}
\vspace*{6pt}}

Unified theories do nothing to address the hierarchy problem we 
encountered in \S~\ref{subsec:hierarchy}. In tomorrow's final 
lecture, we will have a look at approaches to the question, ``Why is 
the electroweak scale small?'' Here are some of the questions raised 
by our quick tour of unified theories:

\begin{questions}{Third}
    \addtocounter{bean}{20}	
    \item  What are the steps to unification? Is there just one, or are 
    there several?
    
    \item Is perturbation theory a reliable guide to coupling-constant 
    unification?
    
    \item Is the proton unstable? How does it decay?
    
    \item What sets the mass scale for the additional gauge bosons in a 
    unified theory? \ldots\ for the additional Higgs bosons?
    
    \item How can we incorporate gravity?
    
    \item Which quark doublet is matched with which lepton doublet?
\end{questions}

\section{Extending the Electroweak Theory \label{sec:beyondew}}
We learned at the end of Lecture~2 (\S~\ref{subsec:hierarchy}) that 
the \ewgg\ electroweak theory does not explain how the scale of electroweak 
symmetry breaking remains low in the presence of quantum corrections. 
In operational terms, the problem is how to control the contribution 
of the integral in \eqn{longint}, given the long range of integration.
In the many years since we learned to take the electroweak theory 
seriously, only a few distinct scenarios have shown promise.

\begin{figure}[b!]
\begin{center}
\includegraphics[width=7.cm]{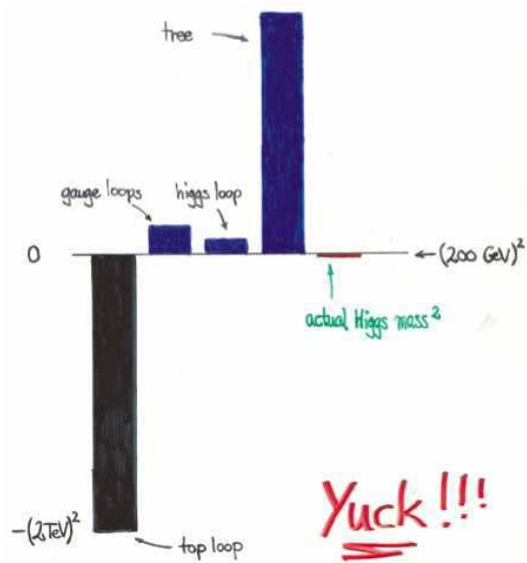}
\caption{Contributions that must balance to great precision if the 
Higgs-boson mass is to take on a reasonable value, assuming that the 
electroweak theory is a good description of physics up to 
$10\tev$~\cite{Schmaltz:2002wx}.\label{fig:schmaltz}}
\end{center}
\end{figure}

We could, of course, ask less of our theory and not demand that it 
describe physics all the way up to the Planck scale or the unification 
scale. But even if we take the reasonable position that the 
electroweak theory is an effective theory that holds up to 
$10\tev$---just an order of magnitude above our present 
experiments---stabilizing the Higgs mass necessitates a 
preternaturally delicate 
balancing act, as shown in Figure~\ref{fig:schmaltz}.
No principle forbids such a fine balance, but---as Martin Schmaltz's 
editorial comment on his figure eloquently conveys---we have come to 
believe that Nature finds solutions more elegant than this one.

The 
supersymmetric solution does have the virtue of elegance 
\cite{Lykken:1996xt, Martin:1997ns, Dawson:1997tz, Murayama:2000fm}.  Exploiting the fact 
that fermion loops contribute with an overall minus sign (because of 
Fermi statistics), supersymmetry balances the contributions of fermion 
and boson loops.  In the limit of unbroken supersymmetry, in which the 
masses of bosons are degenerate with those of their fermion 
counterparts, the cancellation is exact:
\begin{equation}
	\sum_{{i={\rm fermions \atop + bosons}}}C_i 
	\,g^{2}\!\!\!\int_{p^{2}}^{\Lambda^{2}}\!\! dk^2 = 0\;.
\end{equation}
If the supersymmetry is broken (as it must be in our world), the 
contribution of the integrals may still be acceptably small if the 
fermion-boson mass splittings $\Delta M$ are not too large.  The 
condition that $g^2\Delta M^2$ be ``small enough'' leads to the 
requirement that superpartner masses be less than about 
$1\tev$.

In a recently constructed class of theories, called little Higgs
models~\cite{Schmaltz:2002rm}, the Higgs boson is interpreted as a 
pseudo-Nambu--Goldstone boson. A broken global symmetry arranges
that the cancellation of the quantum corrections occurs between fields
of the same spin: fermions cancel fermions and bosons cancel bosons.
The models require new heavy fermionic partners for quarks and leptons,
and also TeV-scale partners for gauge bosons~\cite{Burdman:2002ns, 
Han:2003wu, Azuelos:2004dm}.

A third solution to the problem of the enormous range of integration in 
\eqn{longint} is offered by theories of dynamical symmetry breaking such as 
technicolor \cite{Lane:2002wv, Hill:2002ap}. In technicolor models, the Higgs boson is composite, and 
new physics arises on the scale of its binding, $\Lambda_{\mathrm{TC}} \simeq 
O(1~{\rm TeV})$. Thus the effective range of integration is cut off 
(effectively by form factor effects), and 
mass shifts are under control.

A fourth possibility is that the gauge sector becomes strongly 
interacting. This would give rise to $WW$ resonances, multiple 
production of gauge bosons, and other new phenomena at energies of 1 TeV 
or so.  It is likely that a scalar bound state---a quasi-Higgs 
boson---would emerge with a mass less than about 
$1\tev$~\cite{Chanowitz:1998wi}.

We cannot avoid the conclusion that some new physics must occur on 
the \onetev.\footnote{Since the superconducting phase transition 
informs our understanding of the Higgs mechanism for electroweak
symmetry breaking, it may be useful to look to other collective
phenomena for inspiration.  Although the implied gauge-boson masses
are unrealistically small, chiral symmetry breaking in QCD can induce
a dynamical breaking of electroweak symmetry~{\protect 
\cite{Weinstein:1973gj}}.
(This is the prototype for technicolor models.)  Is it possible that
other interesting phases of QCD---color superconductivity {\protect
\cite{Alford:1998zt, Rajagopal:2000wf, Alford:2001dt}}, for example---might hold lessons
for electroweak symmetry breaking under normal or unusual conditions? 
}

\subsection{Supersymmetry{\protect\footnote{See the excellent courses on 
supersymmetry by Kazakov at the 2000 European School on High-Energy 
Physics~\cite{Kazakov:2000ra}, Roulet at the 2001 Latin-American 
School~\cite{Roulet:2001se}, and Ellis at the 2001 European 
School~\cite{Ellis:2002mx}.}}}

\subsubsection{Why Supersymmetry? \label{subsubsec:susywhy}}
Supersymmetry is a fermion--boson symmetry that arises from the 
presence of \textit{new fermionic dimensions.} The most general 
symmetry of the $S$-matrix is Poincar\'{e} invariance plus internal 
symmetries---plus supersymmetry. The new symmetry relates fermion to 
boson degrees of freedom: roughly speaking, each particle has a 
superpartner with spin offset by $\cfrac{1}{2}$. The observed 
particle spectrum contains no fermion--boson mates that we can 
identify as superpartners. Accordingly, imposing supersymmetry 
requires doubling the spectrum of particles.
The properties of the supersymmetric partners of the \smgg\ particles are 
shown in Table~\ref{tab:susyspec}.
\begin{table}[t!]
    \centering
    \caption{Supersymmetric partners of \smgg\ fermions and gauge 
    bosons.}
    \vspace*{6pt}
    \begin{tabular}{lcccc}
	    \hline\\[-12pt]
Particle & Spin & Color & Charge & $R$-parity \\
\hline
$g$\quad gluon  & 1 & \textbf{8} & 0 & $+1$ \\
$\tilde{g}$ \quad gluino & $\cfrac{1}{2}$ & \textbf{8 }& 0 & $-1$ \\
$\gamma$ \quad photon & 1 & \textbf{1} & 0 & $+1$ \\
$\tilde{\gamma}$ \quad photino & $\cfrac{1}{2}$ & \textbf{1} & 0 & 
$-1$ \\
$W^{\pm},\,Z^{0}$ \quad intermediate bosons & 1 & \textbf{1} & 
$\pm1,\,0$ & $+1$ \\
$\widetilde{W}^{\pm},\,\widetilde{Z}^{0}$ \quad electroweak 
gauginos & $\cfrac{1}{2}$ & \textbf{1} & $\pm1,\,0$ & $-1$ \\
$q$ \quad quark & $\cfrac{1}{2}$ & \textbf{3} & 
$\cfrac{2}{3},\,-\cfrac{1}{3}$ & $+1$ \\
$\tilde{q}$ \quad squark & $0$ & \textbf{3} & 
$\cfrac{2}{3},\,-\cfrac{1}{3}$ & $-1$ \\
$\ell$ \quad charged lepton & $\cfrac{1}{2}$ & \textbf{1} & $-1$ & 
$+1$ \\
$\tilde{\ell}$ \quad charged slepton & 0 & \textbf{1} & $-1$ & 
$-1$ \\
$\nu$ \quad neutrino & $\cfrac{1}{2}$ & \textbf{1} & $0$ & $+1$ \\
$\tilde{\nu}$ \quad sneutrino & $0$ & \textbf{1} & $0$ & $-1$ \\
\hline
\end{tabular}
    \label{tab:susyspec}
\end{table}

In a supersymmetric theory, two Higgs doublets are required to give 
masses to fermions with weak isospin $I_{3}=\cfrac{1}{2}$ and 
$I_{3}=-\cfrac{1}{2}$. Let us designate the two doublets as 
$\Phi_{1}$ and $\Phi_{2}$.  Before supersymmetry is broken, the scalar 
potential has the form
\begin{equation}
	V = \mu^{2}(\Phi_{1}^{2} + \Phi_{2}^{2}) + 
	\frac{g^{2}+g^{\prime2}}{8}(\Phi_{1}^{2} + \Phi_{2}^{2})^{2} + 
	\frac{g^{2}}{2}\left|\Phi_{1}^{*} \cdot \Phi_{2}\right|^{2} \; .
	\label{eq:susyhiggspot}
\end{equation}
By adding all possible soft supersymmetry-breaking terms, we raise the 
possibility that the electroweak symmetry will be broken.  We choose
\begin{equation}
	\begin{array}{c}
		\vev{\Phi_{1}} = v_{1} > 0 \; ,  \\[6pt]
		\vev{\Phi_{2}} = v_{2} > 0 \; ,
	\end{array}
	\label{eq:susyvevs}
\end{equation}
with $v_{1}^{2} + v_{2}^{2} = v^{2}$ and
\begin{equation}
	\frac{v_{2}}{v_{1}} \equiv \tan\beta \; .
	\label{eq:tanb}
\end{equation}

After the $W^{\pm}$ and $Z^{0}$ acquire masses, five spin-zero 
degrees of freedom remain as massive spin-zero particles: the 
lightest scalar $h^{0}$, a heavier neutral scalar $H^{0}$, two charged 
scalars $H^{\pm}$, and a neutral pseudoscalar $A^{0}$.

Supersymmetry also relates the interactions of the superpartners to
those of the known particles. In addition to transcriptions of the 
usual interactions, supersymmetry admits new Yukawa terms,
\begin{equation}
    \mathcal{L}_{\mathrm{SUSY-Yuk}} = 
    \lambda_{ijk}L^{i}L^{j}E^{k} + 
    \lambda^{\prime}_{ijk}L^{i}Q^{j}\bar{D}^{k} + 
    \lambda^{\prime\prime}\bar{U}^{i}\bar{D}^{j}\bar{D}^{k}\;,
    \label{eq:rpv}
\end{equation}
that entail 45 new free parameters. The new interactions in 
\eqn{eq:rpv} induce dangerous lepton- and baryon-number violations. 
For example, expanding the first term we have 
\begin{equation}
    \mathcal{L}_{LLE} = 
    \lambda_{ijk}\,\tilde{\nu}^{i}_{L}e^{i}_{L}\bar{e}^{k}_{R} + \ldots
    \label{eq:lle}
\end{equation}
To banish these, it has become conventional to impose symmetry under
$R$-parity,
\begin{equation}
    R = (-1)^{3B+L+S},
    \label{eq:rpar}
\end{equation}
a multplicative quantum number that is even for the known particles 
and odd for superpartners. In a theory that conserves $R$-parity, the 
superpartners must be produced in pairs and the lightest superpartner 
(LSP) is stable. A neutral LSP is a promising dark matter 
candidate~\cite{Olive:2003iq}.

\begin{figure}[t!]
\begin{center}
\includegraphics[width=10cm]{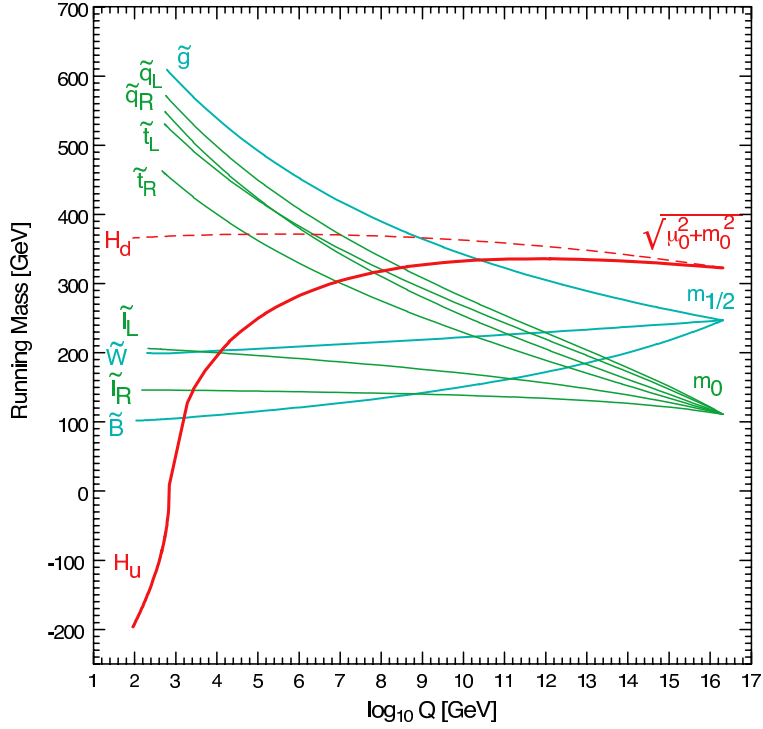}
\caption{Evolution of superpartner masses in  constrained minimal 
supersymmetry, from Ref.\ {\protect \cite{Kane:1994td}}. The 
sign of $M^{2}$ is indicated.
\protect\label{fig:mssmssb}}\end{center}
\end{figure}

An interesting feature of supersymmetric theories is the possibility 
that spontaneous electroweak symmetry might be driven by a heavy top 
quark, provided that $55\gev \ltap m_{t} \ltap 200\gev$. In a 
supersymmetric unified theory, the large top-quark Yukawa coupling 
drives the mass-squared of the Higgs boson responsible for $I_{3} = 
\cfrac{1}{2}$ masses to negative values at low 
energies~\cite{Alvarez-Gaume:1983gj, Ibanez:1983wi}, as shown in 
Figure~\ref{fig:mssmssb} for a 175-GeV top quark.
The resulting Higgs potential is minimized with a nonzero vacuum 
expectation value for the Higgs field---corresponding to hidden 
electroweak symmetry.

\begin{figure}[t!]
\begin{center}
\includegraphics[width=7.5cm]{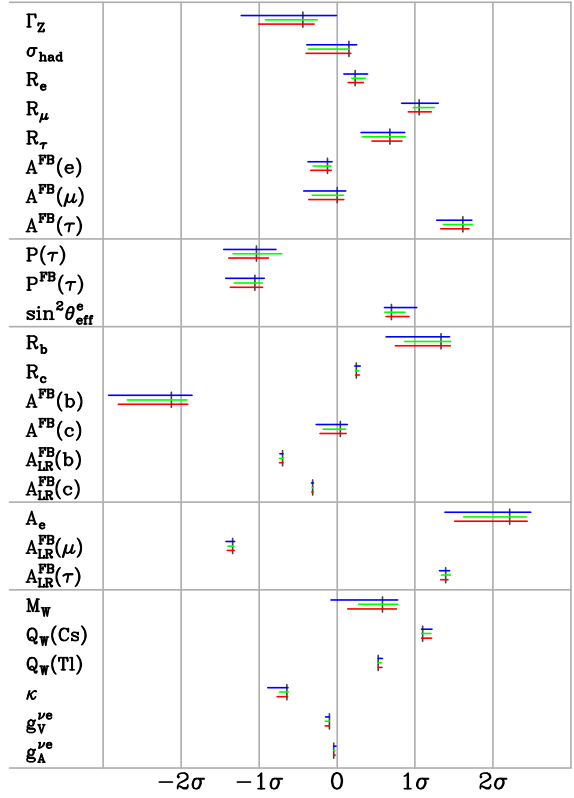}
\caption{The range of best fit predictions of precision observables in
the supergravity model (upper horizontal lines), the $\mathbf{5} \oplus
\mathbf{5^{*}}$ gauge-mediated model (middle lines), the $\mathbf{10}
\oplus \mathbf{10^{*}}$ gauge-mediated model (lower lines), and in the
standard model at its global best fit value (vertical lines), in units
of standard deviation, from Ref.\ {\protect \cite{Erler:1998ur}}.
\protect\label{fig:allobs}}\end{center}
\end{figure}

One of the best phenomenological motivations for supersymmetry on the 
1-TeV scale is that the minimal supersymmetric extension of the standard 
model so closely approximates the standard model itself.  A nice 
illustration of the small differences between predictions of 
supersymmetric models and the standard model is the compilation of 
pulls prepared by Erler and Pierce~\cite{Erler:1998ur},\footnote{See 
also Ref.~\cite{Cho:1999km} for an update.} which is shown 
in Figure \ref{fig:allobs}.  
This is a nontrivial property of new physics beyond the standard model,
and a requirement urged on us by the unbroken quantitative success of
the established theory.

\subsubsection{Coupling Constant Unification in Supersymmetric 
Unification\label{subsubsec:susyu}}
We found in \S\ref{subsec:intlag} that in a unified theory of the 
strong, weak, and electromagnetic interactions based on the gauge 
symmetry \suf, the couplings $1/\alpha_{1}$, $1/\alpha_{2}$, and 
$1/\alpha_{3}$ tend to approach each other at high energies, but fail 
to coincide at a single point we would identify as the unification 
energy $U$. The evolution of the couplings is changed appreciably by 
the influence of supersymmetry's richer spectrum---gluinos, squarks, 
sleptons, the second Higgs doublet, gauginos, and Higgsinos. It would 
be an important indirect encouragement for the development of 
supersymmetric models to find that a supersymmetrized version of our 
example unified theory does unify the couplings, so let us check.
Working again to leading logarithmic approximation, we have
\begin{equation}
    \hbox{SUSY }\mathrm{SU(3)_{\mathrm{c}}:}\quad 1/\alpha_{3}(Q^{2}) = 1/\alpha_{3}(\mu^{2}) + 
    b_{3}\log(Q^{2}/\mu^{2})  \;, \label{eq:susysu3run}
\end{equation}
where $4\pi b_{3} = 9 - 2n_{\mathrm{gen}}$, so that 
$b_{3} = 3/4\pi$; in the normal \suf\ theory, it was $7/4\pi$ 
\eqn{eq:su3run}.

\begin{equation}
    \hbox{SUSY }\mathrm{SU(2)_{\mathrm{L}}:}\quad 1/\alpha_{2}(Q^{2}) = 
    1/\alpha_{2}(\mu^{2}) + b_{2}\log(Q^{2}/\mu^{2})  \;, 
    \label{eq:susysu2run}
\end{equation}
where $4\pi b_{2} = 6-2n_{\mathrm{gen}} -
\cfrac{1}{2}n_{\mathrm{Higgs}}$, so that $b_{2} = -1/4\pi$; without 
supersymmetry, it was $19/24\pi$ \eqn{eq:su2run}.

\begin{equation}
    \hbox{SUSY }\mathrm{U(1)}_{Y}:\quad 1/\alpha_{1}(Q^{2}) =  1/\alpha_{Y}(\mu^{2}) + 
    b_{1}\log(Q^{2}/\mu^{2})  \;, \label{eq:susyu1run}
\end{equation}
with $4\pi b_{1} = -2n_{\mathrm{gen}} - 
\cfrac{3}{10}n_{\mathrm{Higgs}}$, so that $b_{1} = -33/20\pi$; it had 
been $-41/40\pi$ \eqn{eq:u1run}. As usual, $Q^{2}$ is the scale 
of interest and $\mu^{2}$ is the reference scale.

Using the new evolution equations \eqn{eq:susysu3run} --
\eqn{eq:susyu1run} all the way down to $M_{Z}^{2}$, since a
supersymmetric solution to the hierarchy problem demands that
superpartner masses are $\ltap 1\tev$, we estimate the unification
scale $U_{\mathrm{SUSY}} \approx 4\times 10^{16}\gev$ and the unified 
coupling $1/\alpha_{U_{\mathrm{SUSY}}} \approx 24.6$; in
the pure \suf\ case, we found $U \approx10^{15}\gev$ and 
$1/\alpha_{U} \approx 42$. As before, we 
have taken as inputs $\alpha_{3}(M_{Z}^{2}) \approx 1/8.75$,
$\alpha(M_{Z}^{2}) \approx 1/128.9$, and $M_{Z} \approx 91.19\gev$. 
Now for the test: we evaluate the weak mixing parameter at the weak 
scale and find 
\begin{equation}
    \left.x_{W}(M_{Z}^{2})\right|_{\mathrm{SUSY~SU(5)}} \approx 
    0.23\; ,
    \label{eq:susyxwsu5mz}
\end{equation}
whereas \suf\ unification gave $0.21$ \eqn{eq:xwsu5mz}. The 
supersymmetric prediction, \eqn{eq:susyxwsu5mz}, is in very suggestive 
agreement with the measured value, $0.2314 \pm 0.003$ \eqn{eq:xwexp}.

\begin{figure}[t!]
\begin{center}
\includegraphics[width=10cm]{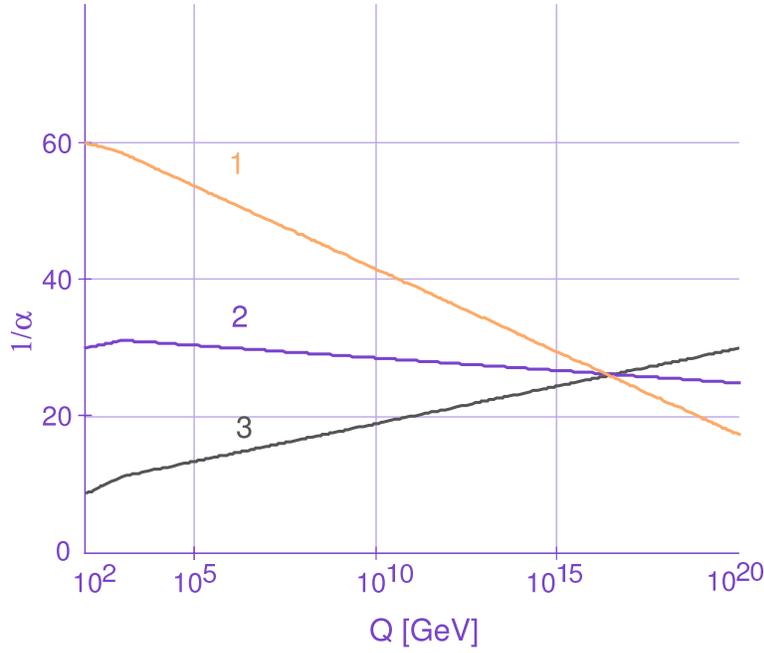}
\caption{Running of $1/\alpha_{1}, 1/\alpha_{2}, 1/\alpha_{3}$ in the
supersymmetric \suf\ unified theory, with the superpartner threshold  
at $1\tev$.  \label{fig:susyalfevolve}}
\end{center}
\end{figure}

To test coupling constant unification graphically, we set the 
superpartner threshold at $1\tev$ and evolve the couplings from 
$M_{Z}^{2}$ up to a high scale. The results displayed in 
Figure~\ref{fig:susyalfevolve}
show that the three values do indeed coincide in the neighborhood of 
$4\times 10^{16}\gev$. This is a highly gratifying outcome for the 
hypothesis that supersymmetry operates on the electroweak scale.

\subsubsection{The Lightest Higgs Boson \label{subsubsec:lhb}}
At tree 
level, we may express all the (pseudo)scalar masses in terms of 
$M_{A}$ and $\tan\beta$, to find
\begin{equation}
	M_{h^{0},H^{0}}^{2} = \frac{1}{2} \left\{ M_{A}^{2} + M_{Z}^{2} \mp \left[ 
	(M_{A}^{2} + M_{Z}^{2})^{2} - 4M_{A}^{2} M_{Z}^{2} \cos^{2}2\beta
	\right]^{1/2}\right\} \; ,
	\label{eq:hHmass}
\end{equation}
and
\begin{equation}
	M_{H^{\pm}}^{2} = M_{W}^{2} + M_{A}^{2} \; .
	\label{eq:chHmass}
\end{equation}
At tree level, there is a simple mass hierarchy, given by
\begin{eqnarray}
	M_{h^{0}} & < & M_{Z}|\cos2\beta|
	\nonumber  \\
	M_{H^{0}} & > & M_{Z}
	\label{eq:susymasshier}  \\
	M_{H^{\pm}} & > & M_{W}
	\nonumber \; ,
\end{eqnarray}
but there are very important \textit{positive} loop corrections to $M_{h^{0}}^{2}$ 
(proportional to $G_{\mathrm{F}}m_{t}^{4}$) that were neglected in the 
earliest calculations.  These loop corrections change the mass 
predictions very significantly.

\begin{figure}[t!]
\begin{center}
\includegraphics[width=10cm]{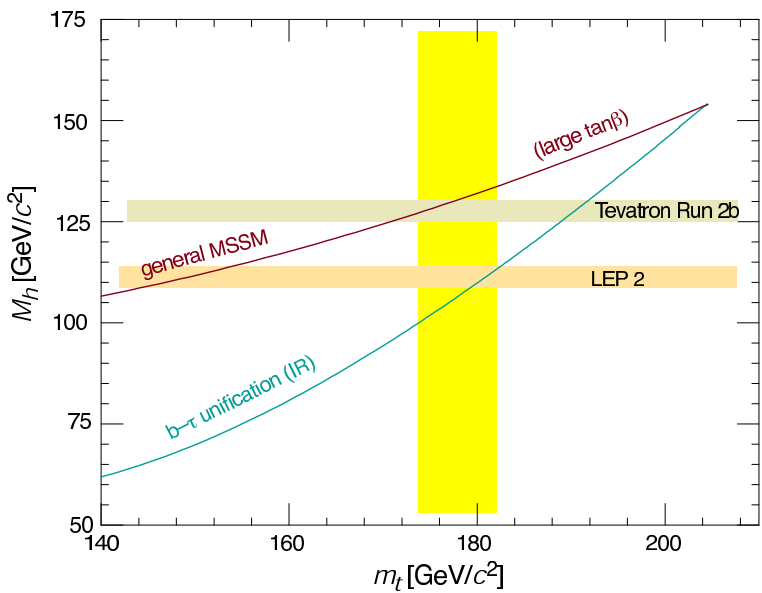}
\caption{Upper bounds on the mass of the lightest Higgs boson, as a function of
the top-quark mass, in two variants of the minimal supersymmetric
standard model.  The upper curve refers to a general MSSM, in the
large-$\tan\beta$ limit; the lower curve corresponds to an
infrared-fixed-point scenario with $b$-$\tau$ unification, from Ref.\
{\protect \cite{Carena:1995bx}}.\label{fig:MHmt}}
\end{center}
\end{figure}

Because the minimal supersymmetric standard model (MSSM) implies 
\textit{upper bounds} on the mass of the lightest scalar $h^{0}$, it 
sets attractive targets for experiment.  Two such upper bounds are 
shown as functions of the top-quark mass in Figure \ref{fig:MHmt}.  
The large-$\tan\beta$ limit of a general MSSM yields the upper curve;
an infrared-fixed-point scheme with $b$-$\tau$ unification produces an
upper bound characterized by the lower curve.  The vertical band shows
the range $m_{t} = 178.0 \pm 4.3\gev$ favored by Tevatron
experiments~\cite{Group:2004rc}.  The LEP~2
experiments~\cite{Kado:2002er, Barate:2003sz} have explored the full
range of lightest-Higgs masses that occur in the infrared-fixed-point
scheme.  High-luminosity running at the Tevatron collider can extend 
the search field and, in the limit, explore much of the range of
$h^{0}$ masses allowed in the MSSM~\cite{Carena:2000yx, Babukhadia:2003zu}. 
The ATLAS~\cite{APtdr} and CMS experiments\cite{RohlfLHCwork, CMSHiggs} at the Large Hadron Collider 
will carry the Higgs-boson search up to $1\tev$. A future 
electron-positron linear collider can carry out incisive studies of 
the lightest Higgs boson and of many superpartners~\cite{Dawson:2004xz}.

\subsubsection{The Stability of Matter \label{subsubsec:stability}}
Supersymmetry doubles the spectrum of fundamental particles.  We know 
that supersymmetry must be significantly broken in Nature, because the 
electron is manifestly not degenerate in mass with its scalar partner, 
the selectron.  It is interesting to contemplate just how different 
the world would have been if the selectron, not the electron, were 
the lightest charged particle and therefore the stable basis of 
everyday matter \cite{Cahn:1996ag}.  If atoms were selectronic, there would be 
no Pauli principle to dictate the integrity of molecules.  As 
Dyson~\cite{DysonStab} and Lieb~\cite{Lieb:2003nc} demonstrated, transforming 
electrons and nucleons from fermions to bosons would cause all 
molecules to shrink into an insatiable undifferentiated blob.  
Luckily, there is no analogue of chiral symmetry to 
guarantee naturally small squark and slepton masses.  So while 
supersymmetry menaces us with an amorphous death, it is 
likely that a full understanding of supersymmetry will enable us to 
explain why we live in a universe ruled by the exclusion principle.

\subsubsection{Challenges for Supersymmetry\label{subsubsec:challenges}}
Supersymmetry is an elegant idea, but if it applies to our world, 
supersymmetry is hidden. Some undiscovered dynamics will be required 
to explain supersymmetry breaking. The attractive notion of ``soft'' 
supersymmetry breaking leads us to the minimal supersymmetric standard 
model (MSSM), with 124 parameters---a large number to track, which 
doesn't instantly present it as progress over the eighteen or so of 
the standard model.

Theorists have invented a number of crafty schemes for supersymmetry 
breaking. The best known is called \textit{gravity mediation,} in 
which supersymmetry breaking at a very high scale is communicated to 
the standard model by gravitational interactions. In \textit{gauge mediation,} 
supersymmetry breaking occurs relatively near to the electroweak 
scale, perhaps below $100\tev$, and that is communicated to the 
standard model by possibly nonperturbative gauge forces. Other 
schemes provide more or less natural solutions to one or another 
requirement, but it is fair to say that none meets all the challenges.

Let us list some of the issues we shall have to face to construct a 
fully acceptable supersymmetric theory. $\Box$ Weak-scale 
supersymmetry (i.e., superpartners on the electroweak scale) protects 
the Higgs-boson mass and keeps it naturally below $1\tev$, but does 
not explain the why the weak scale itself is so much smaller than the 
Planck scale. This is called the $\mu$ problem.\footnote{For a 
compact summary, see the discussion below (5.5) in 
\cite{Martin:1997ns}.} $\Box$ Global supersymmetry must deal with the 
threat of flavor-changing neutral currents. $\Box$ In parallel with 
the standard model, supersymmetric models make reasonably clear 
predictions for the masses of gauge bosons and gauge fermions 
(gauginos), but are more equivocal about the masses of squarks and 
sleptons. $\Box$ Not only is supersymmetry a hidden symmetry in the 
technical sense, it seems to be a very well hidden symmetry, in that 
we have no direct experimental evidence in its favor. A certain number 
of contortions are required to accommodate a (lightest) Higgs-boson 
mass above $115\gev$. $\Box$ We might have hoped that supersymmetry 
would relate particles to forces---quarks and leptons to gauge 
bosons---but instead it doubled the spectrum. $\Box$ Dangerous baryon- and 
lepton-number--violating interactions arise naturally from a 
supersymmetric Lagrangian, but we have only learned to banish them by 
decree. $\Box$ Supersymmetry introduces new sources of $\mathcal{CP}$ 
violation that are potentially too large. $\Box$ We haven't yet found 
a convincing and viable picture of the TeV superworld.

This long list of challenges does not mean that supersymmetry is wrong,
or even that it is irrelevant to the 1-TeV scale.  But SUSY is not
automatically right, either!  If supersymmetry does operate on the 1-TeV scale,
then Nature must have found solutions to all these challenges \ldots
and we will need to find them, too. The discovery of supersymmetry 
will mark a beginning, not an end, of our learning.

\begin{figure}[t!]
\begin{center}
\includegraphics[width=10cm]{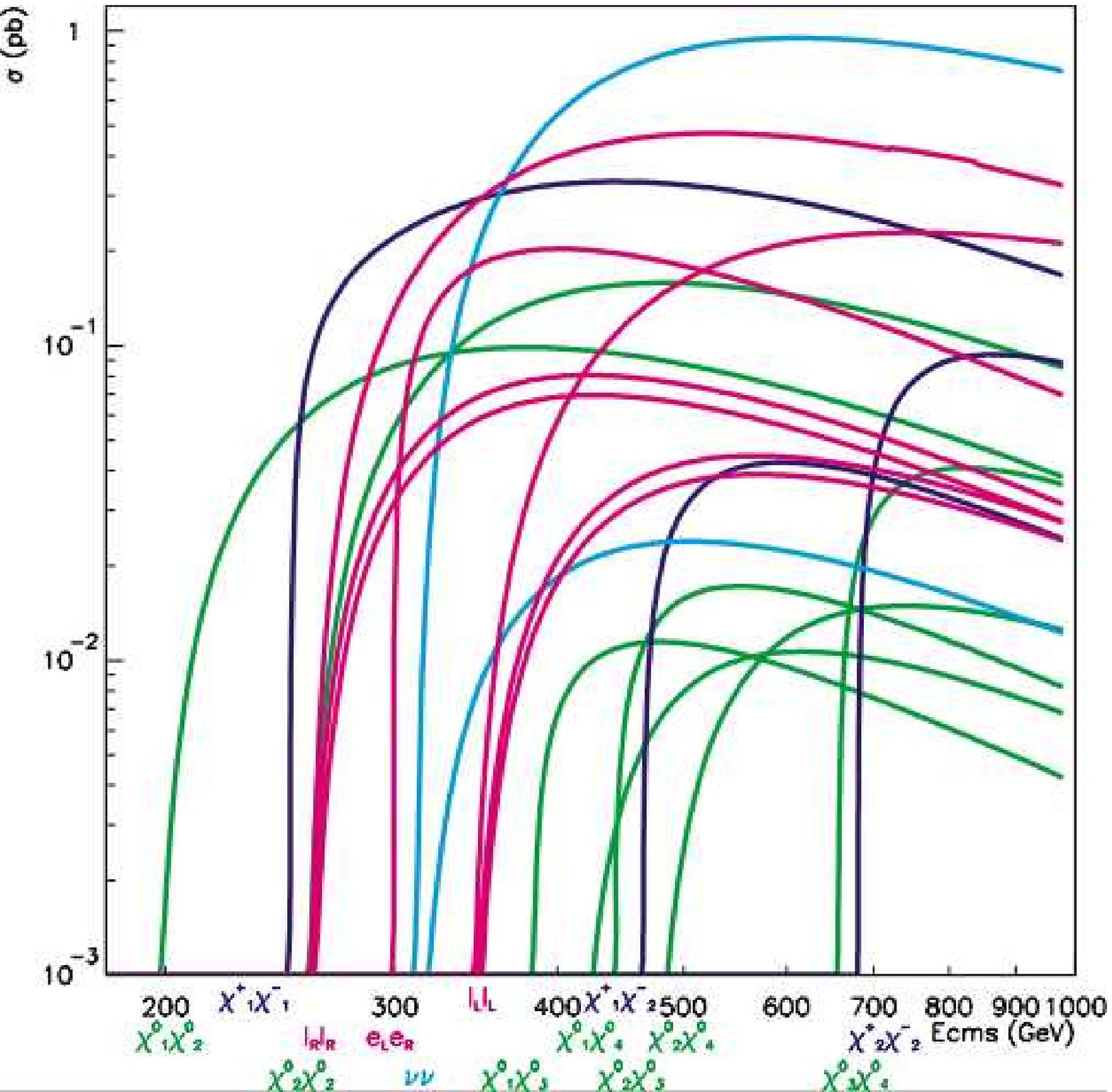}
\caption{Cross sections for electron-positron annihilations to pairs 
of superpartners in a minimal supergravity model, with $\tan\beta = 
3$ (Grahame Blair, unpublished). \label{fig:thresh}}
\end{center}
\end{figure}

If supersymmetry is the solution to electroweak symmetry breaking and 
the hierarchy problem, then we should see concrete evidence of it 
soon, in the Higgs sector and beyond. The thicket of thresholds to be 
explored in electron-positron annihilations into pairs of 
superpartners shown in Figure~\ref{fig:thresh} indicates that
if weak-scale supersymmetry is real, we shall live in ``interesting 
times.''

In my view, supersymmetry is (almost) certain to be true, as a path to 
the incorporation of gravity. Whether supersymmetry resolves the 
problems of the 1-TeV scale is a logically separate question, to which 
the answer is less obvious. \textit{Experiment will decide!}

\subsection{Electroweak Symmetry Breaking: Another Path? 
\label{subsec:TC}}
Dynamical symmetry breaking offers a different solution to the 
naturalness problem of the electroweak theory: in technicolor, there 
are no elementary scalars.  We hope that solving the dynamics that 
binds elementary fermions into a composite Higgs boson and other $WW$ 
resonances will bring addition predictive power.  It is worth saying 
that technicolor is a far more ambitious program than global 
supersymmetry.  It doesn't merely seek to finesse the hierarchy 
problem, it aims to predict the mass of the Higgs surrogate.  Against 
the aesthetic appeal of supersymmetry we can weigh technicolor's excellent 
pedigree.  As we have seen in \S\ref{sub:cache},
the Higgs mechanism of the standard model 
is the relativistic generalization of the Ginzburg-Landau description 
of the superconducting phase transi\-tion.  Dynamical symmetry breaking 
schemes---technicolor and its relatives---are inspired by the
Bardeen--Cooper--Schrieffer theory of superconductivity, 
and seek to give a similar microscopic description of electroweak 
symmetry breaking.  

The dynamical-symmetry-breaking approach realized in technicolor theories
 is modeled upon our understanding of the superconducting phase 
transition. The macroscopic order parameter of the Ginzburg-Landau 
phenomenology corresponds to the wave function of superconducting 
charge carriers, which acquires a nonzero vacuum
expectation value in the 
superconducting state. The microscopic Bardeen-Cooper-Schrieffer 
theory \cite{Bardeen:1957kj} identifies the dynamical origin of the order parameter with 
the formation of bound states of elementary fermions, the Cooper pairs of 
electrons. The basic idea of  technicolor is to replace the 
elementary Higgs boson with a fermion-antifermion bound 
state. By analogy with the superconducting phase transition, the dynamics 
of the fundamental technicolor gauge interactions among technifermions 
generate scalar bound states, and these play the role of the Higgs fields.

The elementary fermions---electrons---and 
gauge interactions---QED---needed to generate the scalar bound states are 
already present in the case of superconductivity. Could a scheme
 of similar economy account
for the transition that hides the electroweak symmetry?
Consider an \smgg\ theory of massless up and 
down quarks. Because the strong interaction is strong, and the electroweak 
interaction is feeble, we may treat the \ewgg\
interaction as a perturbation. For vanishing quark masses, QCD has an exact 
$\mathrm{SU(2)_L\otimes SU(2)_R}$ chiral symmetry. At an energy scale 
$\sim\Lambda_{\mathrm{QCD}},$ the strong interactions become strong, fermion 
condensates appear, and the chiral symmetry is spontaneously broken
to the familiar flavor symmetry:
\begin{equation}
	\mathrm{SU(2)_L\otimes SU(2)_R \to SU(2)_V}\;\; .
\end{equation}
 Three Goldstone bosons appear, one for 
each broken generator of the original chiral invariance. These were 
identified by Nambu~\cite{Nambu:1960xd} as three massless pions.

The broken generators are three axial currents whose couplings to pions are 
measured by the pion decay constant $f_\pi$. When we turn on the 
\ewgg\ electroweak interaction, the electroweak gauge 
bosons couple to the axial currents and acquire masses of order $\sim 
gf_\pi$. The mass-squared matrix,
\begin{equation}
	\mathcal{M}^{2} = \left(
		\begin{array}{cccc}
		g^{2} & 0 & 0 & 0  \\
		0 & g^{2} & 0 & 0  \\
		0 & 0 & g^{2} & gg^{\prime}  \\
		0 & 0 & gg^{\prime} & g^{\prime2}
	\end{array}
		 \right) \frac{f_{\pi}^{2}}{4} \; ,
	\label{eq:csbm2}
\end{equation}
(where the rows and columns correspond to $W^{+}$, $W^{-}$, $W_{3}$, 
and $\mathcal{A}$) has the same structure as the mass-squared matrix 
for gauge bosons in the standard electroweak theory.  Diagonalizing 
the matrix \eqn{eq:csbm2}, we find that $M_{W}^{2} = 
g^{2}f_{\pi}^{2}/4$ and $M_{Z}^{2} = 
(g^{2}+g^{\prime2})f_{\pi}^{2}/4$, so that 
\begin{equation}
	\frac{M_{Z}^{2}}{M_{W}^{2}} = \frac{(g^{2}+g^{\prime2})}{g^{2}} = 
	\frac{1}{\cos^{2}\theta_{W}}\; .
	\label{eq:wzrat}
\end{equation}
The photon emerges massless.

The massless pions thus disappear from the physical spectrum, 
having become the longitudinal components of the weak gauge bosons. 
Unfortunately, the mass acquired by the 
intermediate bosons is far smaller than required for a successful 
low-energy phenomenology; it is only~\cite{Weinstein:1973gj} $M_W\approx 30\mevcc$.

The minimal technicolor model of Weinberg~\cite{Weinberg:1976gm} and 
Susskind~\cite{Susskind:1979ms}  
transcribes the same ideas from QCD to a new setting.  The 
technicolor gauge group is taken to be $\mathrm{SU}(N)_{\mathrm{TC}}$ (usually 
$\mathrm{SU(4)}_{\mathrm{TC}}$), 
so the gauge interactions of the theory are generated by
\begin{equation}
	\mathrm{SU(4)_{\mathrm{TC}}\otimes SU(3)_c \otimes SU(2)_L \otimes U(1)_\mathit{Y}}\; .
\end{equation}
The technifermions are a chiral doublet of massless color singlets
\begin{equation}
\begin{array}{cc}
	\left( \begin{array}{c} U \\ D \end{array} \right)_\mathrm{L} & U_\mathrm{R}, \;
D_\mathrm{R} \; .
\end{array}
\end{equation}
With the electric charge assignments $Q(U)=\frac{1}{2}$ and
$Q(D)=-\frac{1}{2}$, the  
theory is free of electroweak anomalies. The ordinary fermions are all 
technicolor singlets. 

In analogy with our discussion of chiral symmetry breaking in QCD, we 
assume that the chiral TC symmetry is broken,
\begin{equation}
	\mathrm{SU(2)_L\otimes SU(2)_R\otimes U(1)_V\to SU(2)_V\otimes U(1)_V}\; .
\end{equation}
Three would-be Goldstone bosons emerge. These are the technipions
\begin{equation}
\begin{array}{ccc}
\pi_T^+, & \pi_T^0, & \pi_T^-,
\end{array}
\end{equation}
for which we are free to {\em choose} the technipion decay constant as
\begin{equation}
	F_\pi = \left(G_{\mathrm{F}}\sqrt{2}\right)^{-1/2} = 246\gev\; . \label{FPI}
\end{equation}
This amounts to choosing the scale on which technicolor becomes strong.
When the electroweak interactions are turned on, the technipions become the 
longitudinal components of the intermediate bosons, which acquire masses
\begin{equation}
\renewcommand\arraystretch{1.5}
\begin{array}{ccccc}
	M_W^2 & = & g^2F_\pi^2/4 & = & 
{\displaystyle \frac{\pi\alpha}{G_{\mathrm{F}}\sqrt{2}\sin^2\theta_W}} \\
	M_Z^2 & = & \left(g^2+g^{\prime 2}\right)F_\pi^2/4 & = & 
M_W^2/\cos^2\theta_W
\end{array} \; ,
\renewcommand\arraystretch{1}
\end{equation}
that have the canonical standard model values, thanks to our choice 
(\ref{FPI}) 
of the technipion decay constant.

Technicolor shows how the generation of intermediate boson masses 
could arise without fundamental scalars or unnatural adjustments of 
parameters.  It thus provides an elegant solution to the naturalness 
problem of the standard model.  However, it has a major deficiency: it 
offers no explanation for the origin of quark and lepton masses, 
because no Yukawa couplings are generated between Higgs fields and 
quarks or leptons.  Consequently, technicolor serves as a reminder 
that there are two problems of mass: explaining the masses of the 
gauge bosons, which demands an understanding of electroweak symmetry 
breaking; and accounting for the quark and lepton masses, which 
requires not only an understanding of electroweak symmetry breaking 
but also a theory of the Yukawa couplings that set the scale of 
fermion masses in the standard model.  We can be confident that the 
origin of gauge-boson masses will be understood on the 1-TeV scale.  
We do not know where we will decode the pattern of the Yukawa 
couplings; we looked at one approach in \S~\ref{subsec:massprob}.

\begin{figure}[b!]
\begin{center}
\includegraphics[width=4cm]{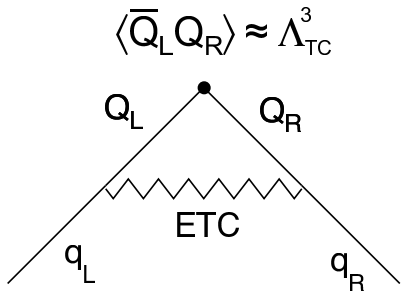}
\caption{In extended technicolor, mass is conveyed to the quarks and 
leptons through the interaction of ETC gauge bosons with the TC 
condensate. \label{fig:etcmass}}
\end{center}
\end{figure}

To generate fermion mass, we may embed technicolor in a larger 
extended technicolor gauge group $G_{\mathrm{ETC}} \supset G_{\mathrm{TC}}$ that 
couples the quarks and leptons to technifermions \cite{Eichten:1980ah, Dimopoulos:1979es}.  
If the 
$G_{\mathrm{ETC}}$ gauge symmetry is broken down to $G_{\mathrm{TC}}$ 
at a scale $\Lambda_{\mathrm{ETC}}$, then the quarks and leptons can 
acquire masses
\begin{equation}
m \sim \frac{g_{\mathrm{ETC}}^{2}F_{\pi}^{3}}{\Lambda_{\mathrm{ETC}}^{2}}\; ,
	\label{eq:mETC}
\end{equation}
through the ``radiative'' mechanism shown in Figure~\ref{fig:etcmass}.

There is no standard technicolor model, in large measure because the 
straightforward implementations of extended technicolor are 
challenged to reproduce the wide range of quark masses while avoiding 
flavor-changing-neutral-current traps. Consider $\abs{\Delta S} = 2$ 
interactions, 
\begin{equation}
	{\mathcal{L}_{\abs{\Delta S} = 2} = \frac{g_{\mathrm{ETC}}^{2} 		
	\theta_{sd}^{2}}{M_{\mathrm{ETC}}^{2}}(\bar{s}\Gamma^{\mu}d)(\bar{s}
	\Gamma^{\prime}_{\mu}d) + \cdots}
\end{equation}
The tiny $K_{\mathrm{L}}$-$K_{\mathrm{S}}$ mass difference, $\Delta 
M_{K} < 3.5 \times 10^{-12}\mev$, implies that ${{M_{\mathrm{ETC}}^{2}}/{g_{\mathrm{ETC}}^{2}
	\abs{\theta_{sd}}^{2}}}$ must be very great, but that makes it hard 
	to generate large enough masses for $c$, $s$, $t$, $b$. Multiscale 
	technicolor~\cite{Lane:1989ej} is a possible approach; it entails many 
	fermions in different technicolor representations, which implies many 
	technipions, among them light $\rho_{\mathrm{T}}$, $\omega_{\mathrm{T}}$, 
	$\pi_{\mathrm{T}}$.  

The generation of fermion mass is where all the experimental threats to
technicolor arise.  The rich particle content of ETC models generically
leads to quantum corrections that are in conflict with precision
electroweak measurements.~\footnote{Some complicated examples, 
 \cite{Dobrescu:1998ci} among others, have been offered of models that avoid
the particle-content catastrophe.} Moreover, if quantum chromodynamics
is a good model for the chiral-symmetry breaking of technicolor, then
extended technicolor produces flavor-changing neutral currents at
uncomfortably large levels.  We conclude that QCD must not provide a
good template for the technicolor interaction.  Keep in mind that, in
addressing the origins of fermion mass, extended technicolor is much
more ambitious than many implementations of global supersymmetry.  For
the current state of model building, see the lectures by
Lane~\cite{Lane:2002wv} and the comprehensive review article by Hill
and Simmons~\cite{Hill:2002ap}.

\subsection{Why is the Planck scale so large? {\protect \footnote{See 
the course by Antoniadis  at the 2001 European School~\cite{Antoniadis:2002jp}.}}}
\begin{figure}[t!]
\begin{center}
\includegraphics[width=10cm]{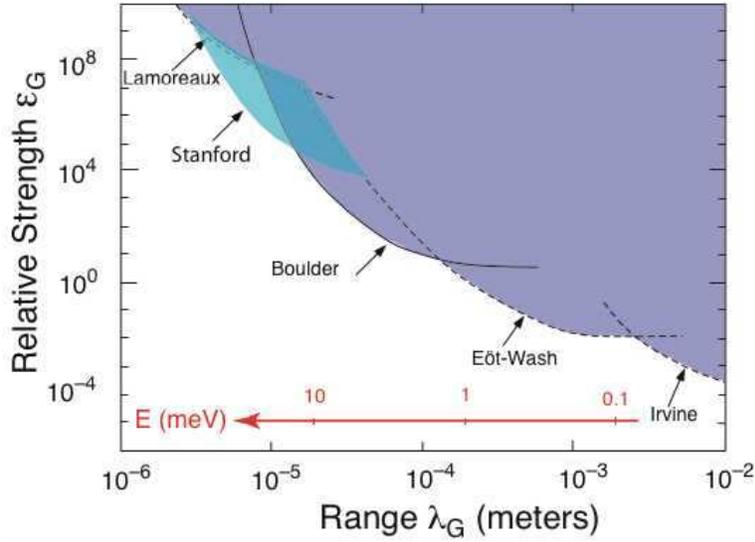}
\caption{95\% confidence level upper limits 
on the strength $\varepsilon_{\mathrm{G}}$ (relative to gravity) versus
the range $\lambda_{\mathrm{G}}$ of a new long-range force
characterized by {\protect{\eqn{eq:nonNewt}}}, from recent experiments
at Boulder~\cite{Long:2002wn}, Stanford~\cite{Chiaverini:2002cb}, and
Washington~\cite{Hoyle:2000cv}.  The region excluded by earlier
work\protect\cite{Hoskins:1985tn, Moscou, Lamoreaux:1997wh} lies above
the heavy lines labeled Irvine, Moscow and Lamoreaux, respectively. 
The red scale running from right to left shows the corresponding 
energy, $E = (\hbar c/\lambda_{\mathrm{G}})$.
\label{fig:nonNgrav}}
\end{center}
\end{figure}

Our understanding of gravity has been handed down to us by Newton and 
Einstein and the gods, whereas the electroweak theory is the work of 
mortals of our own time. It is therefore not surprising that
the conventional approach to new physics has been to extend the 
standard model to understand why the electroweak scale (and the mass 
of the Higgs boson) is so much smaller than the Planck scale.  A novel 
approach, only a few years old, is instead to 
\textit{change gravity} to understand why the Planck scale is so much 
greater than the electroweak scale~\cite{Antoniadis:1990ew, Lykken:1996fj, Arkani-Hamed:1998rs}.  Now, experiment 
tells us that gravitation closely follows the Newtonian force law down 
to distances on the order of $1\mm$.  Let us parameterize deviations 
from a $1/r$ gravitational potential in terms of a relative strength 
$\varepsilon_{\mathrm{G}}$ and a range $\lambda_{\mathrm{G}}$, so that
\begin{equation}
V(r) = - \int dr_{1}\int dr_{2} 
\frac{G_{\mathrm{Newton}}\rho(r_{1})\rho(r_{2})}{r_{12}} \left[ 1+ 
\varepsilon_{\mathrm{G}}\exp(-r_{12}/\lambda_{\mathrm{G}}) \right]\; ,
\label{eq:nonNewt}
\end{equation}
where $\rho(r_{i})$ is the mass density of object $i$ and $r_{12}$ is 
the separation between bodies 1 and 2.
Elegant experiments~\cite{Adelberger:2003zx}
using torsion oscillators and microcantilevers
imply bounds on anomalous gravitational interactions, as shown 
in Figure \ref{fig:nonNgrav}.  Below about a millimeter, the 
constraints on deviations from Newton's inverse-square force law deteriorate 
rapidly, so nothing prevents us from considering changes to gravity 
even on a small but macroscopic scale. Even after this new generation 
of experiments, we have only tested our understanding of 
gravity---through the inverse-square law---up to energies of 10~meV 
(yes, \textit{milli}-electron volts), some fourteen orders of 
magnitude below the energies at which we have studied QCD and the 
electroweak theory. Experiment plainly leaves an opening for 
gravitational surprises.

For its internal consistency, string theory requires an additional six 
or seven space dimensions, beyond the $3+1$ dimensions of everyday 
experience~\cite{Lykken:2000fp}.  Until recently it has been presumed that the extra 
dimensions must be compactified on the Planck scale, with a
compactification radius $R_{\mathrm{unobserved}} \approx
1/M_{\mathrm{Planck}} \approx 1.6 \times 10^{-35}\m$.  One new wrinkle 
is to consider that the \smgg\
standard-model gauge fields, plus needed extensions, reside on 
$3+1$-dimensional branes, not in the extra dimensions, but that 
gravity can propagate into the extra dimensions.

\begin{figure}[tb] 
\begin{center}
    \includegraphics[width=7cm]{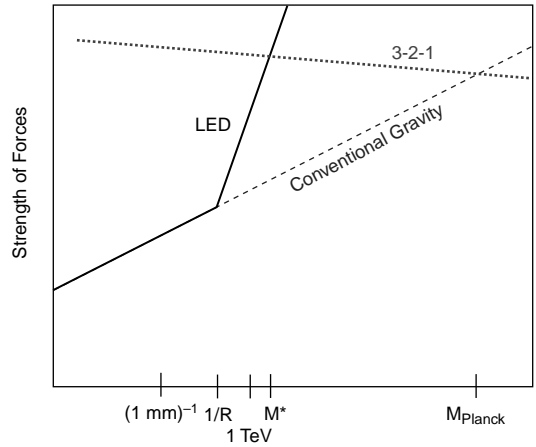}
    \caption{One of these extrapolations (at least!) is false.}
    \label{fig:false}
\end{center}
\end{figure}

How does this hypothesis change the picture?  The dimensional 
analysis (Gauss's law, if you like) that relates Newton's constant to 
the Planck scale changes.  If gravity propagates in $n$ extra 
dimensions with radius $R$, then
\begin{equation}
    G_{\mathrm{Newton}} \sim M_{\mathrm{Planck}}^{-2} \sim M^{\star\,-n-2}R^{-n}\; ,
    \label{eq:gauss}
\end{equation}
where $M^{\star}$ is gravity's true scale.  Notice that if we boldly 
take $M^{\star}$ to be as small as $1\tev$, then the radius of the extra 
dimensions is required to be smaller than about $1\mm$, for $n \ge 
2$.  If we use the four-dimensional force law to extrapolate the 
strength of gravity from low energies to high, we find that gravity 
becomes as strong as the other forces on the Planck scale, as shown 
by the dashed line in Figure \ref{fig:false}.  If the force law 
changes at an energy $1/R$, as the large-extra-dimensions scenario 
suggests, then the forces are unified at an energy $M^{\star}$, as 
shown by the solid line in Figure \ref{fig:false}.
What we know as the Planck scale is then a mirage that results 
from a false extrapolation: treating gravity as four-dimensional down 
to arbitrarily small distances, when in fact---or at least in this 
particular fiction---gravity propagates in $3+n$ spatial dimensions.  
The Planck mass is an artifact, given by $M_{\mathrm{Planck}} = 
M^{\star}(M^{\star}R)^{n/2}$. 

Although the idea that extra dimensions are just around the 
corner---either on the submillimeter scale or on the TeV scale---is 
preposterous, it is not ruled out by observations.  For that reason 
alone, we should entertain ourselves by entertaining the 
consequences.  Many authors have considered the gravitational 
excitation of a tower of Ka\l uza--Klein modes in the extra 
dimensions, which would give rise to a missing (transverse) energy 
signature in collider experiments~\cite{Hewett:2002hv}.

The electroweak scale is nearby; indeed, it is coming within 
experimental reach at the Tevatron Collider and the Large 
Hadron Collider.  Where are the other scales of significance?  In 
particular, what is the energy scale on which the properties of quark 
and lepton masses and mixings are set?  The similarity between the 
top-quark mass, $m_{t} \approx 175\gev$, and the Higgs-field vacuum 
expectation value, $v/\sqrt{2} \approx 174\gev$, encourages the hope 
that in addition to decoding the puzzle of electroweak symmetry 
breaking in our explorations of the \onetev, we might gain insight 
into the problem of fermion mass.  This is an area to be defined over 
the next decade.

\subsection{New Departures \label{subsec:newd}}
In the time available to us in San Miguel Regla, it hasn't been 
possible to examine all the promising attempts to complete the 
electroweak theory. Over the past five years, many stimulating ideas 
have been inspired by the notion that our spacetime is larger than 
four-dimensional. In many ways, the theoretical imagination has been 
liberated, and we have seen many proposals that---whatever their 
eventual fate---have been mind-expanding, showing us ways to think 
about problems that we had not known before.

Randall and Sundrum~\cite{Randall:1999ee} proposed a  
higher-dimensional solution to the hierarchy problem that uses one warped 
extra dimension to generate the weak scale from the Planck scale by 
means of the background metric. The simplest example entails two 
three-branes, one of which contains the standard-model fields. Other 
groups, including~\cite{Csaki:2002ur},
have now explored the idea that the Higgs boson might be 
regarded as an extra-dimensional component of a gauge 
boson. A related approach, which takes the Higgs 
boson as a pseudo-Goldstone boson, gives rise to the ``little Higgs'' 
models~\cite{Arkani-Hamed:2002qy} mentioned in passing at the beginning of 
\S~\ref{sec:beyondew} Universal extra dimensions at a 
compactification scale as low as a few hundred GeV survive current 
experimental constraints~\cite{Appelquist:2000nn}, and might be 
discovered at the Tevatron collider. Supersymmetry and dynamical 
symmetry breaking may be combined; a recent example is the so-called ``fat Higgs'' 
model~\cite{Harnik:2003rs}  to solve supersymmetry's little hierarchy 
problem: constraints from precision electroweak measurements reveal no  
evidence for new physics up to about $5\tev$, but if supersymmetry 
stabilizes the Higgs-boson mass, superpartner masses should be no more 
than about $1\tev$. Another suggestion~\cite{Csaki:2003dt} is that 
Ka\l uza--Klein (extra-dimensional) excitations of gauge fields could 
take the place of a Higgs scalar, at least in the framework of an 
effective field theory at low energies. An extra-dimensional 
generation of the QCD (chiral-symmetry-breaking) mechanism we 
discussed in \S~\ref{subsec:TC} might be all that is required to hide 
the electroweak symmetry through a top-quark seesaw~\cite{Cheng:1999bg}.

This incomplete list indicates the vitality of theoretical 
speculation about the mechanism for electroweak symmetry breaking and 
underscores the urgency of a thorough exploration of the 1-TeV scale. 
We have much, much more to accomplish than to find one particle!

Before summing up, let us consider another round of questions that 
have come to our attention:

\begin{questions}{Fourth}
    \addtocounter{bean}{26}
    \item Why is the world built of fermions, not bosons---i.e., quarks 
    not squarks, leptons not sleptons?

    \item Does gravity follow Newton's force law to very large distances? 
    \ldots\ to very short distances?

    \item Why is gravity weak?

    \item  Is $\mathcal{CPT}$ a good symmetry?
   
    \item  Is Lorentz invariance exact?
   
    \item  Are there extra dimensions?
   
    \item  Is local field theory the ultimate framework?
   
    \item  Can causality be violated?
   
    \item  What is dark matter?
   
    \item  What drives inflation?
   
    \item  What is the origin of dark energy?

\end{questions}

\vskip0.7cm
\section{Outlook \label{sec:outlook}}
In the midst of a revolution in our conception of Nature, we confront
many fundamental questions about our world of diversity and change. I 
find it instructive to organize our concerns around a small number of 
broad themes.

\vspace*{3pt}\noindent\textit{Elementarity.} Are the quarks and leptons 
structureless, or
will we find that they are composite particles with internal
structures that help us understand the properties of the individual
quarks and leptons? If the quarks and leptons do have internal 
structure, of what are they made? What is the compositeness scale, 
and how does it relate to other important scales?

\vspace*{3pt}\noindent\textit{Symmetry.} One of the most powerful lessons of the modern
synthesis of particle physics is that (local) symmetries prescribe
interactions.  Our investigation of symmetry must address the question
of which gauge symmetries exist (and, eventually, why).  Will we find additional 
fundamental forces, reflecting new symmetries?   We have
learned to seek symmetry in the laws of Nature, not (necessarily) in
the consequences of those laws.  Accordingly, we must understand how
the symmetries are hidden from us in the world we inhabit.  For the
moment, the most urgent problem in particle physics is to complete our
understanding of electroweak symmetry breaking by exploring the 1-TeV
scale.  This is the business of the experiments at  the Tevatron
Collider, the Large Hadron Collider, and an $e^{+}e^{-}$ linear 
collider.

\vspace*{3pt}\noindent\textit{Unity.} In the sense of developing explanations that apply not
to one individual phenomenon in isolation, but to many phenomena in
common, unity is central to all of physics, and indeed to all of
science.  At this moment in particle physics, our
quest for unity takes several forms.
$\Box$ 
First, we have the fascinating possibility of gauge coupling
unification, the idea that all the interactions we encounter have a
common origin and thus a common strength at suitably high energy.
$\Box$ 
Second, there is the imperative of anomaly freedom in the electroweak
theory, which urges us to treat quarks and leptons together, not as
completely independent species.  Both of these ideas are embodied, of
course, in unified theories of the strong, weak, and electromagnetic
interactions, which imply the existence of still other forces---to
complete the grander gauge group of the unified theory---including
interactions that change quarks into leptons.
$\Box$ 
The third aspect of unity is the idea that the traditional distinction
between force particles and constituents might give way to a unified
understanding of all the particles.  The gluons of QCD carry color
charge, so we can imagine quarkless hadronic matter in the form of
glueballs.  Beyond that breach in the wall between messengers
and constituents, supersymmetry relates fermions and bosons.
$\Box$ 
Finally, we desire a reconciliation between the pervasive outsider,
gravity, and the forces that prevail in the quantum world of our
everyday laboratory experience.

\vspace*{3pt}\noindent\textit{Identity.} We do not understand the
physics that sets quark masses and mixings.  Although we are testing
the idea that the phase in the quark-mixing matrix lies behind the
observed $\mathcal{CP}$ violation, we do not know what determines that
phase.  The accumulating evidence for neutrino oscillations presents us
with a new embodiment of these puzzles in the lepton sector.  At
bottom, the question of identity is very simple to state: What makes an
electron and electron, and a top quark a top quark?  Will we find new
forms of matter, like the superpartners suggested by supersymmetry?

\vspace*{3pt}\noindent\textit{Topography.} ``What is the
dimensionality of spacetime?''  tests our preconceptions and unspoken
assumptions.  It is given immediacy by recent theoretical work.  For
its internal consistency, string theory requires an additional six or
seven space dimensions, beyond the $3+1$ dimensions of everyday
experience.  Until recently it has been presumed that the extra
dimensions must be compactified on the Planck scale, with a
stupendously small compactification radius $R \simeq
M_{\mathrm{Planck}}^{-1} = 1.6 \times 10^{-35}\m$.
Part of the vision of string theory is that what goes on in even such
tiny curled-up dimensions does affect the everyday world: excitations
of the Calabi--Yau manifolds determine the fermion
spectrum.\footnote{For a gentle introduction to the goals of
string theory, see Ref.\ \cite{beegee}.}
We have recognized recently that Planck-scale compactification is 
not---according to what we can establish---obligatory, and that current
experiment and observation admit the possibility of dimensions not 
navigated by the strong, weak, and electromagnetic interactions that 
are almost palpably large.  A whole range of new experiments will help 
us explore the fabric of space and time, in ways we didn't expect just 
a few years ago~\cite{EDsearch}.

\vskip1cm

\noindent 
I hope that in the course of these lectures you have been prompted to 
ask yourselves many questions, and that you have enjoyed finding at 
least the beginning of ``a lifetime of homework.''
Many of the questions we have come upon together  qualify as great
questions. In the usual way of science,
answering questions great and small can lead us toward the answers to yet broader and more
cosmic questions.  I believe that we are on the threshold of a
remarkable flowering of experimental particle physics, and of
theoretical physics that engages with experiment. We can be quite 
confident, I think, that the way we think about the laws of nature will be very 
different in ten or fifteen years from the conception we hold today. Over the next decade
or two, we may hope to 

		\begin{tabularx}{\linewidth}{X X}
	
{Understand electroweak symmetry breaking}
		
		\textit{Observe the Higgs boson} 
		
		{Measure neutrino masses and mixings}
		
		\textit{Establish Majorana neutrinos ($\beta\beta_{0\nu}$)}
		
		{Thoroughly explore CP violation in $B$ decays} 
		
		\textit{Exploit rare decays ($K$, $D$, \ldots)}
		
		{Observe neutron EDM, pursue electron EDM}
		
		\textit{Use top as a tool}
		
		{Observe new phases of matter}
		
		\textit{Understand hadron structure quantitatively}
		
		{Uncover QCD's full implications}
		
		\textit{Observe proton decay} 
		
		{Understand the baryon excess}
		
		\textit{Catalogue matter and energy of the universe}
		
		{Measure dark energy equation of state}
		
		\textit{Search for new macroscopic forces}  
		
		{Determine GUT symmetry}
	&

	\textit{Detect neutrinos from the universe}
	
	{Learn how to quantize gravity}
	
		\textit{Learn why empty space is nearly weightless} 
		
		{Test the inflation hypothesis} 
		
		\textit{Understand discrete symmetry violation} 
		
		{Resolve the hierarchy problem}
		
		\textit{Discover new gauge forces}

		{Directly detect dark-matter particles} 
		
		\textit{Explore extra spatial dimensions} 
		
		{Understand the origin of large-scale structure} 
		
		\textit{Observe gravitational radiation} 
		
		{Solve the strong CP problem} 
		
		\textit{Learn whether supersymmetry is TeV-scale}
		
		{Seek TeV-scale dynamical symmetry breaking}
		
		\textit{Search for new strong dynamics}
		
		{Explain the highest-energy cosmic rays}
		
		\textit{Formulate problem of identity}
		
\end{tabularx}
\noindent		
\phantom{M}\hfill\ldots and learn the right questions to ask.

\vskip1cm
\noindent

\section*{ACKNOWLEDGEMENTS}
Fermi National Accelerator Laboratory is operated by
Universities Research Association Inc.\ under Contract No.\
DE-AC02-76CH03000 with the U.S.\ Department of Energy.
I thank Rob Harris for providing Figure~\ref{fig:CDF1364}.  Many
colleagues, especially Carl Albright, Uli Baur, Bogdan Dobrescu, Chris
Hill, Andreas Kronfeld, Joe Lykken, Uli Nierste, Yasonuri Nomura, Dave
Rainwater, and Maria Spiropulu, have made valuable contributions to the
double simplex. I thank Bogdan Dobrescu and Olga Mena Requejo for 
perceptive comments and suggestions on the manuscript.

It is a great pleasure to thank  CERN and the Centro Latino Americana 
de F\'{i}sica for organizing this Second Latin-American School of 
High-Energy Physics. I am especially grateful to School 
Director Egil Lillest\o l, School Secretary Danielle M\'{e}tral, and 
Local Director  Alberto Sanchez-Hern\'{a}ndez for their remarkable 
efforts to ensure a superb intellectual and social environment. I 
salute the Discussion Leaders for their energetic contribution to the 
school's atmosphere, and heartily thank the students for their 
curiosity and engagement.

\bibliography{cqSMRproc}

\end{document}